\PassOptionsToPackage{pdfpagelabels=false}{hyperref} 
\documentclass[fleqn,usenatbib]{mnras}

\usepackage{newtxtext,newtxmath,ulem}
\usepackage[T1]{fontenc}
\usepackage{ae,aecompl}

\usepackage{graphicx}	% Including figure files
\hypersetup{draft}
\title[Theoretical clues about dust accumulation and galaxy obscuration at high and low redshift]{Theoretical clues about dust accumulation and galaxy obscuration at high and low redshift}

\author[J. H. Barbosa-Santos et al.]{
J. H. Barbosa-Santos,$^{1}$\thanks{E-mail: jullian.santos@usp.br (JHBS)}
Gast\~ao B. Lima Neto,$^{1}$
and Amancio C. S. Fria\c{c}a$^{1}$
\\
$^{1}$Instituto de Astronomia, Geof\'isica e Ci\^encias Atmosf\'erica, USP, Rua do Mat\~ao, 1226 - Cidade Universit\'aria, 05508-090, S\~ao Paulo, SP, Brazil}

\date{Accepted XXX. Received YYY; in original form ZZZ}

\pubyear{2015}
\begin{document}

\label{firstpage}
\pagerange{\pageref{firstpage}--\pageref{lastpage}}
\maketitle

\begin{abstract}
Since the epoch of cosmic star formation peak at $z \sim 2$, most of it is obscured in high mass galaxies, while in low mass galaxies the radiation escapes unobstructed.
% while low mass ones are mojority unobscured.
During the reionization epoch, the presence of evolved, dust obscured galaxies are a challenge to galaxy formation and evolution models. 
By means of a chemodynamical evolution model, we investigate the star formation and dust production required to build up the bulk of dust in galaxies with initial baryonic mass ranging from $7.5 \times 10^{7}$~M$_\odot$ to $2.0 \times 10^{12}$~M$_\odot$.  
The star formation efficiency was also chosen to represent the star formation rate from irregular dwarf to giant elliptical galaxies. We adopted a dust coagulation efficiency from \citep[][Case A]{dwek1998evolution} as well as a lower efficiency one (Case B), about five times smaller than Case A.
All possible combination of these parameters was computed, summing forty different scenarios.
We find that in high stellar formation systems the dust accretion in ISM rules over stellar production before the star formation peak, making these systems almost insensible to dust coagulation efficiency. In low star formation systems, the difference between Case A and B lasts longer, mainly in small galaxies. Thus, small irregular galaxies should be the best place to discriminate different dust sources.
In our observational sample, taken from the literature, dust-to-gas ratio tends to be more spread only than dust mass, for both stellar mass and star formation rate.
Dust-to-gas vs. dust-to-star diagram is a good tracer for both galaxy and dust evolution, due to the link between gas, star, dust and star formation rate. However, the model do not constrain simultaneously all this quantities.
The new generation facilities (such as JWST, ELT, VLT and SPICA) will be indispensable to constrain dust formation across the cosmic time.
\end{abstract}


%%%%%%%%%%%%%%%%%%%%%%%%%%%%%%%%%%%%%%%%%%%%%%%%%%

%%%%%%%%%%%%%%%%% BODY OF PAPER %%%%%%%%%%%%%%%%%%

\section{Introduction}
\label{sec:Intro}

Cosmic dust is a key component of the interstellar medium (ISM) of galaxies. It provides a catalytic surface to chemical reactions, in special to produce H$_2$ \citep{1963ApJ...138..393G, 1990ARA&A..28...37M}. On the other hand, the star formation rate (SFR) seems to correlate better with H$_2$ density than HI or total gas (H$_2$ + HI) density \citep{1999AJ....118..670R, 2002ApJ...569..157W, 10.1093/mnras/stx889}. This correlation implies that the amount of dust in a galaxy may play an important (maybe central) role to regulate star formation \citep{lupi2017natural}, and can even contribute to star formation quenching by the lack of a site for molecular gas formation \citep{2009ApJ...693..216K}. Also, the typical mass of the final fragments in star-forming clouds can be regulated by dust cooling \citep{whitworth1998gas} and dust grains can also be a major component to accomplish the transition from stellar Population III to II \citep{chiaki2014supernova}. 

Dust is also the greatest contributor to starlight extinction and reddening in the ISM due to absorption and scattering of the stellar light, strongly affecting the optical and ultraviolet (UV) observations \citep{draine1984optical}. The absorbed light by dust is re-emitted in the infrared (IR) as thermal radiation (modified black body), reshaping the galaxy spectral energy distribution (SED) \citep{silva1998modeling} and, for more extreme cases, shifting the SED peak to the Far-IR (FIR). The extinction curve and IR emission depend on the dust composition and distribution in the host galaxy, the grain geometry and size distribution, and the incident radiation field.

How dust is formed by heavy elements and how much mass is locked in grains depends on the metallicity of the host galaxy. The amount of dust may be thus used to infer the evolution of high-$z$ galaxies such as Damped Lyman-$\alpha$ (DLA) systems \citep{vladilo2002chemical,Gioannini2017}. At low-$z$, we observe, for instance in the Milk Way, about half of the heavy elements produced locked in solid particles \citep{ferrara2016problematic}. 

The grains are believed to form mainly in asymptotic giant branch stars (AGBs) and core-collapse supernovae (CCSNe) \citep{dwek1998evolution, tielens1998interstellar}. There are some theoretical works  \citep[e.g.,][]{nozawa2011formation} that suggest that dust may be also produced in SN~Ia, but the observational evidence is weak \citep{gomez2012dust}. Furthermore, grain destruction in both SNe are expected, but this is not completely clear.

The observation of dusty evolved galaxies at $z \gtrsim 6$, in the reionization era, constrains the maximum time spent in dust enhancement to some few hundreds of Myr \citep{cooray2014hermes, knudsen2016merger,magdis2017dust}, which is a short time interval compared to the time evolution of intermediate-mass stars (IMS), and therefore, the enrichment of the ISM. Thus, CCSNe can be more efficient to destroy dust grains than to build it \citep{dwek1998evolution, calura2008cycle}.

Massive star forming regions and starbursts galaxies (SBGs) are often observed in IR, enshrouded in dust cocoons, since the strong UV emission of young stars is reprocessed in dust clouds \citep{silva1998modeling,farrah2008nature,bourne2017evolution}. Those objects are called dust obscured galaxies (DOGs).
Active galaxies nuclei (AGNs) are also frequently obscured \citep{chang2017infrared}, causing sometimes a misclassification between both phenomena. Indeed, both starburst (SB) and AGN can happen at the same time \citep{farrah2002submillimetre,Farrah_2005}. 

DOGs are crucial objects to understand galaxy assembly and evolution, compounding the ultra-luminous infrared galaxies (ULIRGs) \citep{rieke1972infrared,sanders1996luminous} and sub-millimetric galaxies (SMG). They are the most luminous galaxies and the most intense stellar nurseries in the Universe \citep{casey2014dusty} and it is possible that almost all normal and giant galaxies in the local Universe have experienced this phase during their growth \citep{Canalizo_2001}.
The cosmic peak for star formation density, AGN activity and ULIRGs density take place almost concomitantly around $z \sim 2.0$--2.5  \citep{farrah2008nature,casey2014dusty,caputi2007infrared}.

High mass galaxies (even normal star forming) have most of their star formation obscured by dust, reaching $\sim 90$ \% in galaxies with $\log(M$/M$_\odot) = 10.5$, while low mass ones tends to have most of the star formation unobscured \citep{whitaker2017constant}. This pattern seems to be present in galaxies up to $z = 2.5 \sim 3.0$ \citep{whitaker2017constant,magdis2017dust}, when the cosmic star formation rate peaks. The obscured star formation rate in galaxies beyond $z \sim 3.0$, as well as the DOGs cosmic density in the same epoch, are not yet well known and it is possibly underestimated due to observational selection bias, which prioritizes Ly$\alpha$ emitters \citep{knudsen2016merger}. As a consequence, star formation history estimators, even around $z \sim 3.0$, can be misleading and heavy star forming objects can be systematically overlooked \citep{Koprowski_2016,coppin2015scuba}.

In this work, we investigate the dependence of dust and star formation rates on the build up of dust in galaxies.
%SFR and dust production rate on the built up of dust in galaxies . 
We also investigated the dominant process to produce dust during the reionization epoch. We use chemodynamical galaxy simulations for a variety of star formation history and two dust production efficiencies, covering a mass range from dwarf, to giant elliptical galaxies. The simulation results in low- and high-$z$ are compared with observations available in the literature. 

This paper is organised as follows: in section \ref{sec:setup} we describe the galaxy evolution models adopted here and their setup.
%and in section \ref{sec:Models} we give a description of the setup for each galaxy model.  
The models for dust evolution are explained in section~\ref{sec:Dust}, and in section~\ref{sec:Observation} we describe the observational data gathered for this work. The results are presented in section~\ref{sec:Results} and discussed in section~\ref{sec:Discussion}. Finally, in section~\ref{sec:Conclusion} we summarise our conclusions. We adopted a standard $\Lambda$CDM cosmology with $H_0 = 70\,$km~s$^{-1}$~Mpc$^{-1}$, $\Omega_\Lambda = 0.7$, and $\Omega_M = 0.3$. 
%% the most important conclusions of this work is systematised. 

%and the medium case is seem when log(M$/$M$_\odot$) $=$ 9.4, where a half of their star formation is obscured \citep[ver][]{whitaker2017constant}. Galaxies with log(M$/$M$_\odot$) $=$ 9.4 have among a half of its star formation obscured by dust, while galaxies with log(M$/$M$_\odot$) $=$ 10.5 have more than 90 \% obscureced, and low mass galaxies is mainly no obscurced. This relation is observad till \textit{redshift} $z \sim$ 2.5 \citep[see][]{whitaker2017constant}.

\section{Numerical setup}
\label{sec:setup}

Is this work we have adopted the chemodynamical model from \citet{Friaca98} (hereafter FT98), with improvements described in \cite{Lanfranchi2003} and \citet{friacca2017tracing}. 
We carried out a total of forty different simulations by combining different galaxy masses (five values) with star formation efficiency (four models), and dust production efficiency (two models).
In the following, we briefly describe the most relevant features from FT98, while in section~\ref{sec:Dust} we describe our dust production prescription. 
The common parameters for all galaxy models (IMF, stellar yields, etc.) are discussed in section~\ref{sec:chem}, while the parameters of each particular galaxy model are discussed in section \ref{sec:Models}.

\subsection{Chemodynamical model}
\label{sec:chem}
The FT98 model is a multi-zonal, 1D chemodynamical code for spheroidal
galaxies including gas, stars, and a dark matter (DM) halo.
The model allows for mass inflow and outflow. Gas and stars (baryons) exchange material between themselves at any moment, in processes such as star formation, SNe, and stellar winds.

The simulations begin with an initial baryonic mass, $M_{G,0}$, completely in gaseous form, i.e., $M_{Gas} = M_{G,0}$, stellar mass, $M_* = 0$, and metal free ($X = 7.6$, $Y = 0.24$, and $Z = 0$).
The DM halo is assumed static, with mass fixed as $5.6 \times M_{G,0}$. The model adopt a halo density profile $\propto [1+(r/r_h)^2]^{-1}$, where $r$ is the radial distance to the centre, and $r_h$ is the core radius. The simulations are truncated at the tidal radius, $r_t$, fixed as $r_t = 28 r_h$.

The total baryonic mass is the sum of gaseous and stellar masses ($M_G = M_{Gas} + M_*$) and initially follows the DM halo density profile. 
The gas is converted in stars with a specific SFR, $\nu$, described by the equation:
\begin{equation}
\nu(r,t) =  \nu_{0} \left(\frac{\rho}{\rho_0}\right)^{n_{\rm sf}} \times  b\, , \label{TFE}
\end{equation}
\begin{equation}
b = \left\{\!\!
\begin{array}{rl}
\left(1 + t_{\rm dyn} \max[0,\, \nabla \cdot u)]\right)^{-1} \cdot \left(1 +t_{\rm c}/t_{\rm dyn} \right)^{-1} ; & t_c > t_{\rm dyn} \, ,
\\
\left(1 + t_{\rm dyn} \max[0,\, \nabla \cdot u)]\right)^{-1} ; & t_c \leq t_{\rm dyn} \, , \nonumber
\end{array}
\right.
\end{equation}
where $\rho$ is the gas density, $\rho_0$ the initial average gas density inside $r_h$, $\nu_{0}$ the star formation efficiency, $n_{\rm sf}$ the power-law exponent, assumed to be $1/2$ \citep{larson1974dynamical}, $u$ the velocity field. The cooling time is given by $t_{\rm c}= (3/2)k_B T/ \mu m_H \Lambda(T)\rho$, where $k_B$ is the Boltzmann constant, $m_H$ is the hydrogen atomic mass, $T$ is gas the temperature, $\mu$ is the mean molecular mass, and $\Lambda(T)$ is the cooling function. We take the dynamical time as the collapse time-scale, $t_{\rm dyn} = \sqrt{3\pi/16G\rho}$, where $G$ is the gravitational constant. The $b$ factor inhibits star formation in an expanding gas ($\nabla \cdot u > 0$).

The produced stars follow a King distribution \citep{king1962structure}, where the core radius, $r_c$, and the central star density, $\rho_{*0}$ of the distribution are related to the stellar velocity dispersion, $\sigma_*$, by the virial condition $4\pi G \rho_{*0} r_c = 9\sigma^2_*$. 
$M_*$ and $\sigma_*$ follow the classical Faber-Jackson relation $\sigma_* \propto M_{*}^{1/4}$ \citep{faber1976velocity}.

We have adopted the Salpeter stellar initial mass function (IMF) \citet{salpeter1955luminosity}. This IMF is in better agreement with the colour-magnitude diagram of of elliptical galaxies \citep{pipino2004photochemical}, which the FT98 model is tailored to reproduce. 
% being more suited to the spherical symmetry of FT98. 
\citet{calura2009evolution} also apply the Salpeter IMF to dwarf galaxies while investigating the galactic mass-metallicity relation. Since we cover a vast range of $M_{G,0}$ and SFR efficiency 
% representing a wide variety of galaxies 
(see section \ref{sec:Models}), we prefer to use a single IMF to simplify comparisons among individual models.

The chemical enrichment sources are CCSN, SN~Ia, and IMS (assumed in the 0.8--8 M$_\odot$ range and evolves as AGB). The yields adopted here depend on stellar mass and metallicity. \citet{woosley1995evolution} yields was adopted for CCSNe, calculated for stars with mass $M = 12$, 13, 15, 18, 20, 22, 25, 30, 35, and 40 M$_\odot$ and metallicity $Z/Z_\odot$ = 0, $10^{-4}$, $10^{-2}$, $10^{-1}$ and 1. For SNe Ia, we adopted the W7 ($Z=Z_\odot$) and W70 ($Z = 0$) models from \citet{iwamoto1999nucleosynthesis}, leaving no remnant after the explosion. The IMS yields came from \citet{van1997new}, case $\eta_{\rm AGB}  = 4$, with $Z = 0.001$, 0.004, 0.008, 0.02, and 0.4.

The evolution of He, C, N, O, Mg, Si, S, Ca and Fe was computed solving the chemical enrichment equation \citep{matteucci1987chemical} and taking into account the delay caused by the lifetime of stars (but assuming instantaneous mix for stellar ejecta).

The energy output from CCSNe, SNe Ia and quiescent stellar mass loss (AGBs, planetary nebulae and stellar winds) is injected into the gas. The feedback heats the gas, regulates the SFR, and can drive outflow events.

\subsection{Particular models setup}
\label{sec:Models}

Each particular model is described by its baryonic initial mass, $M_{G,0}$, and for its star formation history. For each $M_{G,0}$, a DM core radius, $r_h$, was set following FT98 and \citet{Lanfranchi2003}. In this work, five $M_{G,0}$ models were computed, spanning from $5 \times 10^{7}$ to $2 \times 10^{12}$ M$_\odot$. The $M_{G,0}$ with corresponding $r_h$ values are shown in table~\ref{tab:galaxymod}.

The $M_{G,0}$--$r_h$ relation in this work follows FT98 for $2 \times 10^{11}$~M$_\odot$ and  $2 \times 10^{12}$~M$_\odot$, and with \citet{Lanfranchi2003} for $10^{9}$~M$_\odot$ model, and are similar to \citet{calura2009evolution}. The main difference is for $M_{G,0} = 10^{10}$~M$_\odot$  and $5 \times 10^{7}$ M$_\odot$. Both $M_{G,0} = 10^{9}$~M$_\odot$ and $M_{G,0} = 10^{10}$~M$_\odot$ were modelled by \citet{Lanfranchi2003} with $r_h = 1.0$ kpc and $r_t = 14 r_h$, appropriate to model dwarf galaxies.

We adopted $r_h = 2.5$ kpc for the $M_{G,0} = 10^{10}$~M$_\odot$ model, to take into account more extended objects with more extended star formation history. None of the previously cited works considered such a small galaxy as $M_{G,0} = 5 \times 10^{7}$~M$_\odot$, the lowest mass being $10^8$~M$_{\odot}$ in \citet{calura2009evolution} and a luminous radius of $R_{lum} = 1$~kpc. They used a one-zone model, while our model is more extended with lower density in the outskirt.

\begin{table}
	\centering
	\caption[]{Masses ($M_{G,0}$) and corresponding sizes ($r_h$) used in the
	Simulated galaxy models.}
	\label{tab:galaxymod}
	\begin{tabular}{cccccc} % four columns, alignment for each
		\hline
%		Galaxy Models  \\
%		\hline
		$M_{G,0}$/M$_\odot$ &  $5 \times 10^{7}$ & $10^{9}$ & $10^{10}$ & $2 \times 10^{11}$ &  $2 \times 10^{12}$ \\
		$r_h$ (kpc) & 0.4 & 1.0 & 2.5 & 3.5 & 10.0  \\
		\hline
	\end{tabular}
\end{table}

\begin{table}
	\centering
	\caption[]{Star formation efficiency, $\nu_0$, adopted in this work. Each $\nu_0$ was combined with each $M_{G,0}$ in table \ref{tab:galaxymod}, resulting in twenty different galaxy evolution models.}
	\label{tab:galaxysfr}
	\begin{tabular}{ccccc} % four columns, alignment for each
		\hline
%		Star formation efficiency  \kern-2em  \\
%		\hline
		$\nu_0$ (Gyr$^{-1}$) & 0.1 & 1.0 & 5.0 & 10.0 \\
		\hline
	\end{tabular}
\end{table}

The star formation history is parameterized by $\nu_0$, set as 0.1, 1.0, 5.0 and 10 Gyr$^{-1}$ (table~\ref{tab:galaxysfr}). For elliptical galaxies, FT98 adopted $\nu_0 = 10~$~Gyr$^{-1}$ for $2 \times 10^{11}$ and $2 \times 10^{12}$ M$_\odot$ models. \citet{calura2009evolution} adopted 3, 10  and 20 Gyr$^{-1}$  for $10^{10}$, $10^{11}$ and $10^{12}$ M$_\odot$ models, respectively. \citet{Gioannini2017} adopted  1.0~Gyr$^{-1}$ for a galaxy with 10$^{9}$ M$_\odot$ and \citet{Lanfranchi2003} considered 1.0 and 3.0~Gyr$^{-1}$ for $10^{9}$ M$_\odot$ and 10$^{10}$ M$_\odot$, respectively. \citet{calura2009evolution} set between 0.3--2~Gyr$^{-1}$, for spiral galaxies and 0.001--0.5~Gyr$^{-1}$ irregular dwarfs galaxies, while \citet{Gioannini2017cosmic} adopted 1.0--3.0~Gyr$^{-1}$ for spiral and 0.01 --2.0~Gyr$^{-1}$ for irregular galaxies.

Even known that $M_*$, SFR and metallicity form the "fundamental metallicity relation" \citep[see][for a more general discussion]{mannucci2010fundamental},
we combined all $M_{G,0}$ and $\nu_0$ values to investigate the role of each one. Another advantage is that we can mimic several galaxy types, evolutionary stages and star formation history, allowing the comparison with both low- and, in special, high-$z$ galaxies \citep[][suggested that the Hubble sequence, from early to late type, can be approximate as a sequence of decreasing star formation efficiency]{calura2006cosmic}. For these reasons, we avoided the use of extreme high and low star formation efficiency, and the range 0.1 -- 10 Gyr$^{-1}$ was chosen.

\section{Dust prescription}
\label{sec:Dust}

The dust evolution prescription adopted in this work is mainly based on \citet{dwek1998evolution} (hereafter D98) and \citet{Gioannini2017} formulation. The dust mass balance is affected by three major components: (i) stellar sources and SNe events, (ii) accretion in cold ISM, and (iii) loss in galactic wind and star formation. As star and SN sources can both produce and destroy the dust grains, the description will be based in processes that enhance or diminish the total dust mass.
 
Defining $D_A \equiv \rho_{\rm Dust}(A,r,t)$, the dust mass density of element $A$, at galactic radius $r$ and an instant $t$, the dust evolution balance is described by the equation:
\begin{equation}
\dot{D}_A = \dot{D}_A^{\rm prod} + \dot{D}_A^{\rm acc} - \dot{D}^{\rm lost}_A \, ,
\label{eq:evolutiondust}
\end{equation}
where each term on the right-hand side means, in order, the dust produced, accreted and lost throughout the galaxy evolution. We computed the evolution of carbonaceous and silicate dust grains ($A=$ \{C, Si\}), the most abundant ones. In the following, we review the most relevant features and parameters of Eq.~\ref{eq:evolutiondust}.
%, following the same order of the right-hand side of this equation.

\subsection{Dust production}
\label{sec:Dust-prod}

Following D98, the grains can be produced by stellar winds of ISM during the AGB phase, CCSNe, and SNe Ia events and then delivered to the ISM. The dust production is described by the sum of these three sources, resulting in the following equation:
\begin{equation}
\label{eq:prod-eq}
\dot{D}_A^{\rm prod} = \dot{D}_A^{w} + \dot{D}_A^{\rm CCSNe}  + \dot{D}^{\rm SNIa}_A \, .
\end{equation}

In equation \ref{eq:prod-eq}, the dust sources follow the same stellar threshold than the chemical evolution model (section \ref{sec:chem}). The contributing single stars are the evolved IMS phase, as AGB, and the death of high mass stars, as CCSNe, while binary systems evolves into SNe Ia. The stellar dust yields rely on the relative mass amount between carbon and oxygen and in the dust condensation efficiency, for a $i$ production process, $\delta^i(A)$. 

The $\delta^i(A)$ represents the balance between the $A$ element amount available to compose grains and the amount that will in fact end as grain. For a complete production, without destruction, $\delta^i(A)=1$, otherwise  $\delta^i(A)<1$. This quantity can rely on the mass of the progenitor, its metallicity and, for SNe, the surrounding ISM density \citep[see][for a more detailed discussion]{Piovan2011}.

An IMS with mass $M$ produces a dust mass $M_{\rm Dust}(A,M)$ relying in the carbon and oxygen numbers of atoms ($N_C$ and $N_O$, respectively). The available carbon will mainly form CO and the surplus will be converted in grains. If the $N_C$ is higher than the $N_O$, silicates cannot be formed. But if has more $N_O$ than $N_C$, the excessive oxygen will form silicates and carbonaceous grain will not be formed.

Then, for $M \leq 8$~M$\odot$ we have and $\delta^w(A)$ the condensation efficiency for stellar winds:
\begin{enumerate}

\item Case C/O$ > 1$ in the ejected material
\begin{eqnarray}
	M_{\rm Dust}(C,M) &  = & \delta^w(C)\ \left[M_{\rm ej}(\rm C,M)-\frac{3}{4}\ M_{ej}(O,M)\right] \nonumber \\
 M_{\rm Dust}(Si,M) & = & 0 \label{1a}
\end{eqnarray}
\item Case C/O $<$ 1 in the eject material
\begin{eqnarray}
	M_{\rm Dust}(C,M) &  = & 0 \nonumber \\
 M_{\rm Dust}(Si,M) & = & \delta_{\rm cond}^w(Si)\ M_{ej}(\rm Si,M) \, . \label{1b}  
\end{eqnarray}

\end{enumerate}

In CCSNe, $M_{\rm Dust}(A,M)$ relies only in the mass of the $A$ element ejected in the explosion and in the condensation efficiency $\delta^{CC}(A)$. The dust production in stars with $M > 8$~M$_\odot$ is:
\begin{eqnarray}
 M_{\rm Dust}(C,M)  & = & \delta^{CC}(C) \ M_{\rm ej}(C,M)  \nonumber \\
 M_{\rm Dust}(Si,M) & = & \delta^{CC}(Si)\ M_{\rm ej}(Si,M) \, .\label{eq:SNe-prod}
\end{eqnarray}

The dust production formulation for SNe Ia is quite similar to \ref{eq:SNe-prod}, changing $\delta^{CC}(A)$ to $\delta^{Ia}(A)$. Hereafter the dust production will be described by the condensation efficiencies $\Delta_A \equiv (\delta^w(A)$, $\delta^{Ia}(A)$, $\delta^{CC}(A))$. 

Many values of $\Delta_C$, as dust evolution recipes, are available in the literature 
%and are submitted to different constrains 
\citep{ferrarotti2006composition, calura2008cycle, zhukovska2008evolution, Piovan2011, Gioannini2017}. D98 uses the constants $\Delta_C = (1.0,\, 0.5,\, 0.5)$ for carbon, and $\Delta_{Si} = (1.0,\, 0.8,\, 0.8) $. This value was chosen to reproduce the dust mass in the Galaxy at the present age, but it is a controversial choice. \citet{zubko2004interstellar} find that this formulation can lead to a shortage of iron (and maybe silicon) in grain form. The $\Delta_A$ from \citet{Piovan2011} is sensible to both stellar mass and metallicity, implying in a cosmic evolution of $\Delta_A$ and more sensibility to IMF, while \citet{Gioannini2017} consider SNe Ia as a negligible dust source.

\begin{table}
	\centering
	\caption[]{Dust condensation efficiency adopted in this work. Each adopted $\Delta_A$ was combined with each particular galaxy evolution model, summing forty runned models for this work.}
	\label{tab:Delta-values}
	\begin{tabular}{cccc} % four columns, alignment for each
		\hline
		& \multicolumn{3}{c}{Condensation efficiency}   \\
		\hline
		Case A (D98) & $\delta^w(A)$ & $\delta^{Ia}(A)$ & $\delta^{CC}(A))$ \\
		\hline
		$\Delta_C$    & 1.0 & 0.5 & 0.5 \\
		$\Delta_{Si}$ & 1.0 & 0.8 & 0.8 \\
		\hline
		Case B  & $\delta^w(A)$ & $\delta^{Ia}(A)$ & $\delta^{CC}(A))$\\
		\hline
		$\Delta_C$    & 0.1 & 0.0 & 0.1 \\
		$\Delta_{Si}$ & 0.1 & 0.0 & 0.1 \\
		\hline
	\end{tabular}
\end{table}

Despite the controversy about dust production in SNe Ia, we adopted here two sets of constants $\Delta_A$: the D98 set (Case A) and a low efficient set (Case B), that is described by $\Delta_{\rm C} = (0.1,\, 0.0,\, 0.1 )$ and $\Delta_{\rm Si} = (0.1,\, 0.0,\, 0.1)$. How the main interest here is to investigate dust evolution in galaxies spanning a large range of mass and redshifts, we prefered to adopt a classical recipe. 

\citet{Gioannini2017} dust evolution model adopted $\delta^{i}(A)$ from \citet{Piovan2011}. Their $\delta^{i}(A)$ relies in the progenitor stellar mass, metallicity, and also in ISM density for CCSNe. We choose to adopt a constants set of $\delta^{i}(A)$ to reduce the dust production dependence with metallicity and IMF. It is worth to stress that \citet{Piovan2011} $\delta^{i}(A)$ is, in average, lower than Case A (the possible exception is the CCSNe in a environmental density $n_H=0.1$~cm$^{-1}$, but \citet{Gioannini2017}, for example, use $\delta^{CC}(A)$ for  $n_H=1.0$~cm$^{-1}$, higher than the Case B. Dust production Cases A and B parameters are given in table~\ref{tab:Delta-values}.
%
%\citep[][for example, uses the $\delta^{CC}(A)$ from $n_H=1.0$~cm$^{-1}$ case]{Gioannini2017} and higher than the Case B. Dust production Case A and B are highlighted in table \ref{tab:Delta-values}.

\subsection{Dust Accretion}
\label{ssec:Accr}

The grains produced by stellar sources are processed in the ISM, altering their mass, size and composition \citep{asano2013determines}. The most important processes for dust grain growth are coagulation and accretion. While the former changes only the grain size distribution, due to grain-grain interaction, the latter enhances the total mass locked in dust form, capturing elements available in the ISM gas. In fact, accretion efficiency is sensible to the grain size distribution, but following \citet{Gioannini2017}, here we adopt the single grain size approximation, making dust insensible to coagulation.

Following \citet{hirashita2000dust}, grain accretion from ISM gas is effective only in cool gas medium and its efficiency depends of the dust amount and on the accretion time-scale, $\tau_{\rm acc}$. The dust accretion rate is described by:
\begin{equation}
\label{eq:accretion}
\dot{D}_A^{\rm acc} = \frac{D_A}{\tau_{\rm acc}} \, .
\end{equation}
This time-scale relies on the cool gas fraction and on the amount of $A$ element available for accretion, given by:
\begin{equation}
\tau_{\rm acc}= \tau_g/(X_{cl} \, \chi_A) \, ,
\end{equation}
where $X_{cl}$ is the cool gas fraction and $\chi_A=(1-f_A)$, where $f_A$ is the ratio between the dust and the amount of element $A$ in the gas. The cool gas mass fraction is not the same in galaxies, D98 and \citet{silva1998modeling} suggest $X_{cl} = 0.5$. \citet{lianou2016dustier} elliptical sample has a mean $X_{cl} \sim 0.4$ and \citet{10.1093/pasj/psx041} found about 0.7, for interacting galaxies, and 0.5 for disk isolated ones. We assume a constant $X_{cl} = 0.5$ in the simulations, following D98.

The characteristic dust growth time-scale, $\tau_g$, is given by:
\begin{equation}
\label{eq:tau_g}
\tau_g = 2.0\times 10^7  \left[ \left(  \dfrac{a}{0.1\mu {\rm m}} \right)  \left( \dfrac{n_H}{100 {\rm cm}^{-3}}\right)^{-1} \left( \dfrac{T}{50 {\rm K}}\right)^{-1/2}  \left( \dfrac{Z}{0.02} \right)^{-1}   \right] \mbox{yr},
\end{equation}
for gas number density, $n_H$, mean grain radius, $a$, and cool gas temperature, $T$. We assume the same values as \cite{Gioannini2017}, 100~cm$^{-3}$, $0.1 \mu$m and 50 K, respectively, while the metallicity, $Z$, evolves during galaxy evolution.

\subsection{Dust Destruction}
\label{ssec:Dest}

%As well as grain can accrete and coagulate in ISM, it also can be destroyed in ISM.
Dust grains can also be destroyed in the ISM.
Shattering process is important to dust grain size distribution, but it does not affect the dust mass balance \citep{asano2013determines}. Following D98, we consider only dust destruction by sputtering in SNe events.

The dust composed by element $A$ at galactocentric distance $r$, and at instant $t$ is destroyed due to grain-grain collision, $m_{\rm dest}(A,r,t)$, described as:
\begin{equation}
\left[\frac{d\rho(A,r,t)}{dt} \right]_{\rm SNR}=m_{\rm dest}(A,r,t) \ {\cal R}_{SN}(r,t) =\dfrac{D_A}{\tau_{\rm SNR}(r,t)} \, .
\end{equation} 
In this equation, ${\cal R}_{SN}(r,t)= {\cal R}_{SNIa}(r,t) + {\cal R}_{CCSNe}(r,t)$, is the combined SNIa and CCSNe rate in pc$^{-3}$ Gyr$^{-1}$. The time-scale for dust destruction in SN events is given by $\tau_{\rm SNR}$.

We assume that $m_{\rm dest}$ is proportional to the dust-to-gas ratio and to $M_{\rm SNR}$, the total mass swept by one supernova event. The proportionality parameter is the dust destruction efficiency $\epsilon$: 
\begin{equation}
m_{\rm dest}(A,r,t) = \left(\frac{D_A}{\rho_{\rm ISM}(A,r,t)}\right) \epsilon M_{\rm SNR} \label{mdest}
\end{equation} 
and the time-scale to grain destruction is then given by:

\begin{equation}
\tau_{\rm SNR}(A,r,t) =  \left(\epsilon\ M_{\rm SNR}\right)^{-1}\ \left[\frac{\rho_{\rm ISM}(r,t)}{ {\cal R}_{SN}(r,t)}\right]
\end{equation}
the $\epsilon$ parameter really weakly in the ISM density and $M_{\rm SNR}$ do not rely in the density, so the quantity $\epsilon M_{\rm SNR}$ can be assumed constant in the model.  
Dust is also depleted in SF process, coupled with the gas consumed in the process. SF do not changes metallicity due to dust metals been converted in gas phase again. 

Finally, dust loss is also due to galactic wind as a result of SN outflowns.
%The other important source of dust loss is the galactic wind. In a chemodynamical approach, SNe feedback generate outflows, expelling gas and dust from the galaxy. 
Models with high star formation efficiency will have strong outflow events, which reduce both dust and gas mass.

\begin{figure*}
\begin{center}
%\setcaptionmargin{1cm}
\includegraphics[width=1.0 \columnwidth,angle=0]{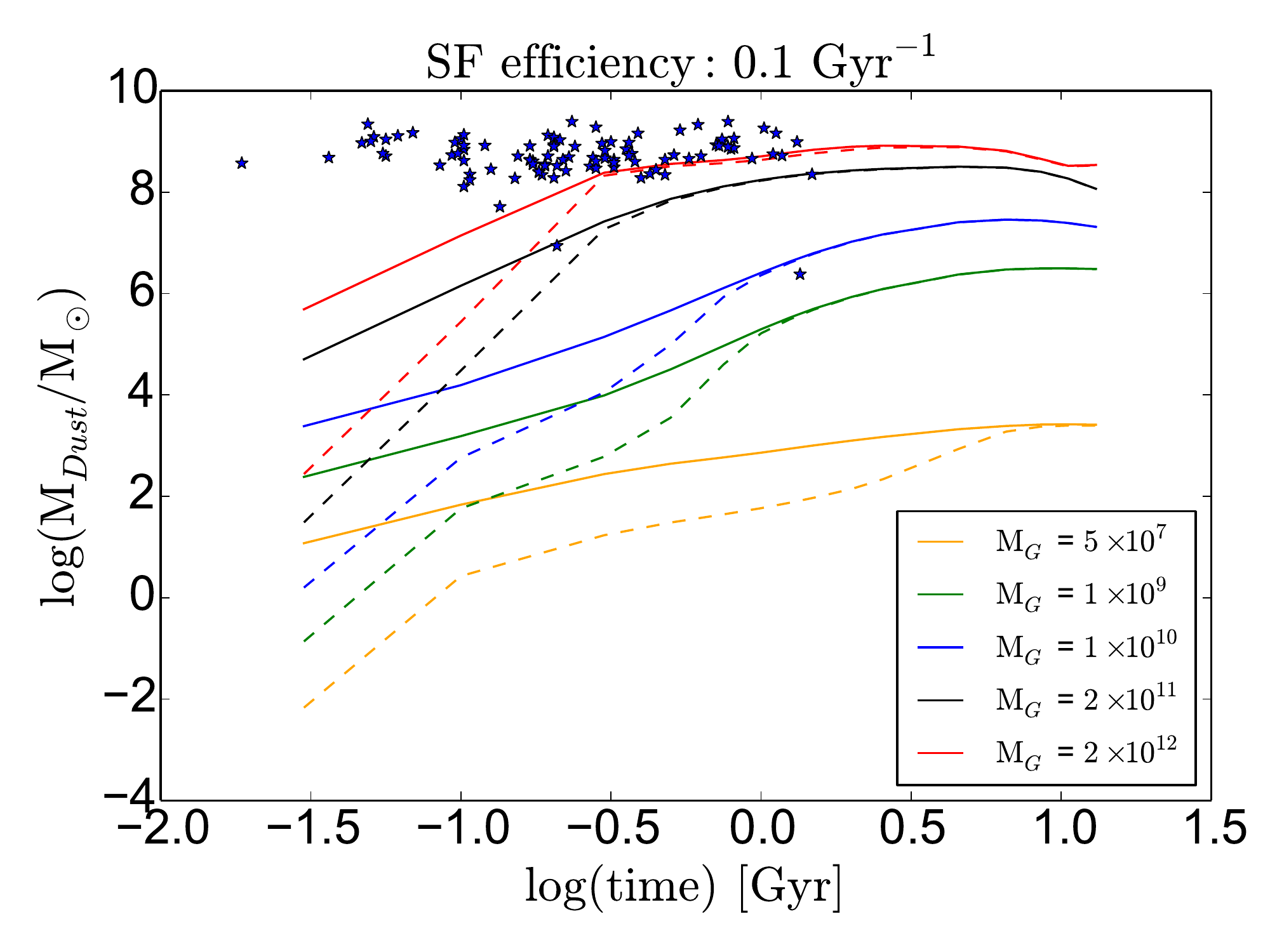}
\includegraphics[width=1.0 \columnwidth,angle=0]{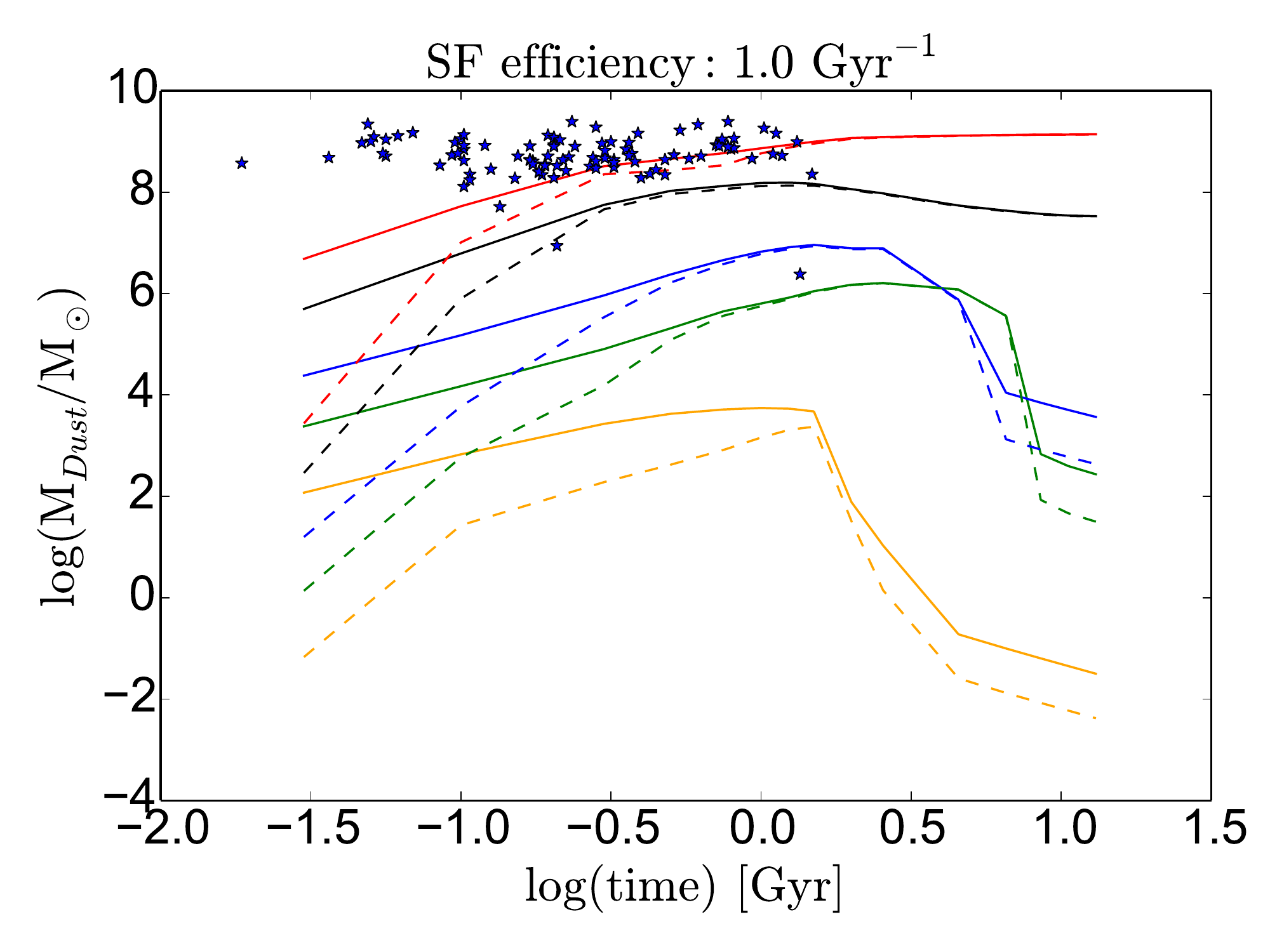}
\includegraphics[width=1.0 \columnwidth,angle=0]{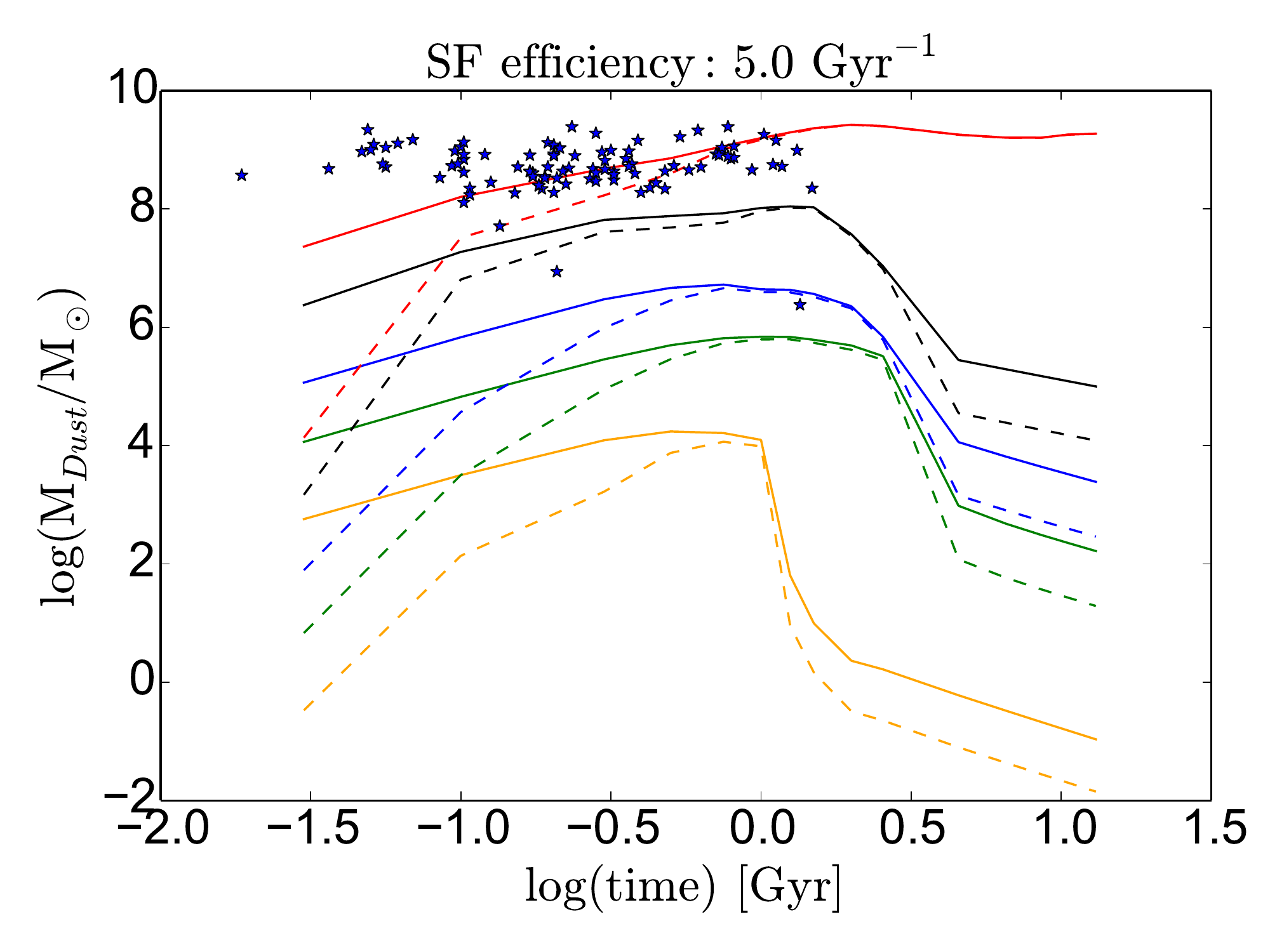}
\includegraphics[width=1.0 \columnwidth,angle=0]{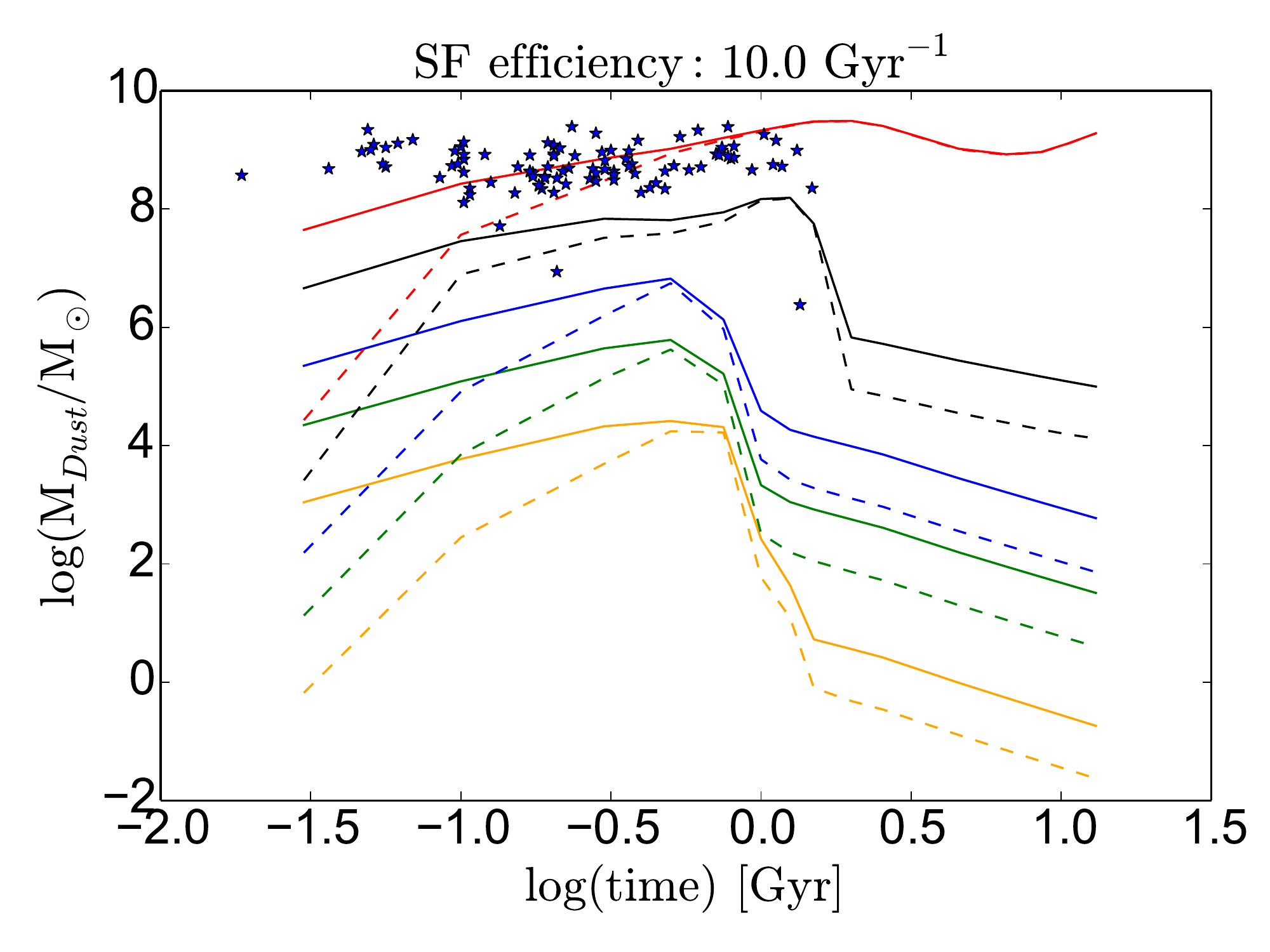}
\caption[]{Dust mass evolution predicted by the model. Each panel corresponds to a star formation efficiency of the model and the model tracks are colour-coded according the initial galaxy mass:  $5 \times 10^7 \, {\rm M}_\odot$ (yellow),  $1 \times 10^9 \, {\rm M}_\odot$ (green), $1 \times 10^{10}  \, {\rm M}_\odot$ (blue),  $2 \times 10^{11} \, {\rm M}_\odot$ (black), and $2 \times 10^{12} \, {\rm M}_\odot$ (red). Solid and dashed lines stand for  Case A and Case B dust production formulation, respectively. Blue stars represent the SMG sample  of \citet{da2015alma}.} 
\label{fig:dust-time}
\end{center}
\end{figure*}

\begin{figure*}
\begin{center}
%\setcaptionmargin{1cm}
\includegraphics[width=1.0 \columnwidth,angle=0]{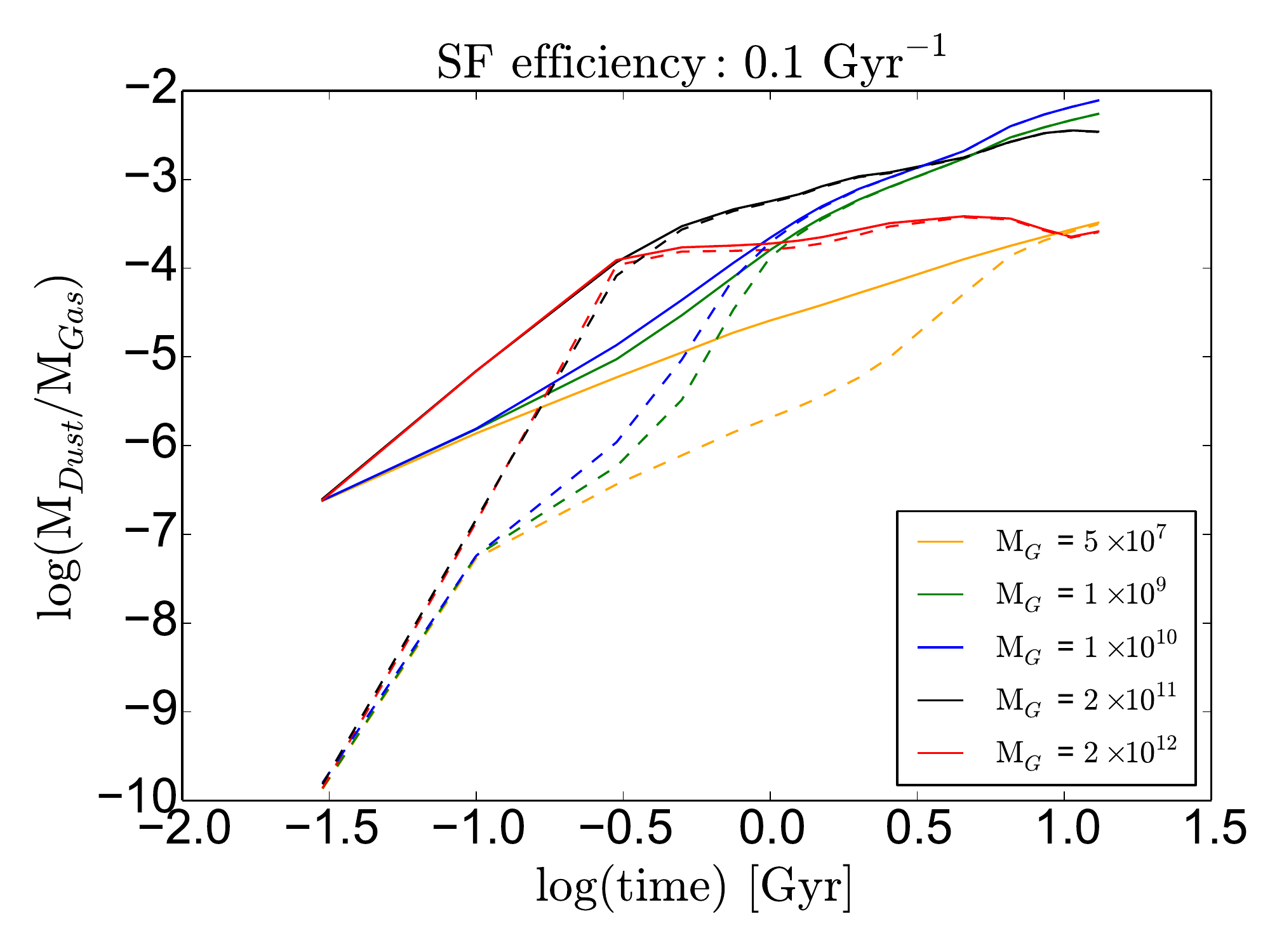}
\includegraphics[width=1.0 \columnwidth,angle=0]{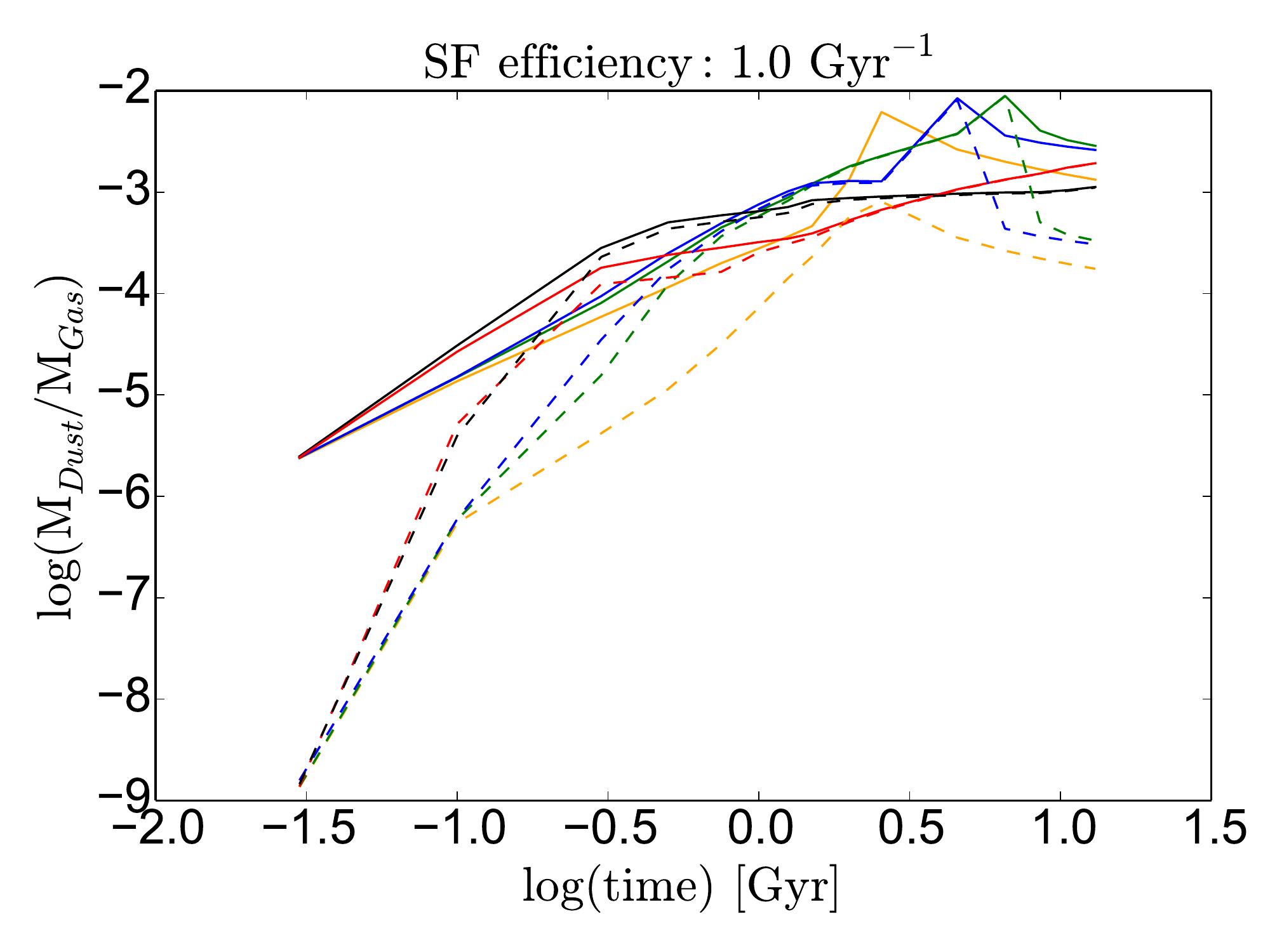}
\includegraphics[width=1.0 \columnwidth,angle=0]{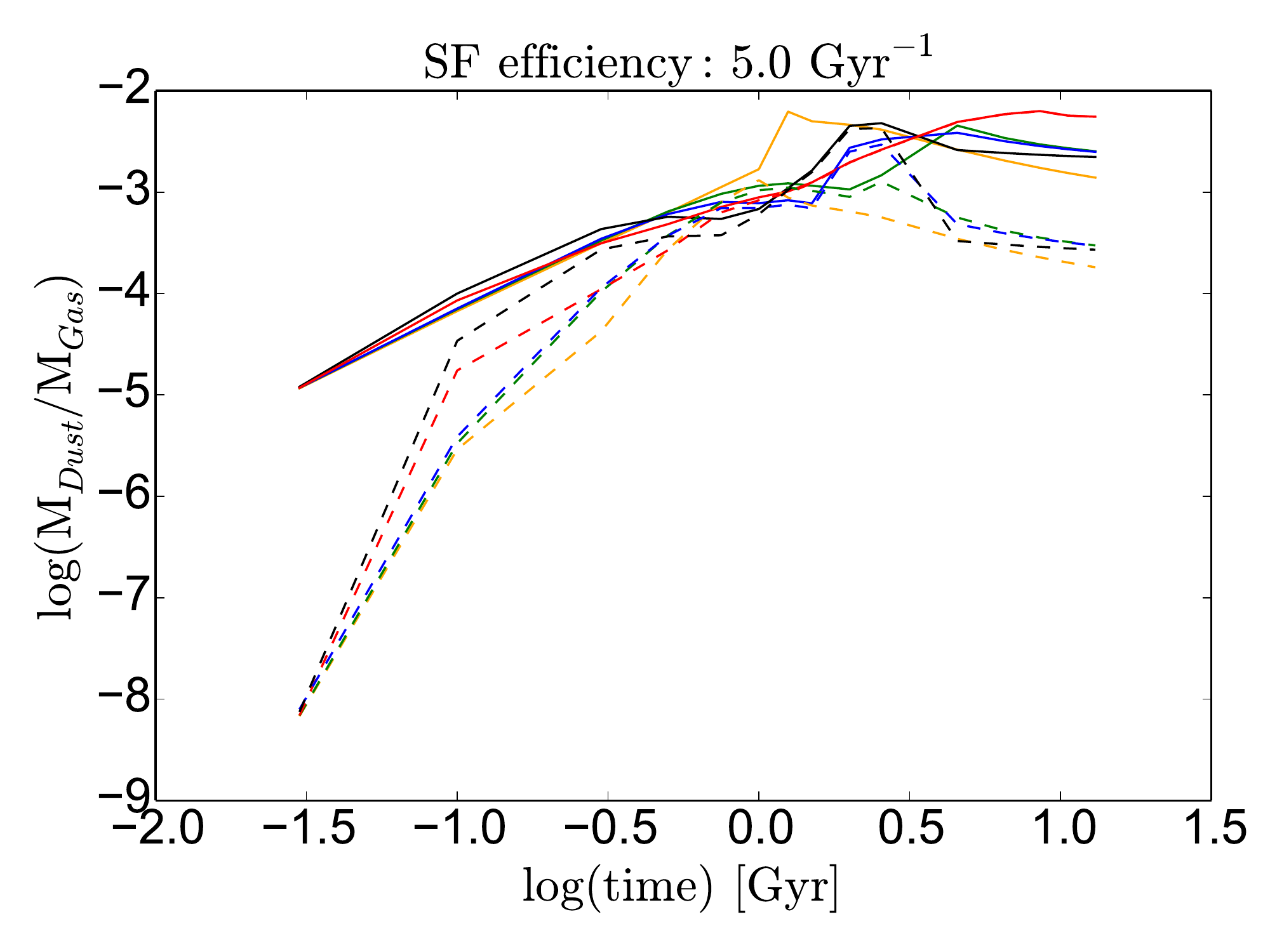}
\includegraphics[width=1.0 \columnwidth,angle=0]{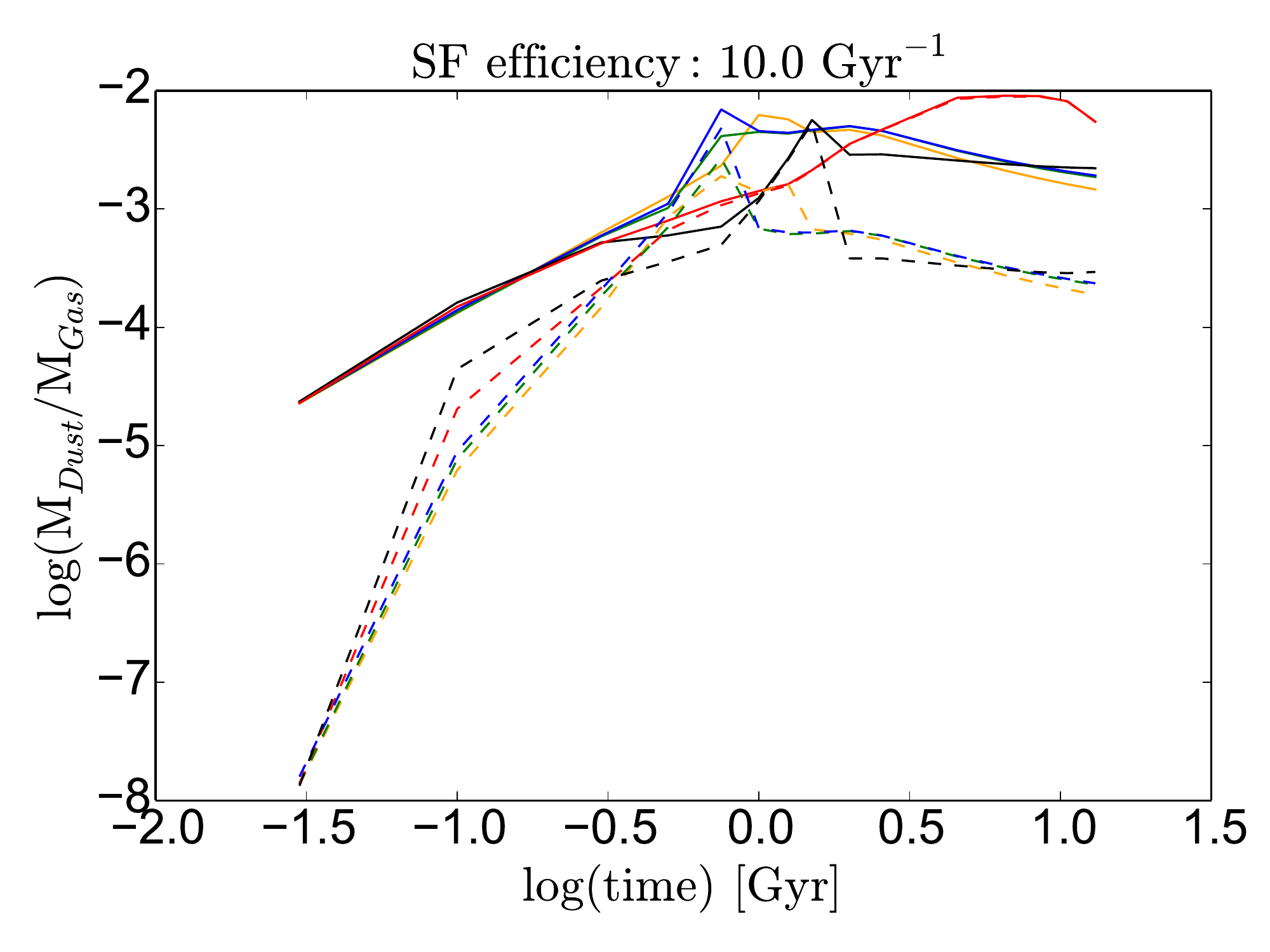}
\caption{Dust-to-gas mass ratio evolution predicted by the model. Line colours and styles of the model tracks have the same meanings as in Fig~\ref{fig:dust-time}.} 
\label{fig:dust/gas-time}
\end{center}
\end{figure*}

\section{Observational data}
\label{sec:Observation}

We have built a sample from data available in the literature in order to compare, whenever possible, with the simulations presented here. We have select data with available parameters such as $M_{\rm Dust}$, $M_*$, $M_{\rm Gas}$ and SFR, spanning a wide range of galactic mass and redshift. In this section we will explain the main features of the data collected, separating in low- and high-$z$ samples.  

\subsection{Low redshift sample}
\label{subsec:low-z}

\citet{remy2014gas} provide $M_{\rm Dust}$ and  $M_{\rm Gas}$ for galaxies from the KINGFISH survey \citep{2011PASP..123.1347K}, the Dwarf Galaxy Survey \citep[DGS][]{1538-3873-126-945-1079} and a selected sample from \citet{galametz2011probing}, named G11.

KINGFISH, DGS and G11 galaxies are all low-$z$ objects with diverse morphology. Their log(O/H) + 12 metallicity tracer covers a range of 2 dex.
\citet{remy2015linking} give $M_*$ and SFR for KINGFISH and DGS.

The early type sample was taken from \citet{lianou2016dustier}.
% a linha abaixo deve ir na legenda da figura onde aparece
% and it is indicated by ``+''. 
This sample was drawn from the Herschel Reference Survey \citep{boselli2010herschel}. $M_{*}$, SFR and $M_{\rm Dust}$ were estimated with SED fitting models via MAGPHYS \citep{2008MNRAS.388.1595D} and PCIGALE (v0.9.0) \citep{2005MNRAS.360.1413B, 2009A&A...507.1793N, 2014ASPC..485..347R}. MAGPHYS uses only stellar templates, while PCIGALE includes AGN templates in their fits. Since it is a substantial difference between them and we included both in our analysis.
%% Info para legenda
%% The MAGPHYS data are tagged in green and PCIGALE in blue.

From \citet{de2016herschel}, based on the  \textit{Herschel}-ATLAS Phase-1 Limited-Extent Spatial Survey \citep{clark2015herschel}, we adopted a selection of objects with high gas mass fraction ($> 80\%$). This sample lies inside a volume limited by $0.0035 <z< 0.01$, with galaxies in several evolutionary phases and, in general, low SFR (the maximum value is log~SFR $\sim 0.6$ for a galaxy with $\log M_{*}/$M$_\odot \sim 10.16$).
SFR and $M_{\rm Dust}$ were calculated using MAGPHYS. 
% Para legenda
%These data is always symbolized by red dots.

\subsection{High redshift sample}
\label{subsec:high-z}

We have adopted a submillimeter galaxy (SMG, a high-$z$ ULIRG-like galaxy) sample from the ALESS survey \citep{hodge2013alma,karim2013alma}. The quantities $z_{\rm phot}$, $M_*$, $M_{\rm Dust}$, SFR, and a mass-weighted stellar population age was computed by \citet{da2015alma}, using MAGPHYS. The SMG sample covers a $z_{\rm phot}$ range from 1.58 to 5.82, covering much of the galaxy assembly epoch. 
%Dust obscured galaxies are crucial objects to understand galaxy evolution during the height of their formation.
The sample mean stellar mass is about $M_* \sim 9 \times 10^{10}$~M$_\odot$, the mean SFR is $281 M_*/$yr, the mean age is 0.24~Gyr, and $M_{\rm Dust} = 5.6 \times 10^{8}$ M$_\odot$. 

The SMGs do not make a homogeneous population, even at the same redshift. In $z_{phot} \sim 2$ (at the cosmic SFR peak), half of this sample consists of starburst galaxies (SBG), with SFR more than three times above the typical star-forming galaxy main sequence \citep[the $M_*$--SFR relation, see][for details]{2014ApJS..214...15S}, while the other half consists basically of high mass main sequence galaxies. But the $z_{phot} \sim 3.5$ sub-sample tends to have higher SFR and stellar masses, although the number of galaxies above the main sequence is less than a third, which suggests fast evolution of these objects across the cosmic time. 
%% Para a Legenda
%% SMGs sample is always represented by blues stars in the plots. 

%To derive dust information in high-$z$ objects is, generally, a hard and expensive task, since the number of available objects generally drops while $z$ grows \citep{0004-637X-854-1-36}.
We also included two Lyman-break galaxies (LBGs) at $z \sim 3$ \citep{magdis2017dust}, D49 and M28, as well as two galaxies from the reionization epoch, A1689-zD1 at $z \sim 7.5$ \citep{knudsen2016merger} and A2744\_YD4 at $z \sim 8.3$ \citep{laporte2017dust}. The reionization DOGs impose an important constraint to the lower efficiency required for dust and stellar evolution.

Since FT98 galactic model evolves from a primordial cloud, the presence of galaxies from high- and low-$z$ is fundamental to constrain dust production across the cosmic time and the main processes that lead to obscuration by dust.

\section{Results}
\label{sec:Results}

\begin{figure*}
\begin{center}
%\setcaptionmargin{1cm}
\includegraphics[width=\columnwidth,angle=0]{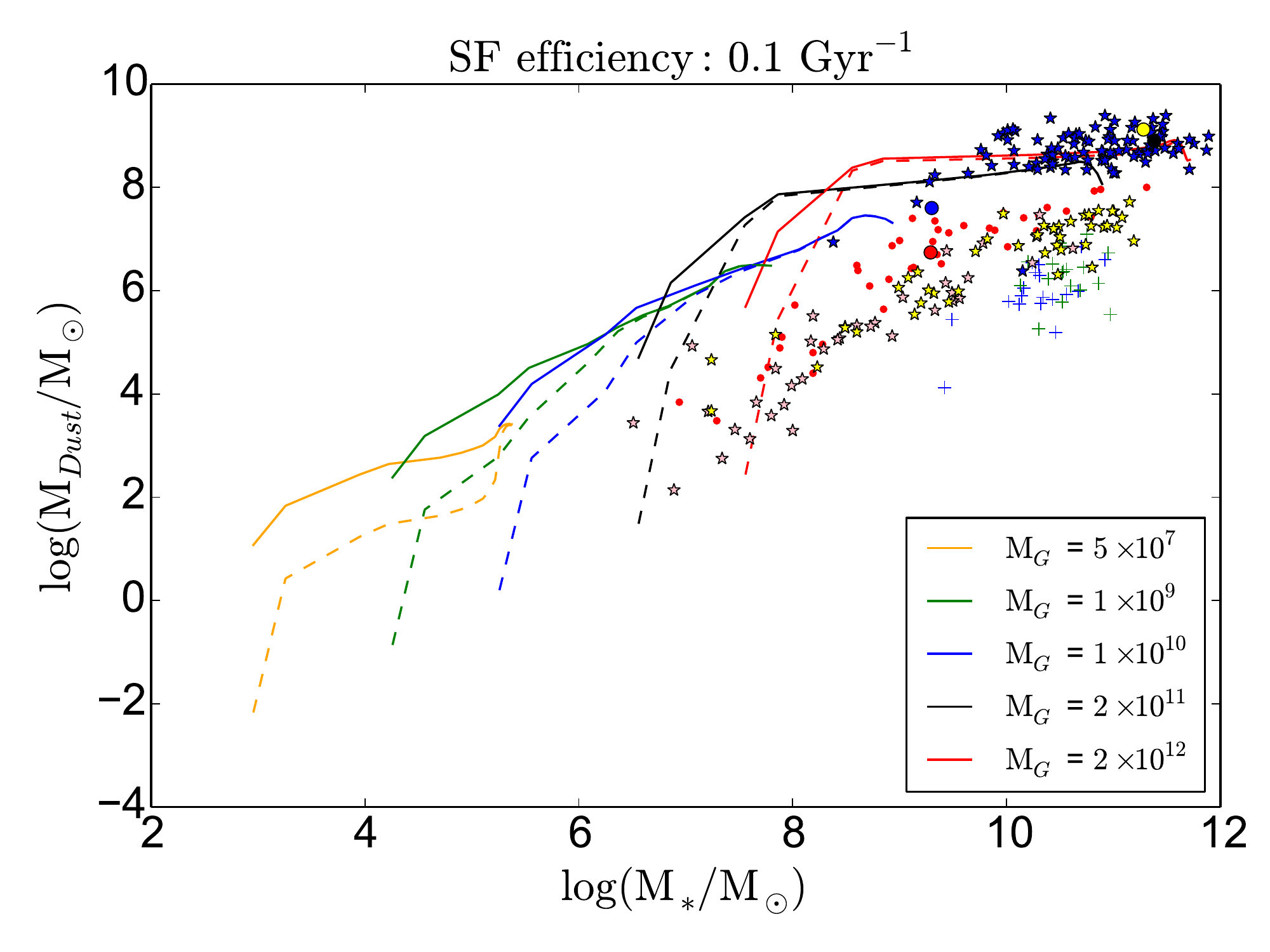}
\includegraphics[width=\columnwidth,angle=0]{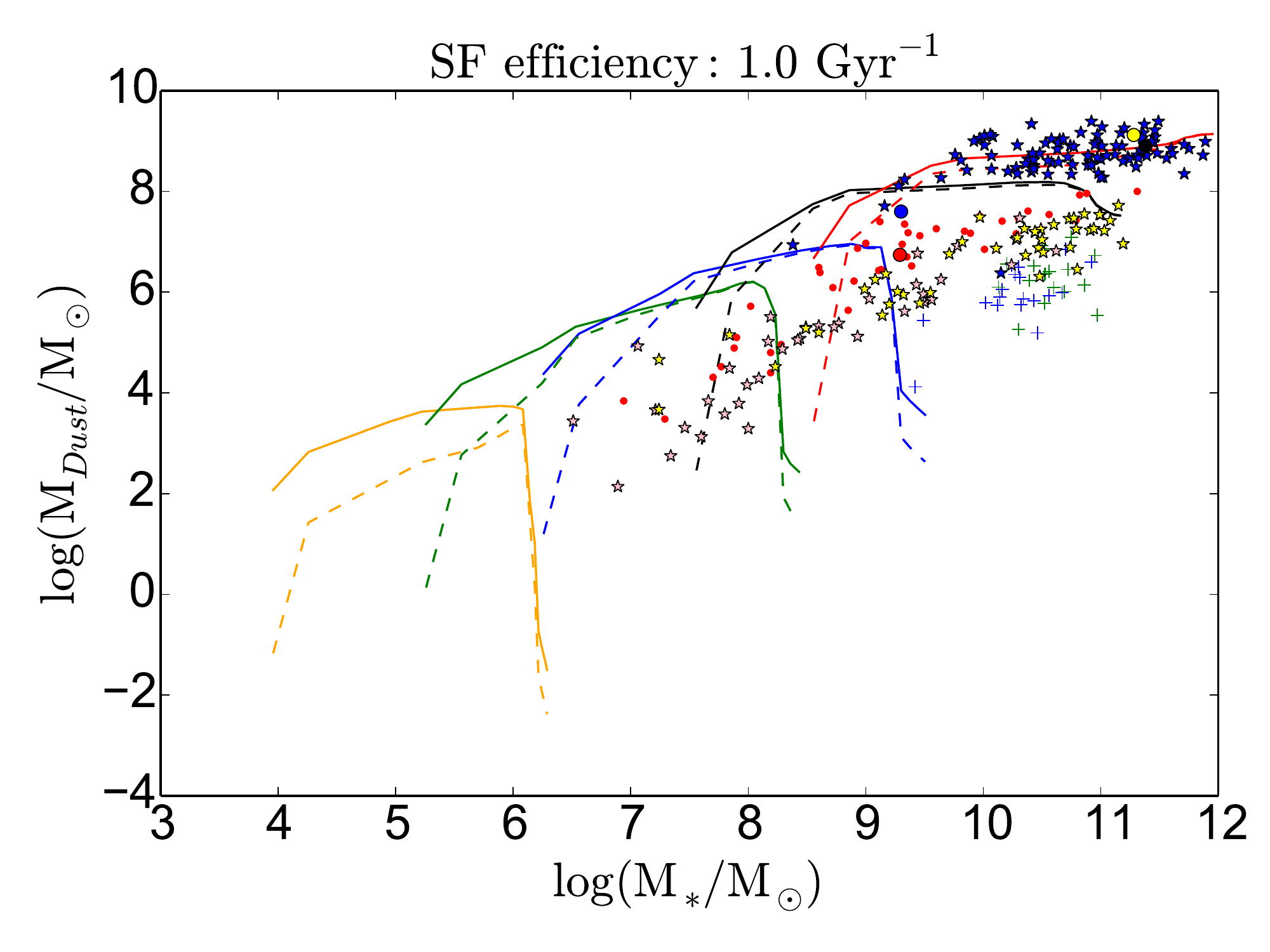}
\includegraphics[width=\columnwidth,angle=0]{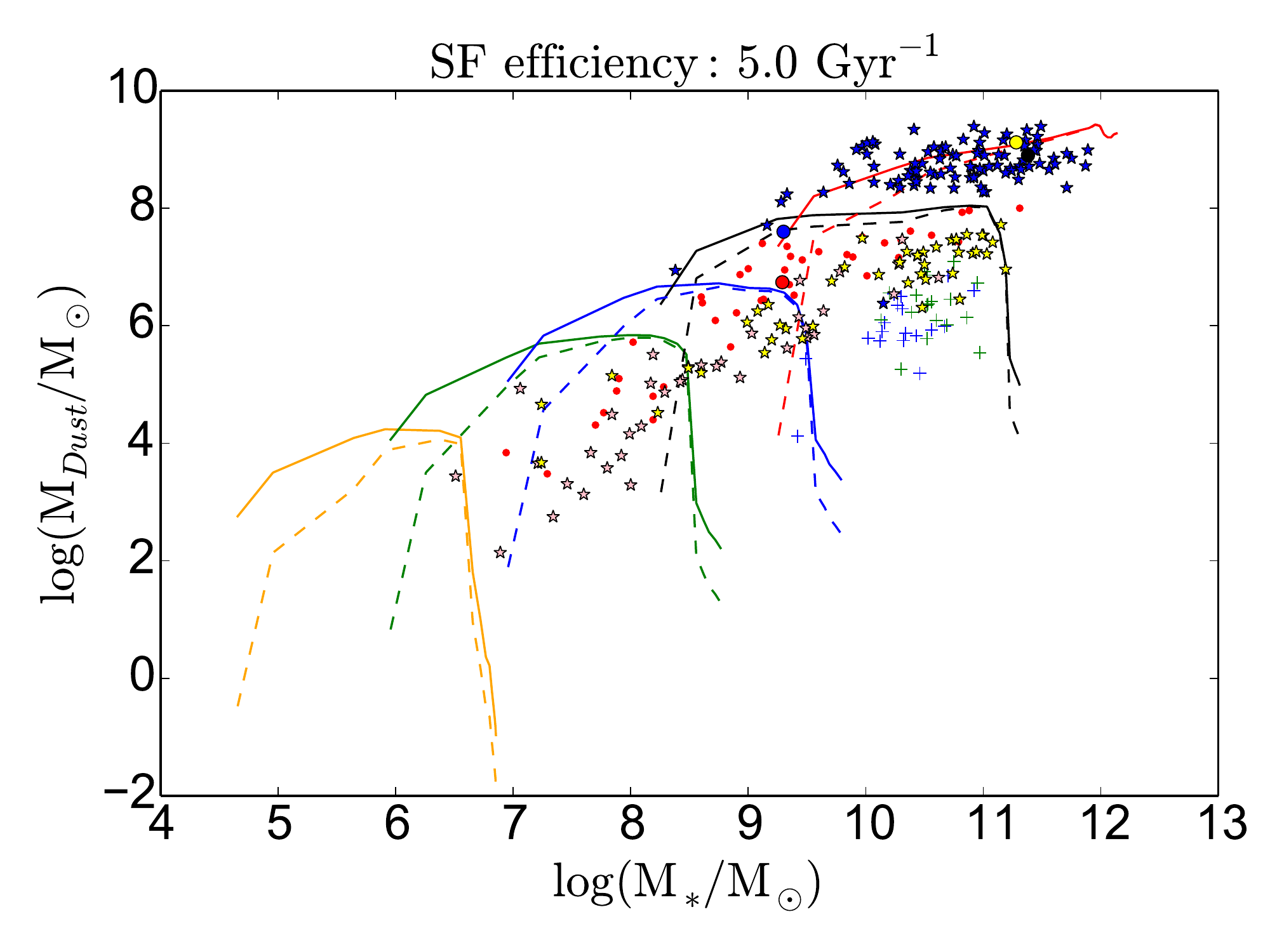}
\includegraphics[width=\columnwidth,angle=0]{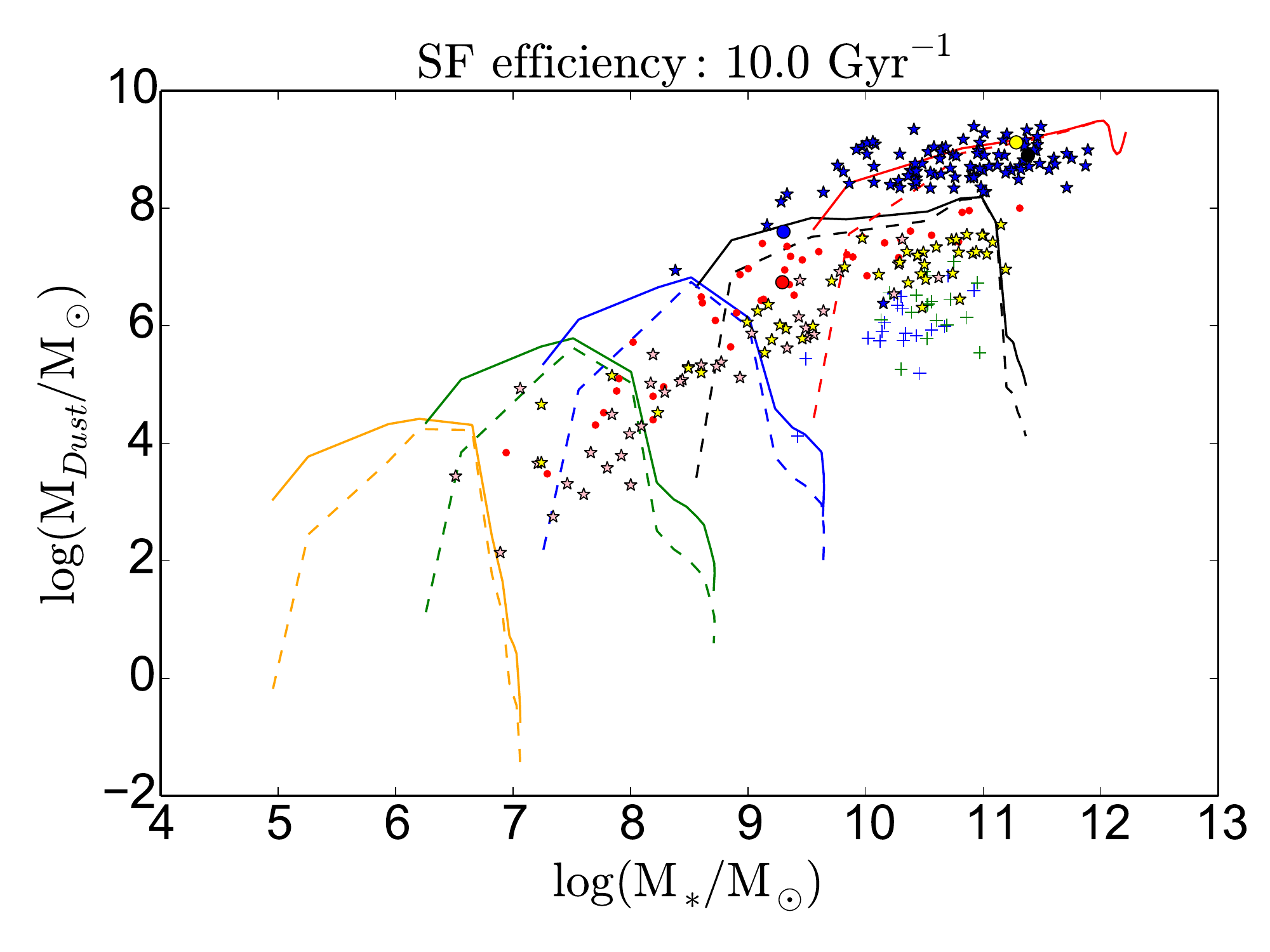}
\caption[]{Dust mass versus stellar mass predicted by the model. Line colours and styles of the model tracks have the same meanings as in Fig.~\ref{fig:dust-time}. The SMG sample of \citet{da2015alma} is represented by blue stars,
the data for galaxies of \citet{remy2014gas,remy2015linking},
by yellow and pink stars, of \citet{de2016herschel}, by small red dots, of ellipitical galaxies of \citet{lianou2016dustier}, by blue crosses, the data for the LBGs D49 and M28 of \citet{magdis2017dust}, by the yellow and black large dots, respectively, and for the reionization epoch DOGs A1689-zD1 of \citet{knudsen2016merger} and A2744\_YD4 of  \citet{laporte2017dust}, by the large blue and red dots, respectively.} 
\label{fig:DUST-Star}
\end{center}
\end{figure*}

\begin{figure*}
\begin{center}
%\setcaptionmargin{1cm}
\includegraphics[width=1.0 \columnwidth,angle=0]{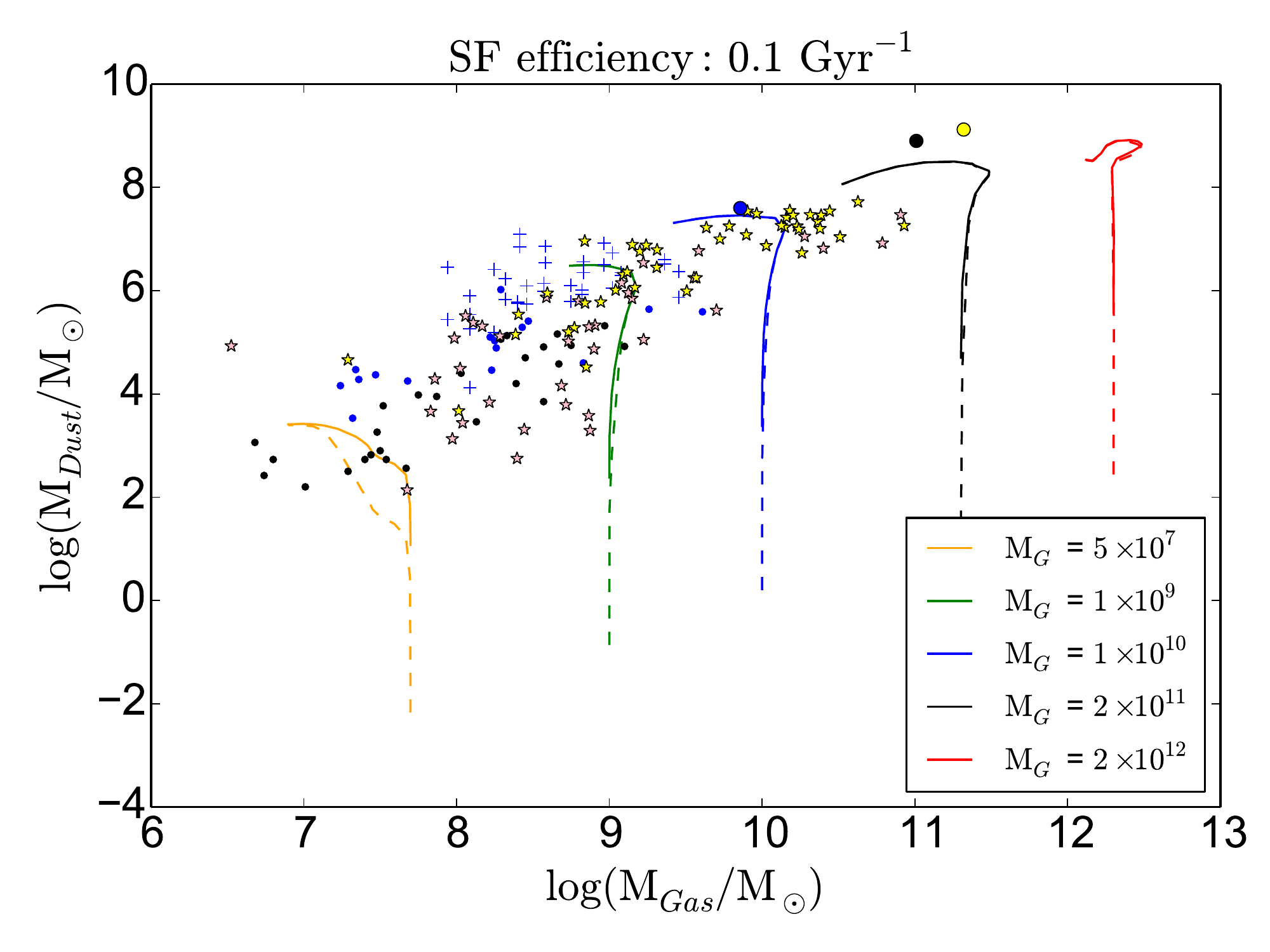}
\includegraphics[width=1.0 \columnwidth,angle=0]{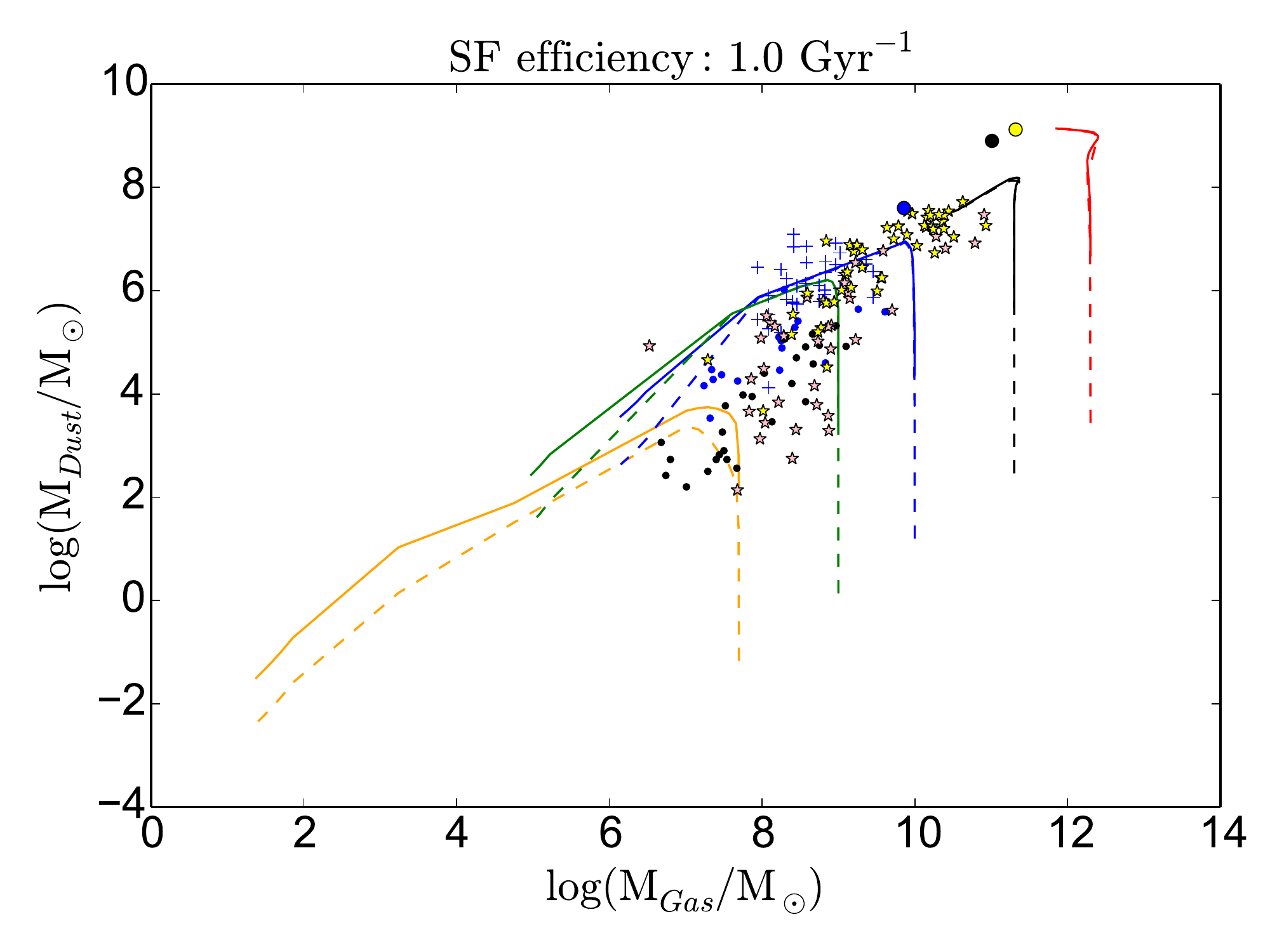}
\includegraphics[width=1.0 \columnwidth,angle=0]{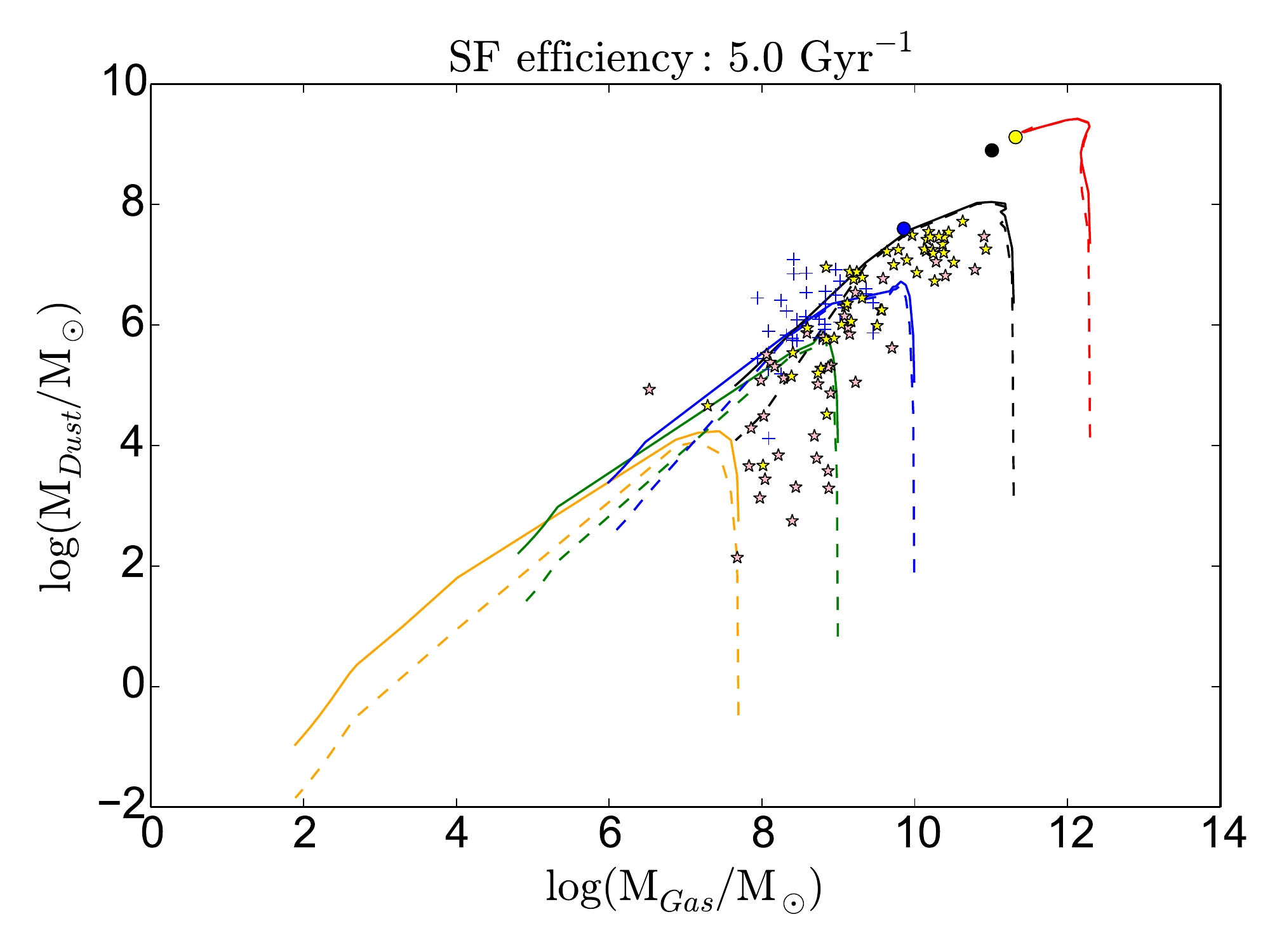}
\includegraphics[width=1.0 \columnwidth,angle=0]{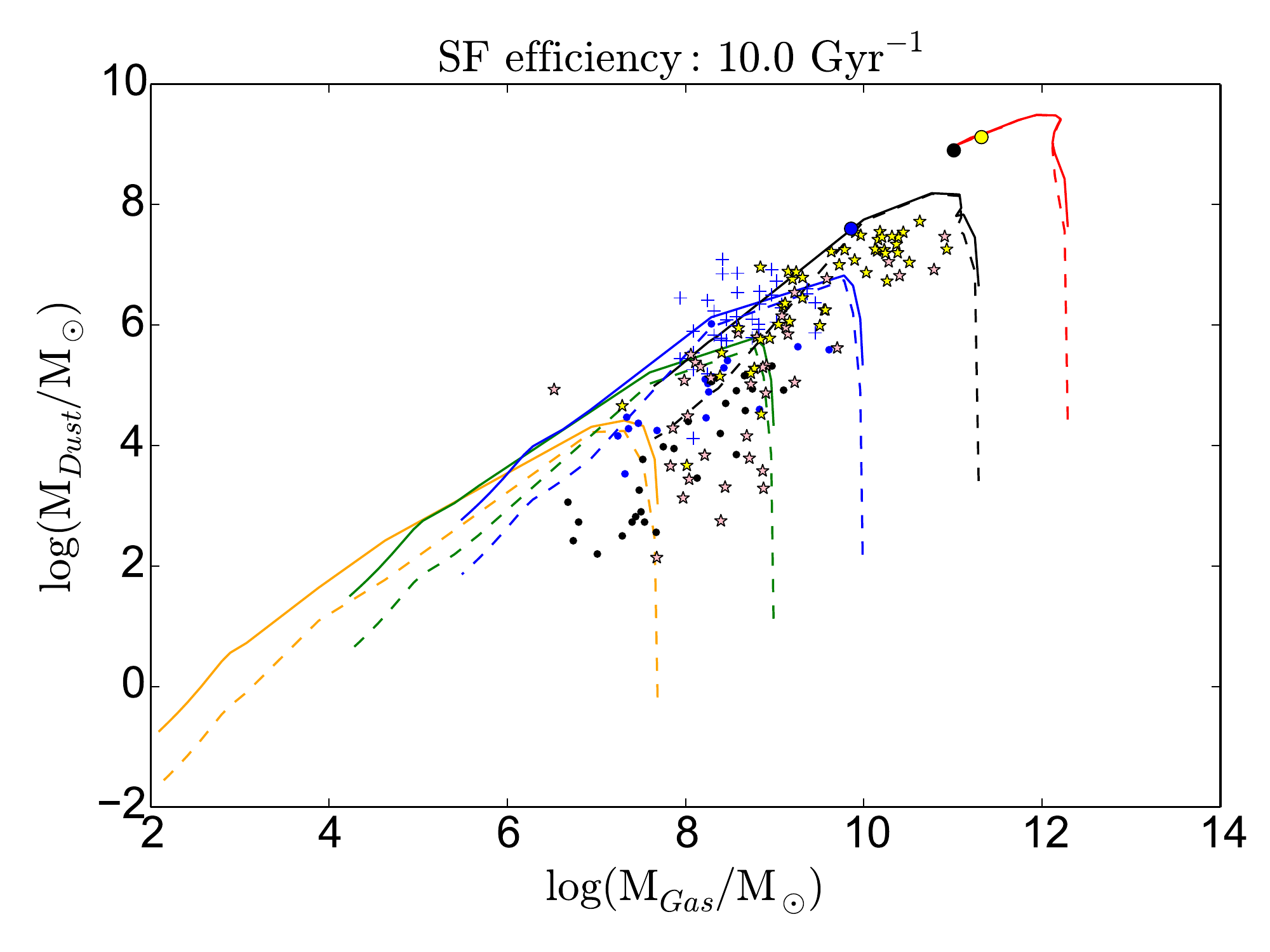}
\caption[]{Dust mass versus gas mass predicted by the model. Line colours and styles of the model tracks have the same meanings as in Fig.~\ref{fig:dust-time}. The data for elliptical galaxies of \citet{remy2014gas,remy2015linking} is represented by yellow and pink stars, of \citet{lianou2016dustier}, by blue crosses, and for the ionization epoch
DOG A1689-zD1 of \citet{knudsen2016merger}  by the large blue dot.} 
\label{fig:DUST-x-GAS}
\end{center}
\end{figure*}

\begin{figure*}
\begin{center}
%\setcaptionmargin{1cm}
\includegraphics[width=\columnwidth,angle=0]{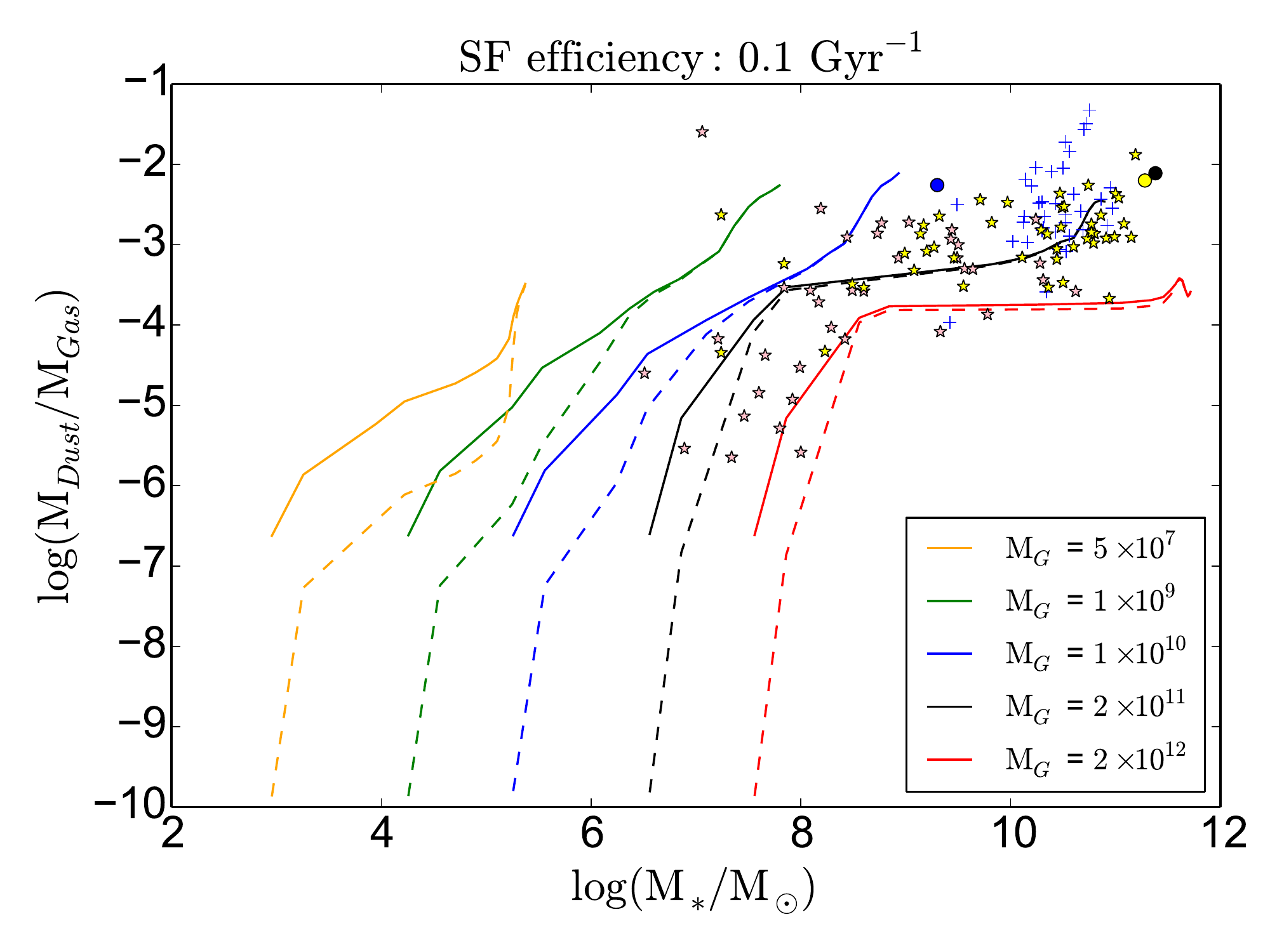}
\includegraphics[width=\columnwidth,angle=0]{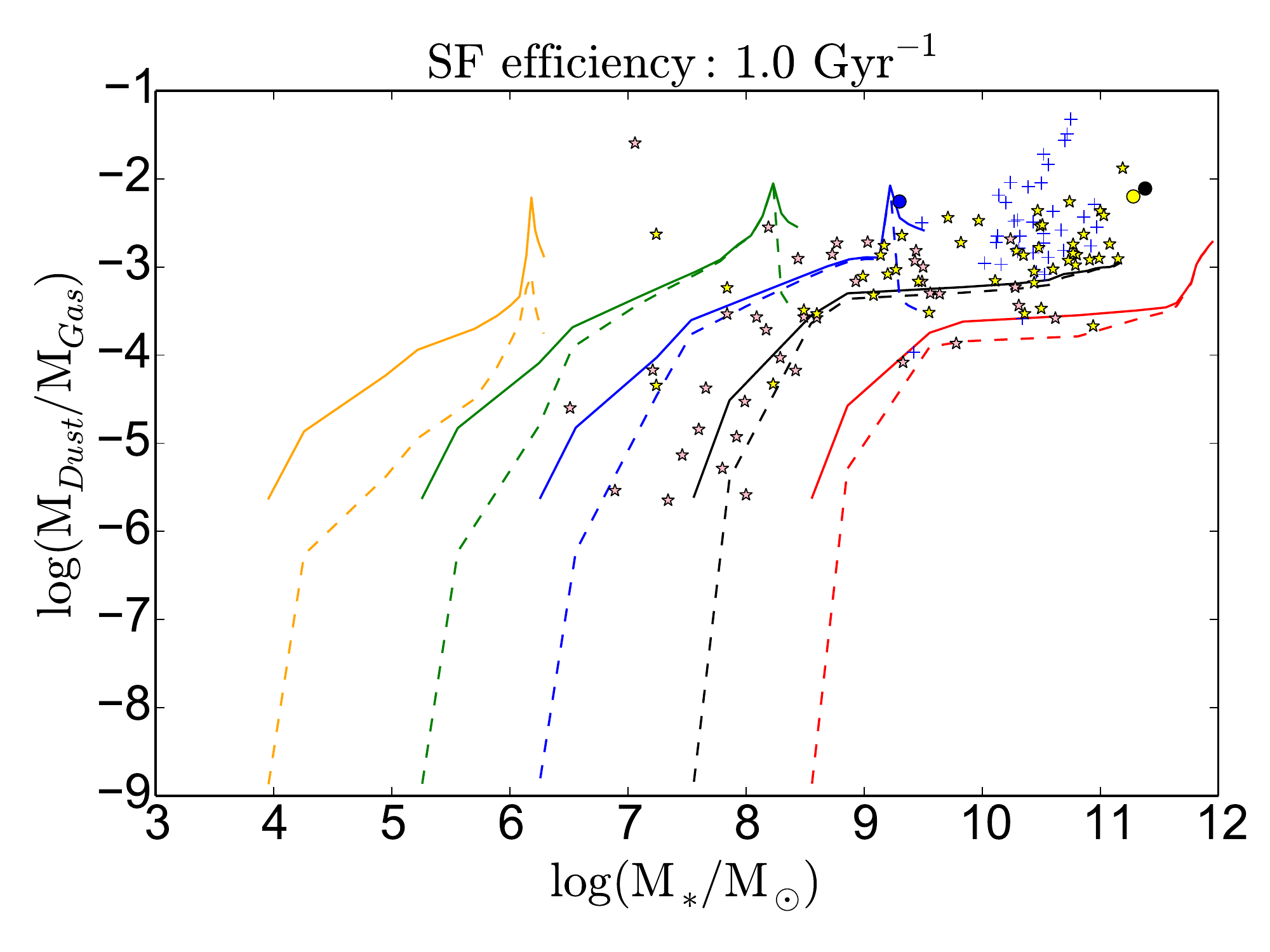}
\includegraphics[width=\columnwidth,angle=0]{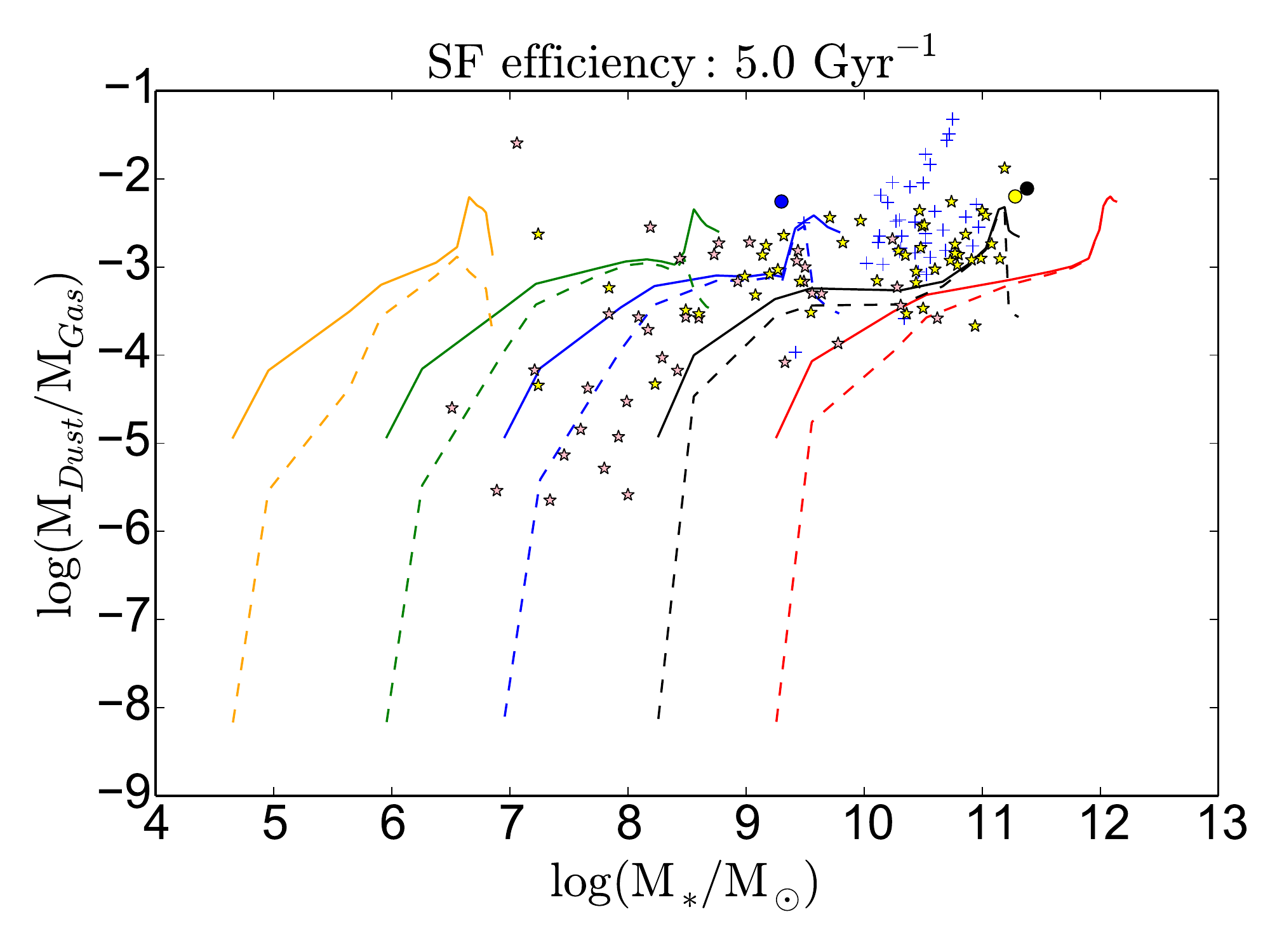}
\includegraphics[width=\columnwidth,angle=0]{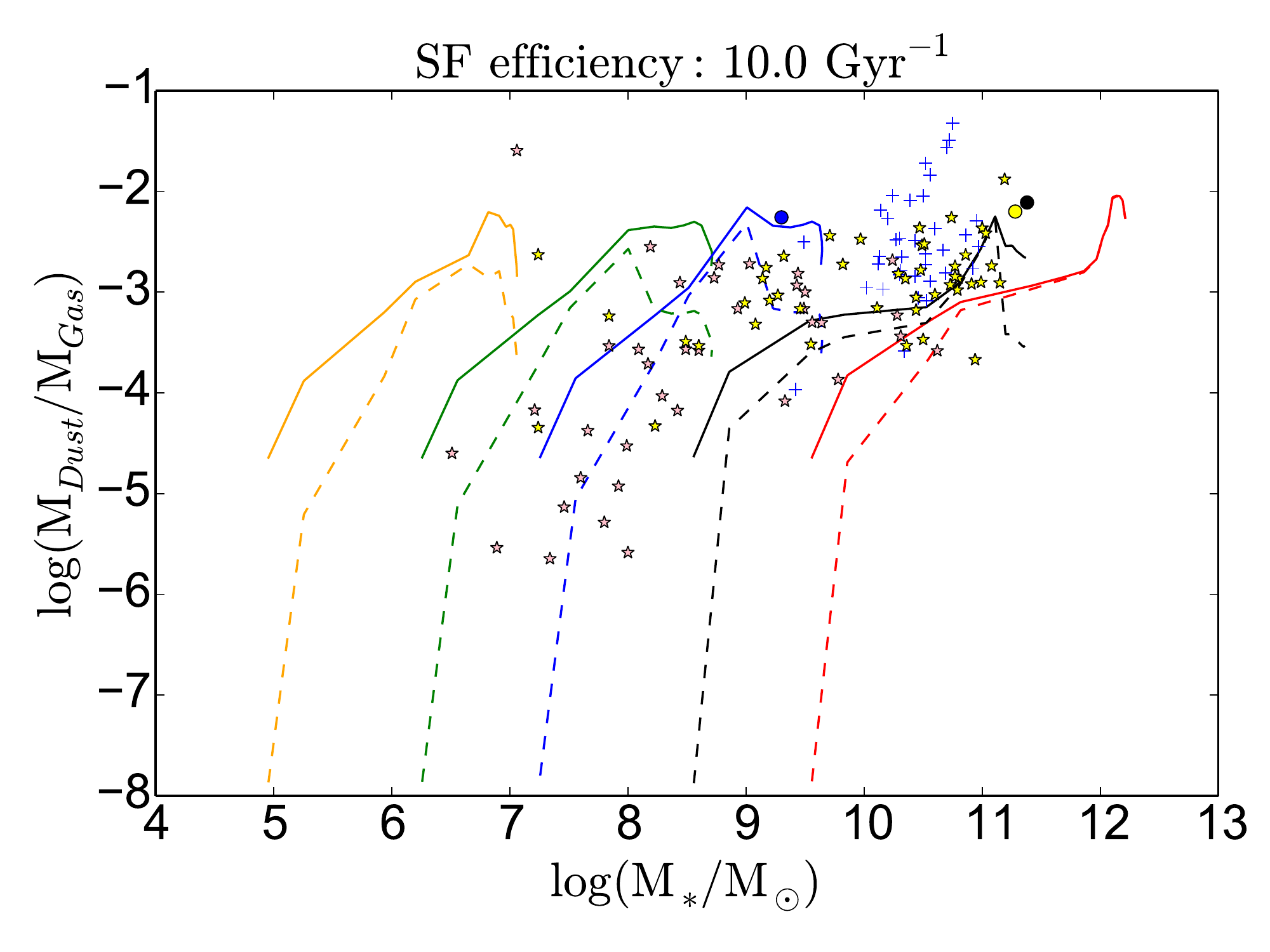}
\caption{Dust-to-gas ratio versus stellar mass predicted by the model. Line
colours and styles of the model tracks have the same meanings as in Fig.~\ref{fig:dust-time}. The data for galaxies of \citet{remy2014gas,remy2015linking} is represented by yellow and pink stars, elliptical galaxies of \citet{lianou2016dustier}, by blue crosses, the data for the LBGs D49 and M28 of \citet{magdis2017dust}, by the yellow and black large dots, respectively, and for the reionization epoch DOG A1689-zD1 of \citet{knudsen2016merger} by the large blue dot.}
\label{fig:DUST_GAS-Star}
\end{center}
\end{figure*}

\begin{figure*}
\begin{center}
%\setcaptionmargin{1cm}
\includegraphics[width=\columnwidth,angle=0]{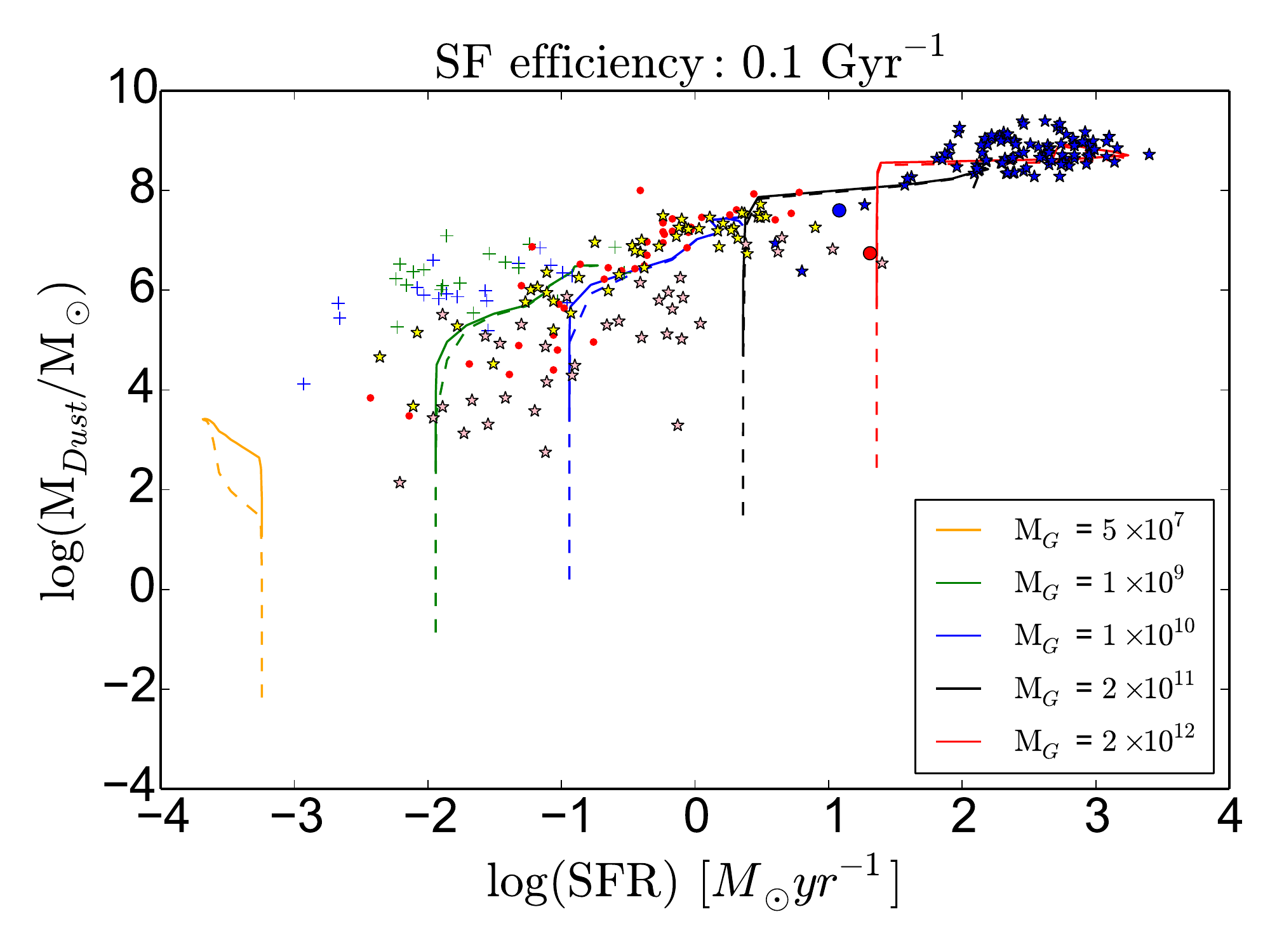}
\includegraphics[width=\columnwidth,angle=0]{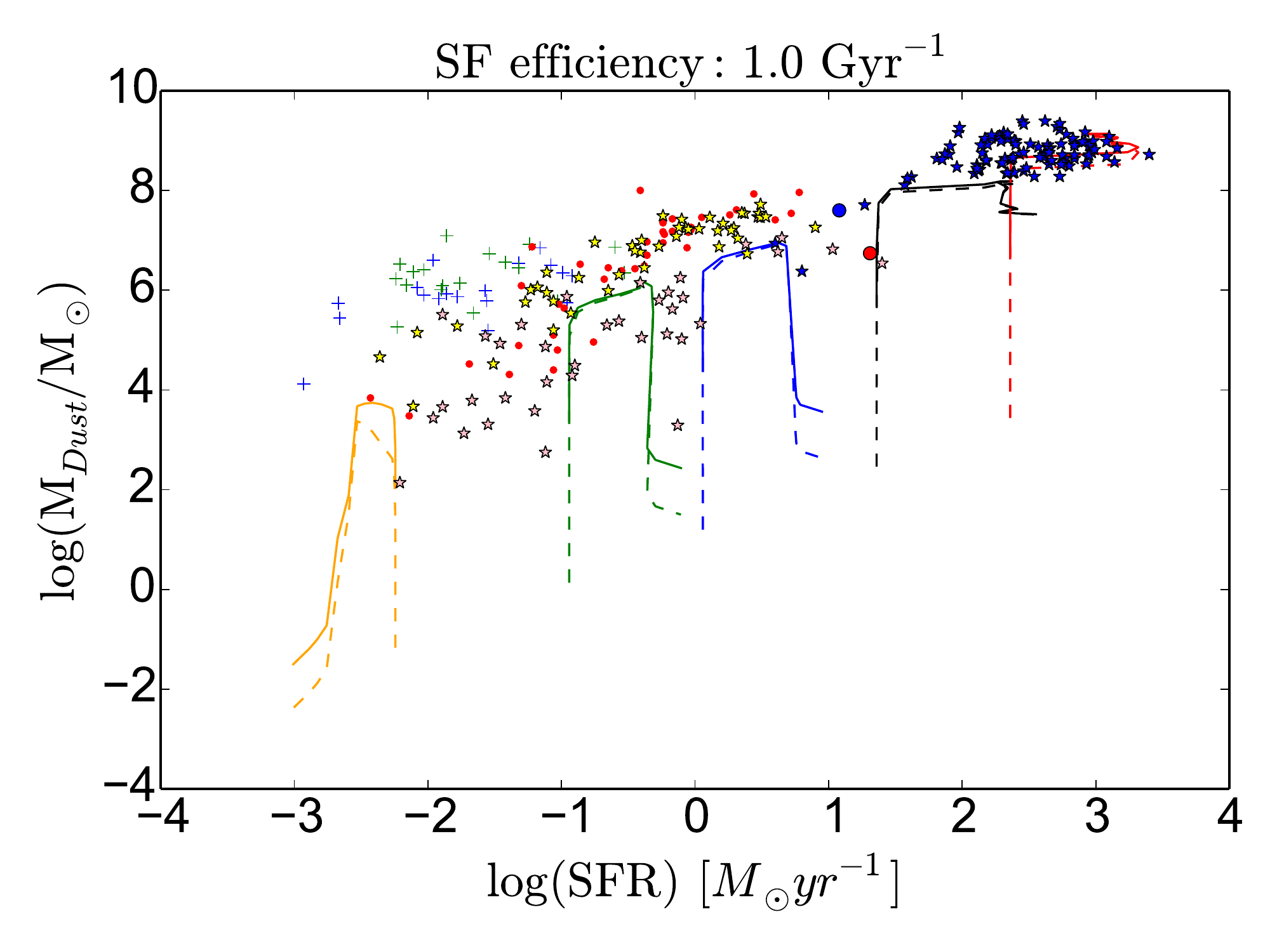}
\includegraphics[width=\columnwidth,angle=0]{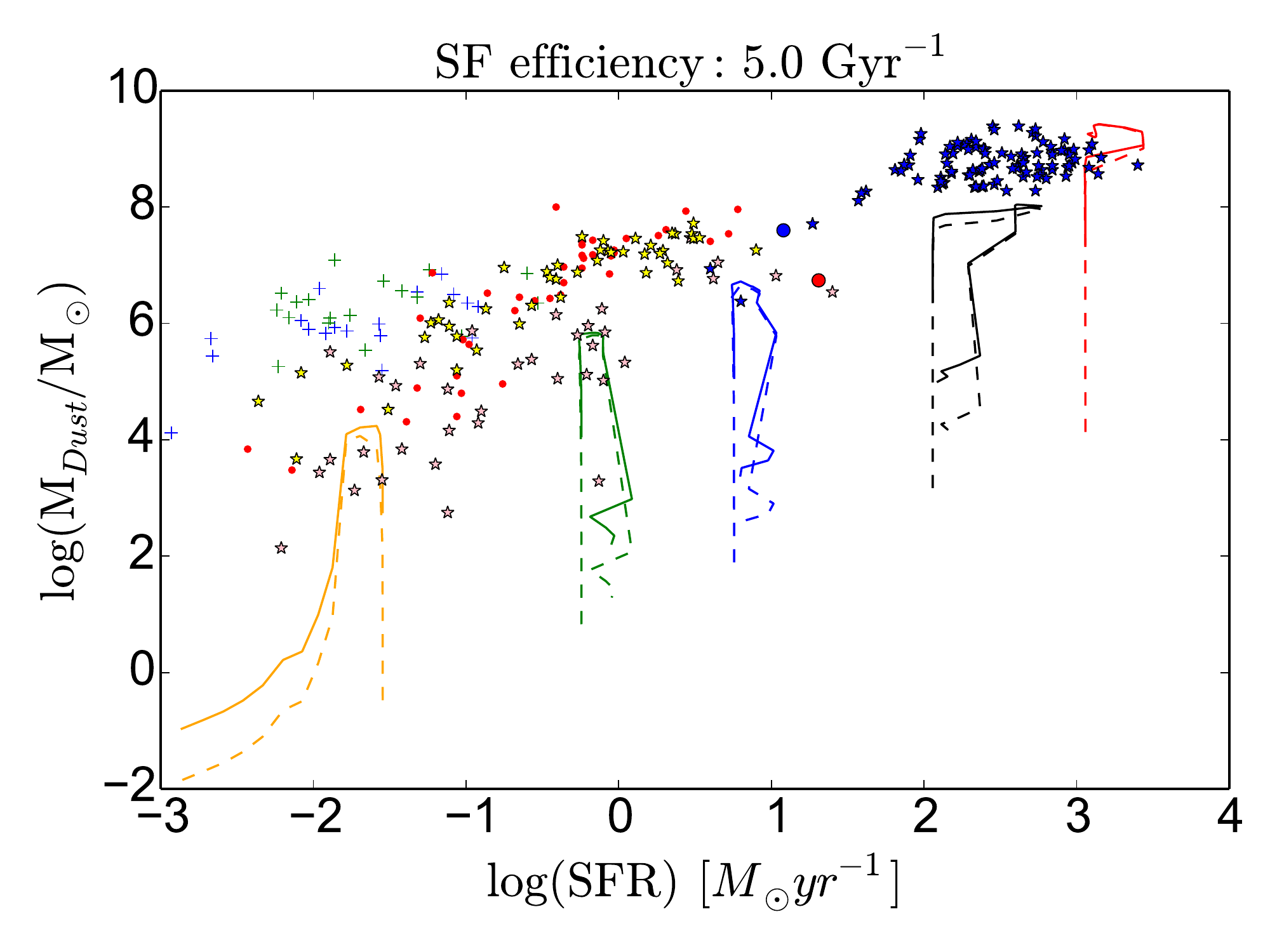}
\includegraphics[width=\columnwidth,angle=0]{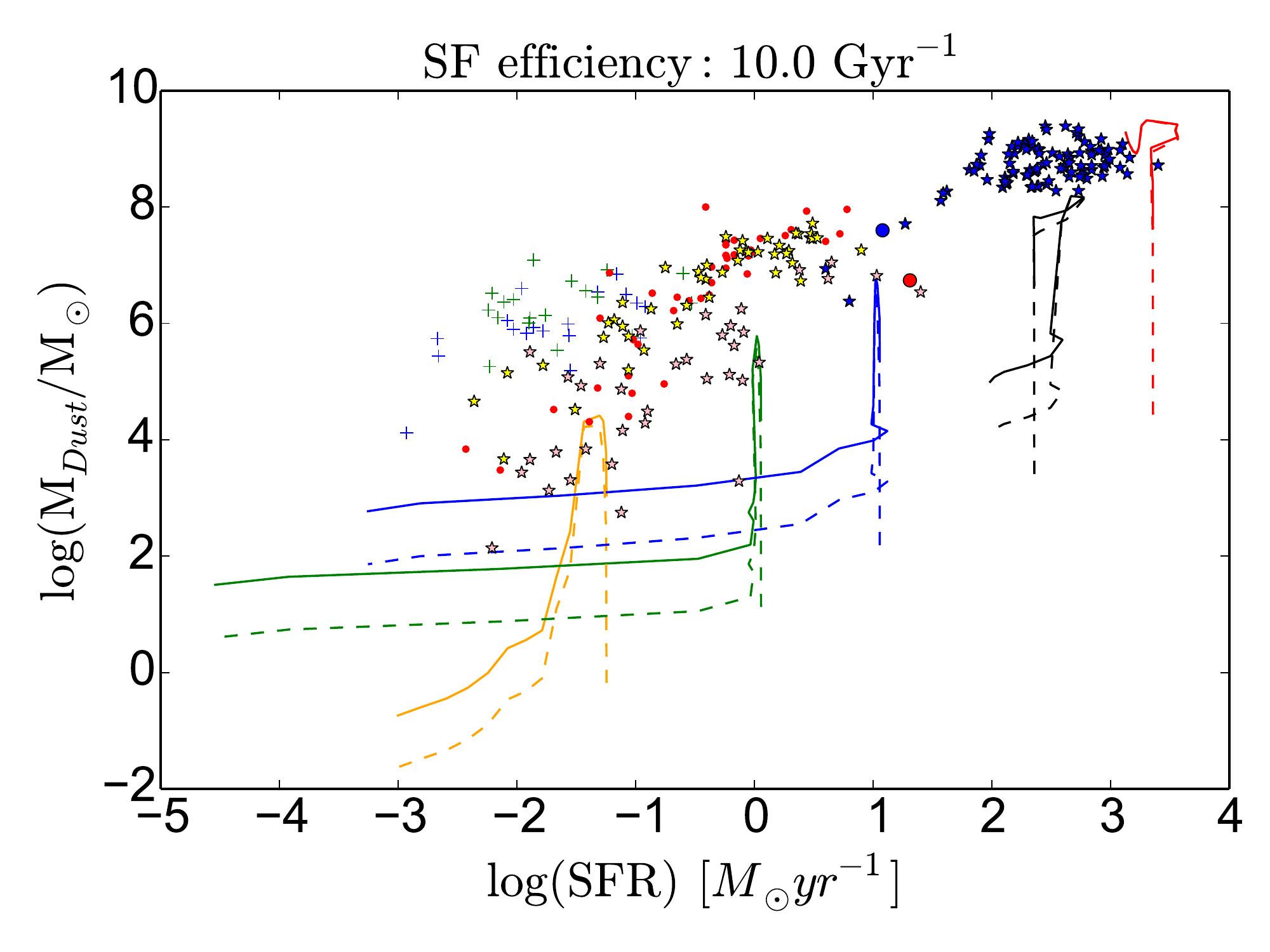}
\caption{Dust mass versus star formation ratio predicted by the model. Line
colours and styles of the model tracks have the same meanings as in Fig.~\ref{fig:dust-time}. The SMG sample of \citet{da2015alma} is represented by blue stars, the data for galaxies of \citet{remy2014gas,remy2015linking}, by yellow and pink stars, of \citet{de2016herschel}, by
small red dots, elliptical galaxies of \citet{lianou2016dustier}, by blue crosses, and for reionization epoch DOGs A1689-zD1 of  \citet{knudsen2016merger}, and A2744\_YD4 of  \citet{laporte2017dust}, by the large blue and red dots, respectively.} 
\label{fig:dust-sfr}
\end{center}
\end{figure*}

\begin{figure*}
\begin{center}
%\setcaptionmargin{1cm}
\includegraphics[width=\columnwidth,angle=0]{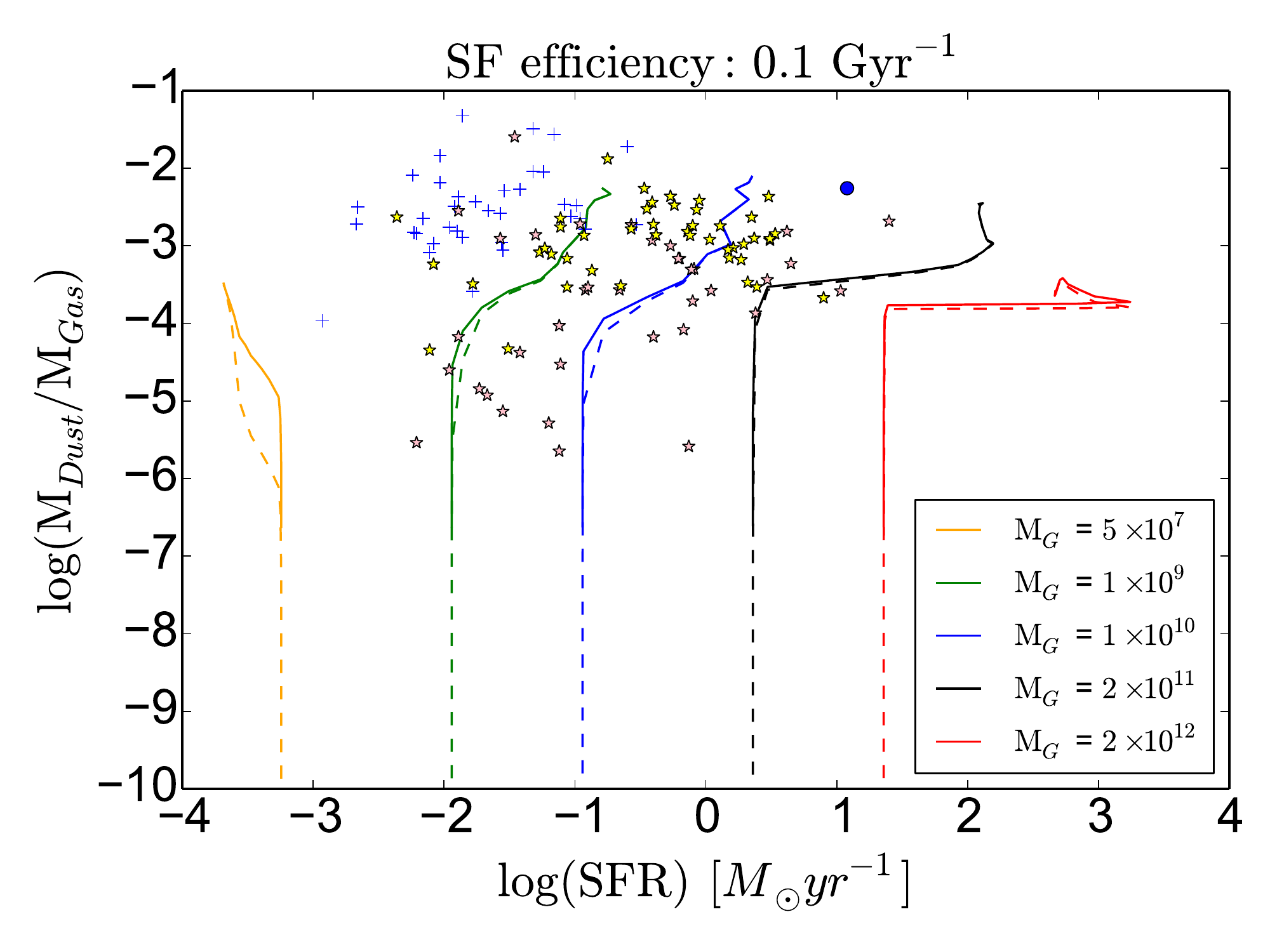}
\includegraphics[width=\columnwidth,angle=0]{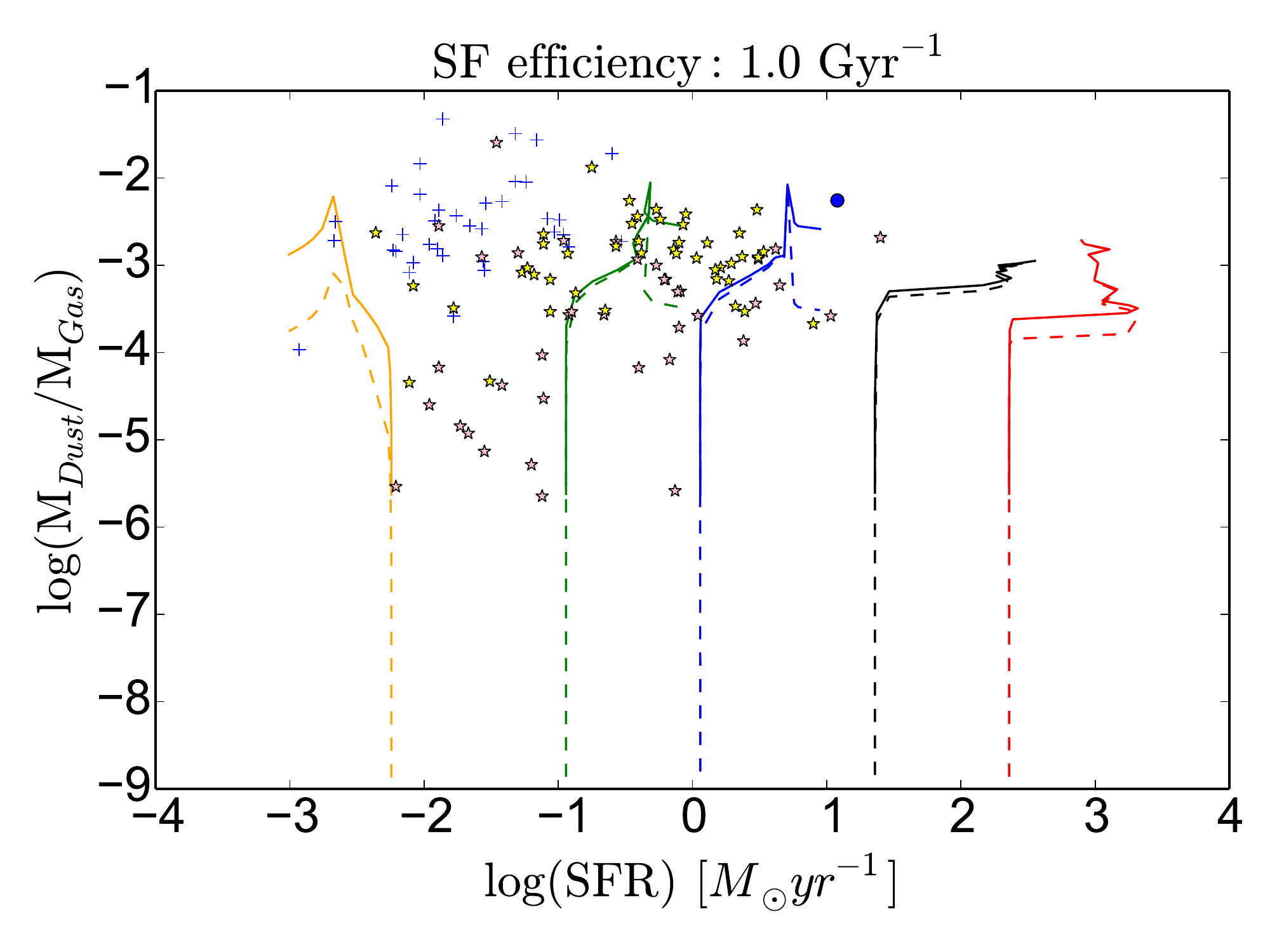}
\includegraphics[width=\columnwidth,angle=0]{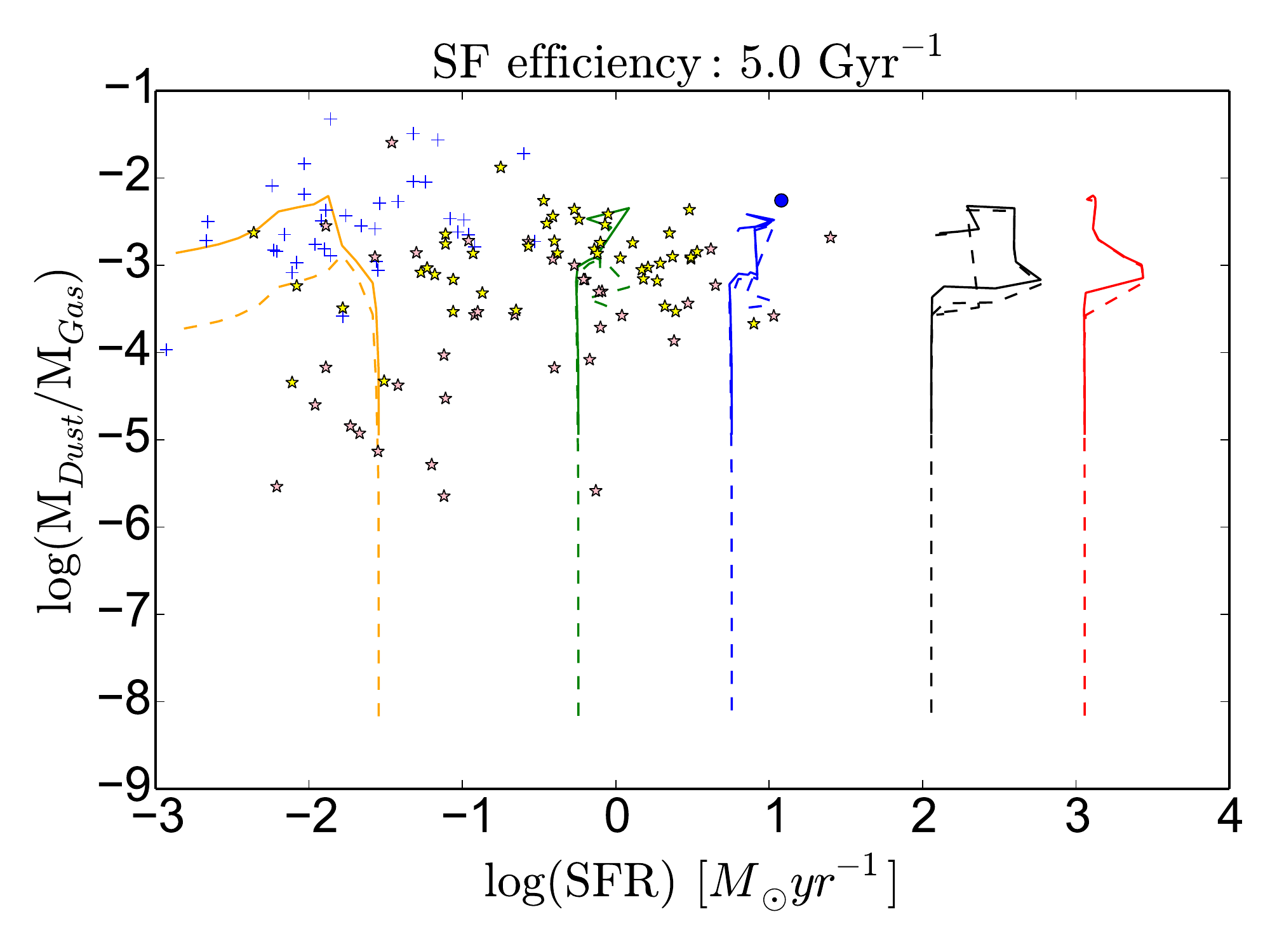}
\includegraphics[width=\columnwidth,angle=0]{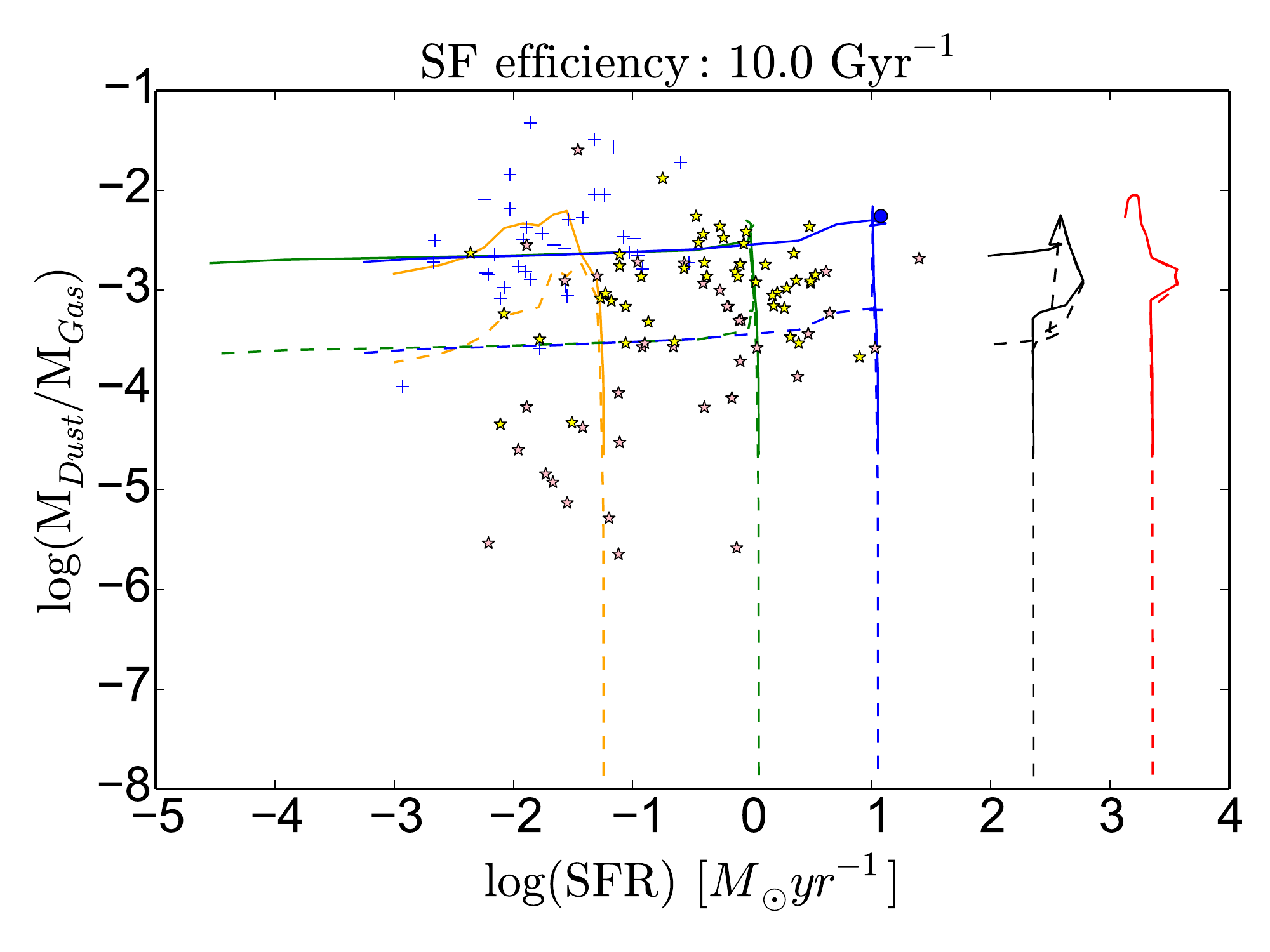}
\caption{Dust-to-gas ratio versus stellar formation rate predicted by the model. Line colours and styles of the model tracks have the same meanings as in Fig.~\ref{fig:dust-time}. The data for elliptical galaxies of \citet{remy2014gas,remy2015linking} is represented by yellow and pink stars, of \citet{lianou2016dustier}, by blue crosses, and for the ionization epoch
DOG A1689-zD1 of \citet{knudsen2016merger}  by the large blue dot.} 
\label{fig:dust/gas-sfr}
\end{center}
\end{figure*}

\begin{figure*}
\begin{center}
%\setcaptionmargin{1cm}
\includegraphics[width=\columnwidth,angle=0]{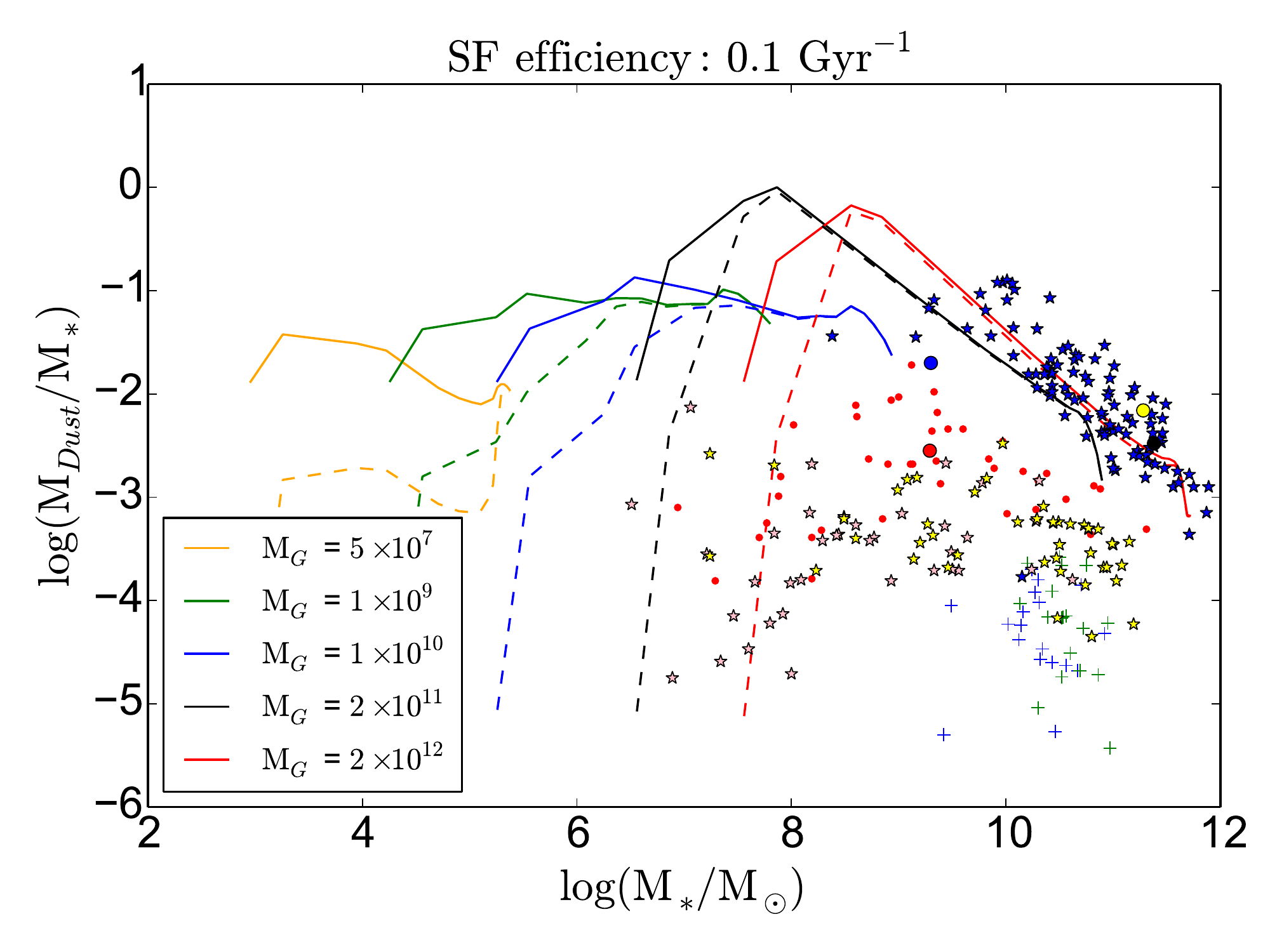}
\includegraphics[width=\columnwidth,angle=0]{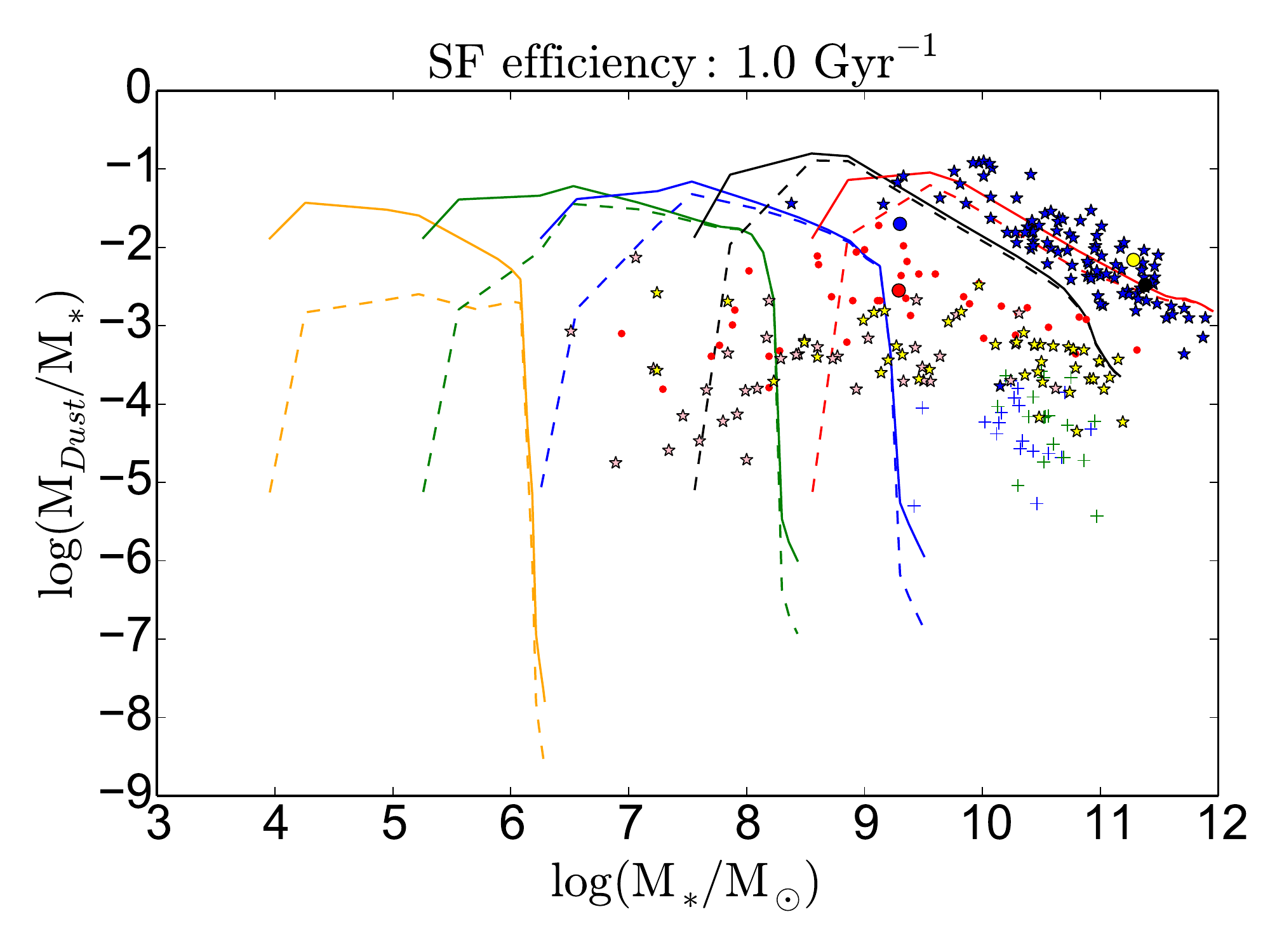}
\includegraphics[width=\columnwidth,angle=0]{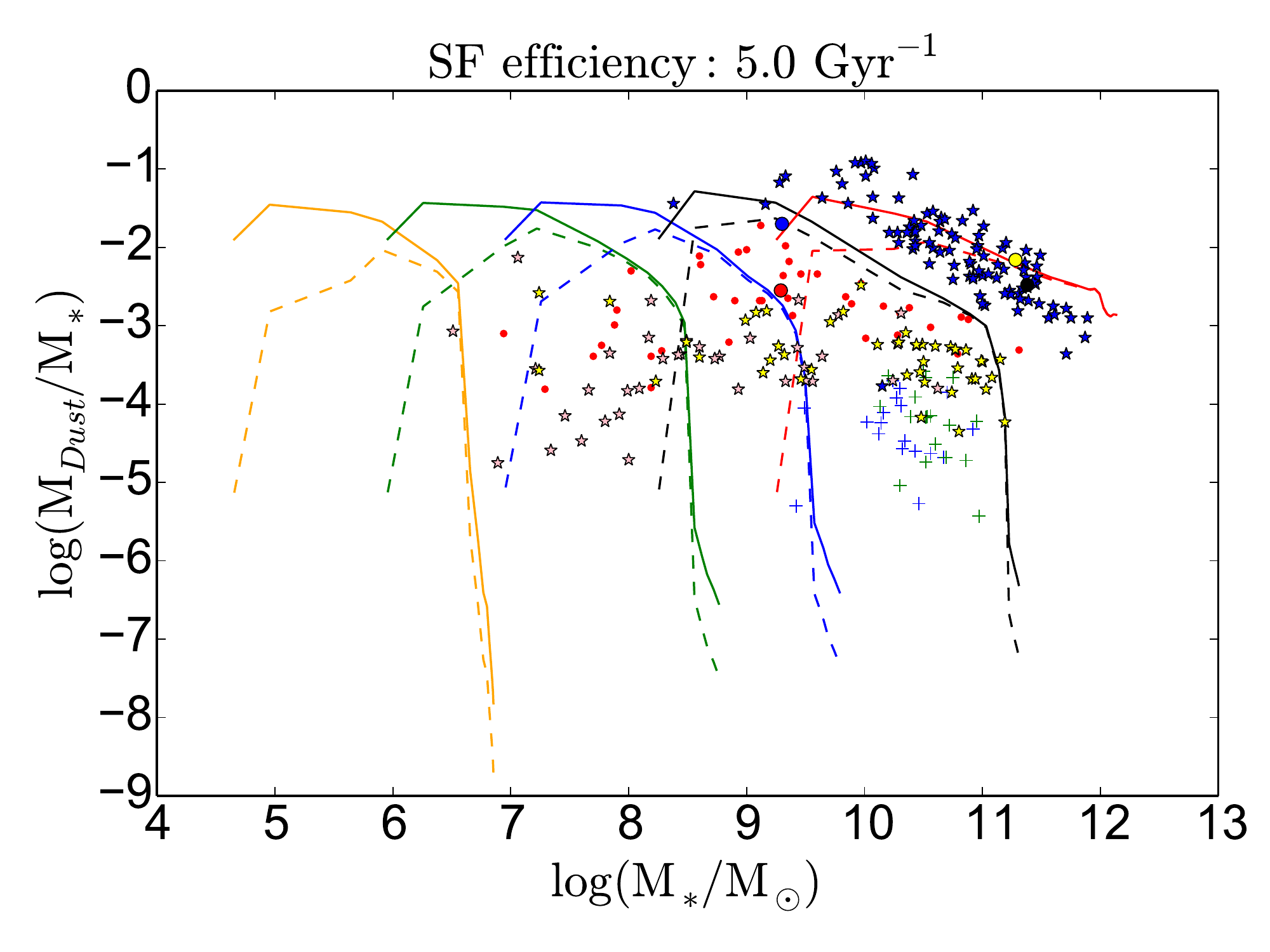}
\includegraphics[width=\columnwidth,angle=0]{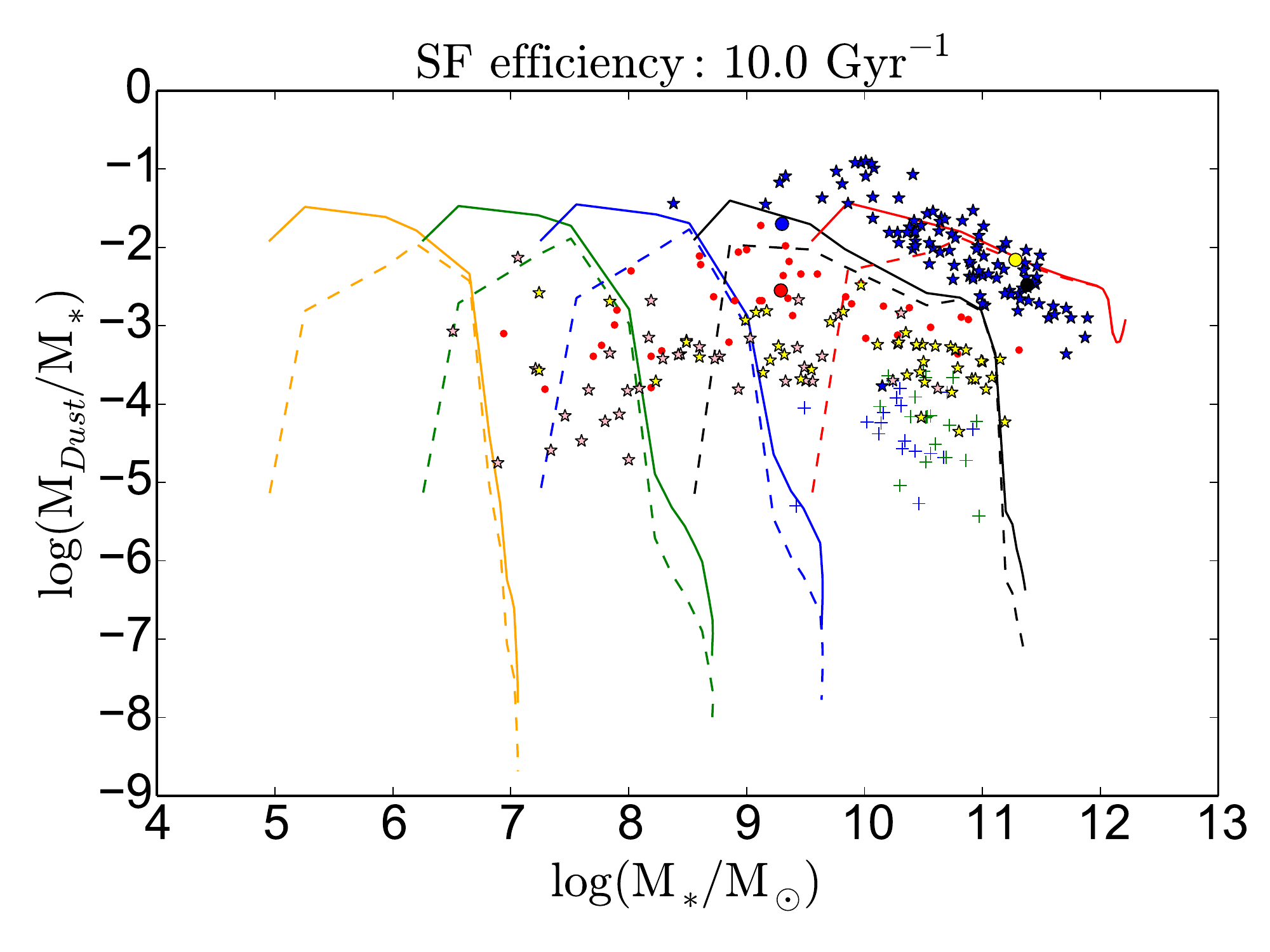}
\caption{Dust-to-star ratio versus stellar mass predicted by the model. Line colours and styles of the model tracks have the same meanings as in Fig.~\ref{fig:dust-time}. The SMG sample of \citet{da2015alma} is represented by blue stars,
the data for galaxies of \citet{remy2014gas,remy2015linking}, by yellow and pink stars, of \citet{de2016herschel}, by small red dots, of ellipitical galaxies of \citet{lianou2016dustier}, by blue crosses, the data for the LBGs D49 and M28 of \citet{magdis2017dust}, by the yellow and black large dots, respectively, and for the reionization epoch DOGs A1689-zD1 of \citet{knudsen2016merger} and A2744\_YD4 of  \citet{laporte2017dust}, by the large blue and red dots, respectively.} 
\label{fig:DUST_Star-Star}
\end{center}
\end{figure*}

The chemodynamical time evolution for gaseous, stellar and dust mass (as well as their sum) are shown in appendix~\ref{App:galaxy-ev} for all computed galaxy models. The model is not a ``closed box'', allowing in- and outflows ($M_G$ is not constant in the simulation). Some models exhibit an abrupt drop in $M_{\rm Gas}$ and $M_{\rm Dust}$ (for both Cases) due to galactic winds led by SNe events ($M_G$ also diminish due to the winds). The bottom-right panel of figures~\ref{fig:Galcomp-3} and \ref{fig:Galcomp-5} illustrate this drop in $M_{\rm Gas}$, $M_{\rm Dust}$ and $M_G$.
% We do not include them in the main text since stars and gas evolution is not the main goal of this work, but rather how they drive the galaxy and dust evolution.
%, we provide them attached to the text.

The time evolution of the dust mass, for all computed models, is shown in Fig.~\ref{fig:dust-time}.
Each panel shows the dust evolution for all $M_{G,0}$ and $\Delta_A$ computed with the same star formation efficiency $\nu_0$. The $\nu_0$ values are signalled on the top of the panels and are organized in ascending order, from left to right and from top to down. Each $M_{G,0}$ model is tagged by colour in figure~\ref{fig:dust-time} and $\Delta_A$ is set as a continuous line for Case A (D98) and dashed in Case B. With the exception of the appendix \ref{App:galaxy-ev}, the model representation in the figures is always the same. The only age estimate available we have is for the SMG sample from \citet{da2015alma}.
%and it is displayed as blue star.

The models with $\nu_0 = 0.1$~Gyr$^{-1}$ (top left panel of Fig.~\ref{fig:dust-time}) do not show significant outflows  and the star formation is almost a continuous process (see appendix~\ref{App:galaxy-ev}); therefore, the galaxy do not lose any significant amount of dust during 13~Gyr of evolution.

For all $M_{G,0}$ models, the $\Delta_A$ Case B converges to Case A due to grain accretion, but the more massive galaxies systematically converge faster than the less massive ones. For the $2 \times 10^{11}$~M$_\odot$ and $2 \times 10^{12}$~M$_\odot$ Cases A and B coincide in a time-scale of $\sim 0.3$~Gyr. The $M_{G,0} = 10^{9}$~M$_\odot$ and $10^{10}$ M$_\odot$ models take $\sim 1.0$~Gyr, and the $5 \times 10^{7}$~M$_\odot$ model takes $\sim 8$~Gyr.

As expected, for higher $\nu_0$ values, outflows expel the gas and dust mass, in a few Gyrs. 
Figure~\ref{fig:dust-time} top-right panel, with $\nu_0 = 1.0$~Gyr$^{-1}$, show that all models with $M_{G,0} \leq 10^{10}$~M$_\odot$ have high dust mass loss between $\sim 1$ and  $\sim 8$~Gyr. The $2 \times 10^{12}$~M$_\odot$ does not lose dust, while the $2 \times 10^{11}$~M$_\odot$ model loses less than 1~dex. For $\nu_0 = 5.0$~Gyr$^{-1}$ (lower left panel) and $\nu_0 = 10.0$~Gyr$^{-1}$ (lower right panel), all models, except $M_{G,0} = 2 \times 10^{12}$~M$_\odot$, suffer strong outflows, 
%\rouge{de novo outflow....}
but in the last panel the $2 \times 10^{12}$~M$_\odot$ the weak outflow is followed by an infall. 
%\rouge{mesmo problema do outflow. Qual \'e a evid\^encia?? {\bf R:} {\it Nos plots do anexo, acha que preciso explicar isso antes?} Acho que seria \'util, n\~ao? Pelo menos uma frase...} 
Due to the deep potential well, the SNe cannot remove the gas and dust from the galaxy. The time needed to eject dust also drops for $\nu_0 = 10.0$ Gyr$^{-1}$, being $\sim 1.0$~Gyr for the less massive models and $\sim 2.0$~Gyr for $2 \times 10^{11}$~M$_\odot$. 

The time taken for both $\Delta_A$ cases to produce an equal amount of dust is also shorter for higher $\nu_0$ values, due to the stellar sources and more efficient accretion in high metallicity systems (see equation \ref{eq:tau_g}). For all models with strong outflows, Case B reaches Case A near the maximum dust mass, and then it abruptly drops after the quenching of SF. The dust production in Case A prevents a more abrupt fall, ending with $\sim 1$~dex more dust mass than Case B. 

Figure~\ref{fig:dust-time} shows that the SMG population (small blue stars) is well represented by the $2 \times 10^{12}$~M$_\odot$ Case A model, between $\sim 0.3$~Gyr and 1.5~Gyr (the SMG stellar population age is in range  $\sim 0.02$~Gyr and 1.5~Gyr), for all $\nu_0$. The inferior threshold belongs to $\nu_0 = 0.1$~Gyr$^{-1}$ and this limit is lower for higher $\nu_0$ values, reaching 1~Gyr for 10.0 Gyr$^{-1}$ model. For an age lower than 0.1~Gyr all $\nu_0$ models underestimate the dust amount when compared with SMG. This difference is probably due to a higher $\nu_0$ required for this objects, indicating a very fast evolution for them, corroborating with the scenario where SMGs are progenitors to present day elliptical galaxies \citep{article,10.1111/j.1365-2966.2006.10673.x}. \citet{Gioannini2017} considered $\nu_0$ between 10.0~Gyr$^{-1}$ and 20.0~Gyr$^{-1}$ for elliptical galaxies and \citet{calura2009evolution} used 25.0~Gyr$^{-1}$ for $M_{G,0} = 2 \times 10^{12}$~M$_\odot$, 
but, for the purpose of this work a more conservative $\nu_0$ is more adequate (see section \ref{sec:Models}).

To evaluate the obscuration in each model,
%and also to compare each M$_{G,0}$ model, 
a key quantity is the dust-to-gas ratio, shown in figure~\ref{fig:dust/gas-time}. 
Unfortunately, observational data with dust and gas masses, together with age estimate, are not available in the literature. The time spent for Case B models to reach Case A is basically the same as in figure \ref{fig:dust-time}, but figure \ref{fig:dust/gas-time} shows different evolutionary pattern than figure \ref{fig:dust-time}.
 
In the top-left panel of Fig.~\ref{fig:dust/gas-time} ($\nu_0 =$ 0.1~Gyr$^{-1}$), the $M_{G,0} = 2 \times 10^{11}$ and $2 \times 10^{12}$~M$_\odot$ models are indistinguishable until 0.3 Gyr (when Case B reaches Case A for both $M_{G,0}$). After 0.3~Gyr, both slopes get shallower, but in the $2 \times 10^{12}$~M$_\odot$ model the shallowness is more pronounced. The models with $10^{9}$ and $10^{10}$~M$_\odot$ follow almost the same track, but the $10^{10}$~M$_\odot$ track is slightly higher. For both models, the Cases A and B converges at $\sim 1$~Gyr, the same time that their dust-to-gas ratio turns higher than the $2 \times 10^{12}$~M$_\odot$ model. At $\sim 4$~Gyr, the $10^{9}$ and $10^{10}$~M$_\odot$ models also surpass $2 \times 10^{11}$~M$_\odot$. The less massive model has approximately a constant slope for Case A and reaches the more massive model after 10~Gyr of evolution (when the Cases A and B convergence).

For all $\nu_0$ and $M_{G,0}$ combinations, the difference between Cases A and B reaches almost 3~dex at the beginning of the simulation. This difference grows smaller during the star formation period, being almost null at the peak (generally at dust-to-gas peak), due to the coupling of ISM accretion and gas depletion. During the passive phase, dust Cases A and B diverges, differing by 1~dex. For higher $\nu_0$, Case B reaches Case A faster, while for higher $M_{G,0}$ the coincidence of both Cases occurs later. %The 0.1 Gyr$^{-1}$ and 5 $\times$ 10$^{7}$ M$_\odot$ model the Case B only reach Case A after $\sim$ 10 Gyr and the model with 2 $\times$ 10$^{12}$ M$_\odot$ the Case B reaches Case A after $\sim$ 2 Gyr and both remains the same during all the galaxy evolution. 

In figure~\ref{fig:dust/gas-time} we see that models that undergo strong outflows episodes have a bump in the dust-to-gas ratio, for both dust production Cases, due to the coupling of SNe dust production and SNe feedback. The bump precedes the quenching of star formation, happens earlier in high $\nu_0$ models, for fixed $M_{G,0}$, and is sensitive to both $M_{G,0}$ and $\nu_0$. In the $\nu_0 = 0.1$~Gyr$^{-1}$ panel, no model shows a bump. Models with $M_{G,0} \leq 10^{10}\, \mathrm{M}_{\odot}$ show for $\nu_0 = 1.0$~Gyr$^{-1}$, and $M_{G,0} \geq 2 \times 10^{11}\, \mathrm{M}_{\odot}$ show for $\nu_0 = 5.0$~Gyr$^{-1}$.% The bump is also reached sooner for high $\nu_0$ values.

For $\nu_0 = 5.0$~Gyr$^{-1}$ and $10.0$~Gyr$^{-1}$, the dust-to-gas ratio in Case A is nearly insensitive to $M_{G,0}$ for the first 1~Gyr and 0.3~Gyr of galaxy evolution, respectively.
Case B also shows a similar pattern, but the evolutionary tracks have a higher dispersion among themselves.
Before the dust-to-gas bump, the amount of dust is almost insensitive to $\Delta_A$ for all models. After the maximum, the dust-to-gas ratio drops and becomes sensitive to $\Delta_A$ and this continues during the entire passive phase. Since in the $2 \times 10^{12}$~M$_\odot$ models star formation is never quenched, they are always insensitive to $\Delta_A$ after Case A and B converge.

Figures~\ref{fig:dust-time} and \ref{fig:dust/gas-time} model the time spent to build up the dust bulk of galaxies. The object A1689-zD1, $z  =  7.5$, has $M_{\rm Dust} = 4.0 \times 10^7  \, \mathrm{M}_\odot$, $M_{*} = 2.0 \times 10^9 \, \mathrm{M}_\odot$ and a molecular gas reservoir of 7.2 $\times$ 10$^9$ M$_\odot$ \citep{knudsen2016merger}, resulting in a dust-to-mass ratio (assuming only molecular gas) of $5.5 \times 10^{-3}$, while A2744\_YD4, in $z =8.38$, has $M_{\rm Dust} = 5.5 \times 10^{6} \, \mathrm{M}_\odot$ and $M_{*} = 1.95 \times 10^9 \, \mathrm{M}_\odot$ \citep{laporte2017dust}. For the adopted cosmology, the time available for each galaxy to evolve is 0.7 Gyr and 0.6 Gyr, respectively, and this impose a strong constrain in both galaxy evolution model concerning dust production prescriptions.
The most suitable initial mass to describe these systems is $10^{10}$~M$_\odot$, since A1689-zD1 has $M_G = 9.2 \times 10^9$~M$_\odot$, and the stellar mass from this galaxy is quite similar to A2744\_YD4. 
In figure \ref{fig:dust-time}, in $M_{G,0} = 10^{10}$~M$_\odot$ and $\nu_0 = 10.0$~Gyr$^{-1}$ model, the dust mass reaches A2744\_YD4 value in $\sim 0.5$~Gyr, but the maximum value of $M_{\rm Dust}$ for this model is $6.3 \times 10^6$~M$_{\odot}$, about six times smaller than A1689-zD1 dust mass and takes 0.5 Gyr to reach the peak. In figure \ref{fig:dust/gas-time}, the same model takes about $\sim 0.4$~Gyr to reach A1689-zD1 dust-to-gas ratio. These results are insensitive to $\Delta_A$.

Since stellar mass is possibly the most remarkable quantity in galaxy evolution, in figure \ref{fig:DUST-Star} we show the relation between stellar and dust mass while in figure \ref{fig:DUST_GAS-Star} the dust-to-gas ratio and stellar mass relation. Figure~\ref{fig:DUST-Star} contains data from \citet{lianou2016dustier, de2016herschel, remy2014gas} and \citep{remy2015linking, magdis2017dust, da2015alma, knudsen2016merger} and \citep{laporte2017dust}.  For all $\nu_0$ the dust mass models tend to lie above the data during the star forming phase, but for the $\nu_0 = 5.0$~Gyr$^{-1}$ case all $M_{\rm Dust}$ tracks show a knee that passes over the data points. The exception are SMGs and two Lyman-breaks galaxies, that are in agreement with $M_{G,0} = 2 \times 10^{12}$~M$_\odot$ models, for all $\nu_0$. SMGs and these two LBGs have similar properties as infrared luminosity, $M_{\rm Dust}$, $M_{\rm Gas}$, $M_*$ and $z$. In fact, previous studies with the chemodynamical model used here, have shown that the early evolution of the formation of massive spheroids reproduces the properties of SMGs \citep{archibald2002coupled, rosa2004detectability} and LBGs \citep{friacca1999lyman}. The sample from \citet{lianou2016dustier} shows low $M_{\rm Dust}$ for a given $M_*$, generally with $M_* >  10^{10}$~M$_\odot$, and is compatible with the passive phase of our models. While $\nu_0$ becomes higher, the tracks approach the data points, being the closest for $\nu_0 = 10.0$~Gyr$^{-1}$. A1689-zD1 and A2744\_YD4 also have too much dust for $\nu_0 = 10.0$~Gyr$^{-1}$ and $M_{G,0} = 10^{10}$~M$\odot$ models, for both dust Cases, but the former lies always near the $2 \times 10^{11}$~M$\odot$ models, and the last reionization galaxy agrees with $\nu_0 = 5.0$~Gyr$^{-1}$.

We also show the relation between dust and gas masses in figure \ref{fig:DUST-x-GAS}. For $\nu_0 = 0.1$~Gyr$^{-1}$ models (with exception of $M_{G,0} = 5 \times 10^{7}$~M$\odot$), the gas mass increases just before the depletion into stars, while higher $\nu_0$ models do not show pronounced gas mass enhancement. For all galaxy models, the evolutionary track shows two patterns, the first is related to the beginning of galaxy evolution, with $M_{\rm Gas} \sim M_{\rm G,0}$, for any $M_{\rm Dust}$, while the other expresses the $M_{\rm Gas}$ depletion into stars or the gas eject in outflow episodes. The transition between them forms a knee that lies close to the observational data. 

For all models, the difference between Case A and B dust masses reaches more than 2~dex at the beginning of galaxy evolution, but it has almost vanished when the tracks reach the knee. For models that do not undergo a strong outflow, Cases A and B remain almost the same after the knee, while the models that undergo strong outflows exhibit a difference lower than 1~dex. The high-$z$ galaxies lie in the high gas and dust mass locus and the LBGs have the higher gas and dust mass of our sample.

Figure \ref{fig:DUST_GAS-Star} has data from \citet{lianou2016dustier, remy2014gas,remy2015linking,magdis2017dust} and \citet{knudsen2016merger}. The shape of evolutionary tracks do not change significantly while $\nu_0$ grows, but they shift toward the high $M_*$ direction and the lower dust-to-gas ratio limit goes up. For $\nu_0 = 0.1$~Gyr$^{-1}$, the major part of the data is well represented by the $M_{G,0} = 2 \times 10^{11}$~M$_\odot$ models and the lower limit of dust-to-gas ratio ($\sim 2 \times 10^{-7}$ in Case A and $\sim 10^{-10}$ in Case B) is lower than the lower data value ($\sim 10^{-6}$). For $\nu_0 = 10.0$~Gyr$^{-1}$, almost all data lies between the tracks $10 ^9$~M$_\odot$  and $2 \times 10^{11}$~M$_\odot$ and the dust-to-gas ratio lower limit up to $\sim 2 \times 10^{-5}$, in Case A, and $10^{-8}$ in Case B. Case B is better to explain the dust-to-gas mass ratio of low dust galaxies, possible due to evolution in $\Delta_A$.

A1689-zD1 has dust-to-gas ratio higher than galaxies with the same stellar mass (see figure \ref{fig:DUST_GAS-Star}), probably due to its high dust amount (in figure \ref{fig:DUST-Star} it is near of low mass SMG), and is, again, best represented by $M_{G,0} = 10^{10}$~M$\odot$, for any $\nu_0$, and it is insensitive to $\Delta_A$. For $\nu_0 = 0.1$~Gyr$^{-1}$, the dust produced in $10^{10}$~M$\odot$ models takes 13 Gyr to reach the $M_{\rm Dust}$ observed in A1689-zD1 and a half of its $M_{*}$. For $\nu_0 = 5.0$~Gyr$^{-1}$ track peak is less than 1 dex lower than the data, while for 1.0~Gyr$^{-1}$ and 10.0 Gyr$^{-1}$ the tracks are quite compatible with A1689-zD1, but the former requires $\sim 3$~Gyr to reach its dust-to-mass ratio, while the latter takes $\sim 0.5$~Gyr.

Similarly, the two LBGs are closer to $M_{G,0} = 2 \times 10^{11}$~M$\odot$ models, for both $\Delta_A$ and all $\nu_0$. For 0.1~Gyr$^{-1}$, the tracks do not reach the $M_*$ estimated for this objects, even after 13~Gyr, and the track slightly drops when $M_* \sim 10^{11}$~M$_\odot$, which is the general pattern for star formation quenching. For 1.0~Gyr$^{-1}$, a $\sim 3$~dex difference in dust-to-gas ratio is seen between the track and the data. In the two higher $\nu_0$ models, the track ends near the galaxies data points, with the dust-to-gas peak slightly displaced to lower $M_*$. The $\nu_0 = 5.0$~Gyr$^{-1}$ models take $\sim 2$~Gyr to reach the peak, while the $\nu_0 = 10.0$~Gyr$^{-1}$ require $\sim$ 1.5 Gyr (see figure \ref{fig:dust/gas-time}). It is worth to stress that $M_G$ (molecular gas + $M_*$) of these galaxies are $4 \times 10^{11}$ and $2.5 \times 10^{11}$~M$_\odot$, for D49 and M28, respectively, and, therefore, we expect a fast evolution for this objects. The age of the universe in $z \sim 3$ is $t_H \approx 2.1$~Gyr, therefore the highest $\nu_0$ models are more suitable to explain both dust mass and dust-to-gas ratio of high-$z$ massive galaxies.

In general, passive elliptical galaxies show high dust-to-mass ratio due much more to lack of gas than to their dust amount. In fact, the data from \citet{lianou2016dustier}, in figure \ref{fig:DUST-Star} shows that these galaxies have small $M_{\rm Dust}$ for a given $M_*$, but due to their extremely small $M_{\rm Gas}$, they show high dust-to-gas ratio.

The relation of $M_{\rm Dust}$ and dust-to-gas ratio with SFR are shown in figures~\ref{fig:dust-sfr} and \ref{fig:dust/gas-sfr}. It is interesting to notice that $M_{\rm Dust}$ is well represented by low star formation galaxies, while the dust-to-gas ratio is better modelled by high star formation ones.

In figure \ref{fig:dust-sfr}, the SMGs show good agreement with $M_{G,0} = 2 \times 10^{12}$~M$_\odot$ evolutionary track, but while $\nu_0$ grows, the star formation rate tends to be overestimated for $\nu_0 = 10 $~Gyr$^{-1}$. The model with $2 \times$ 10$^{11}$ M$_\odot$ and $\nu_0 =$ 10 Gyr$^{-1}$ agrees with SMG star formation rate, although underestimate dust mass. For $\nu_0 =0.1$~Gyr$^{-1}$ models, the evolutionary tracks match the sample from KINGFISH and from Herschel survey, while DGS generally show less dust than the tracks, being better represented by $\nu_0 =$ 1.0~Gyr$^{-1}$ models. A1689-zD1 is more likely 2 $\times$ 10$^{11}$ M$_\odot$, for $\nu_0 =$ 0.1 Gyr$^{-1}$ and 10$^{10}$ M$_\odot$, but underestimating dust mass, for $\nu_0 =$ 10 Gyr$^{-1}$, been the last more suitable due to the stellar mass correspondence (see figure \ref{fig:DUST-Star}). Elliptical galaxies lies systematically above the star forming sample (KINGFISH and HERSCHEL) for the same SFR. Interesting to note that the SFR tends to be constant during the its peak.

The dust-to-gas ratio and SFR relation, figure  \ref{fig:dust/gas-sfr}, are shown together with KINGFISH, Herschel, elliptical galaxies and A1689-zD1. For this figure, as in figure \ref{fig:DUST_GAS-Star}, near the SFR knee, the model $\nu_0 = 10$~Gyr$^{-1}$ and $M_{G,0} = 10^{10}$~M$_\odot$ shows good agreement with A1689-zD1, KINGFISH and DGS show high dispersion for dust-to-gas in the range $\sim 2 \times 10^{-6}$ and $5 \times 10^{-2}$, and SFR in the range $\sim 6 \times 10^{-3}$ and 10~M$_\odot$yr$^{-1}$ (KINGFISH shows little more dust-to-gas than DGS). The elliptical galaxies lie, again, in high dust-to-gas and low SFR region.

As \citet{calura2016dust} we use dust-to-stellar mass ratio as probe to galaxy evolution (figure \ref{fig:DUST_Star-Star}), since this ratio means a true measurement of the global dust production efficiency, or a real balance between the dust produced and the dust mass loss during galaxy evolution. The elliptical galaxies, from \citet{lianou2016dustier}, have always low dust-to-stellar mass ratio (generaly between 10$^{-5}$ and 10$^{-4}$). \citet{remy2014gas} and \citet{remy2015linking} tends to have dust-to-stellar mass ratio lower than \citet{de2016herschel} gas rich sample, even for the same stellar mass, while SMG \citep{da2015alma} generally has the larger ratio. The LBGs lie together with SMG sample in figure \ref{fig:DUST_Star-Star}, and A1689-zD1 \citep{knudsen2016merger} has dust-to-stellar mass ratio lower than SMG sample and higher than A2744\_YD4 \citep{laporte2017dust}, while the former lies into the high rate of \citet{de2016herschel} sample and the last lies in the lower rate.

The simulation tracks from figure \ref{fig:DUST_Star-Star} lie systematically above the observational sample during the broad of star formation, except by SMG sample, that has good agreement for all star formation efficiency and mainly for the 1.0 Gyr$^{-1}$ , but local normal galaxies seem that it is more suitable to high SFR scenario. Models with strong outburst have a similar pattern and similar dust-to-stellar mass ratio, and after the star formation quenching the track are able to explain elliptical galaxies for objects with M$_*$ in the range 1 $\times$ 10$^{10}$ M$_\odot$ and 2 $\times$ 10$^{11}$ M$_\odot$. The data do not represents the simulation with 5 $\times$ 10$^{7}$ M$_\odot$. The difference between dust Case A and B is well pronounced in the star forming epoch, mainly for low mass e low star formation efficiency, when the stellar dust source domains over accretion.  

\section{Discussion}
\label{sec:Discussion}

\begin{figure*}
\begin{center}
%\setcaptionmargin{1cm}
\includegraphics[width=1. \columnwidth,angle=0]{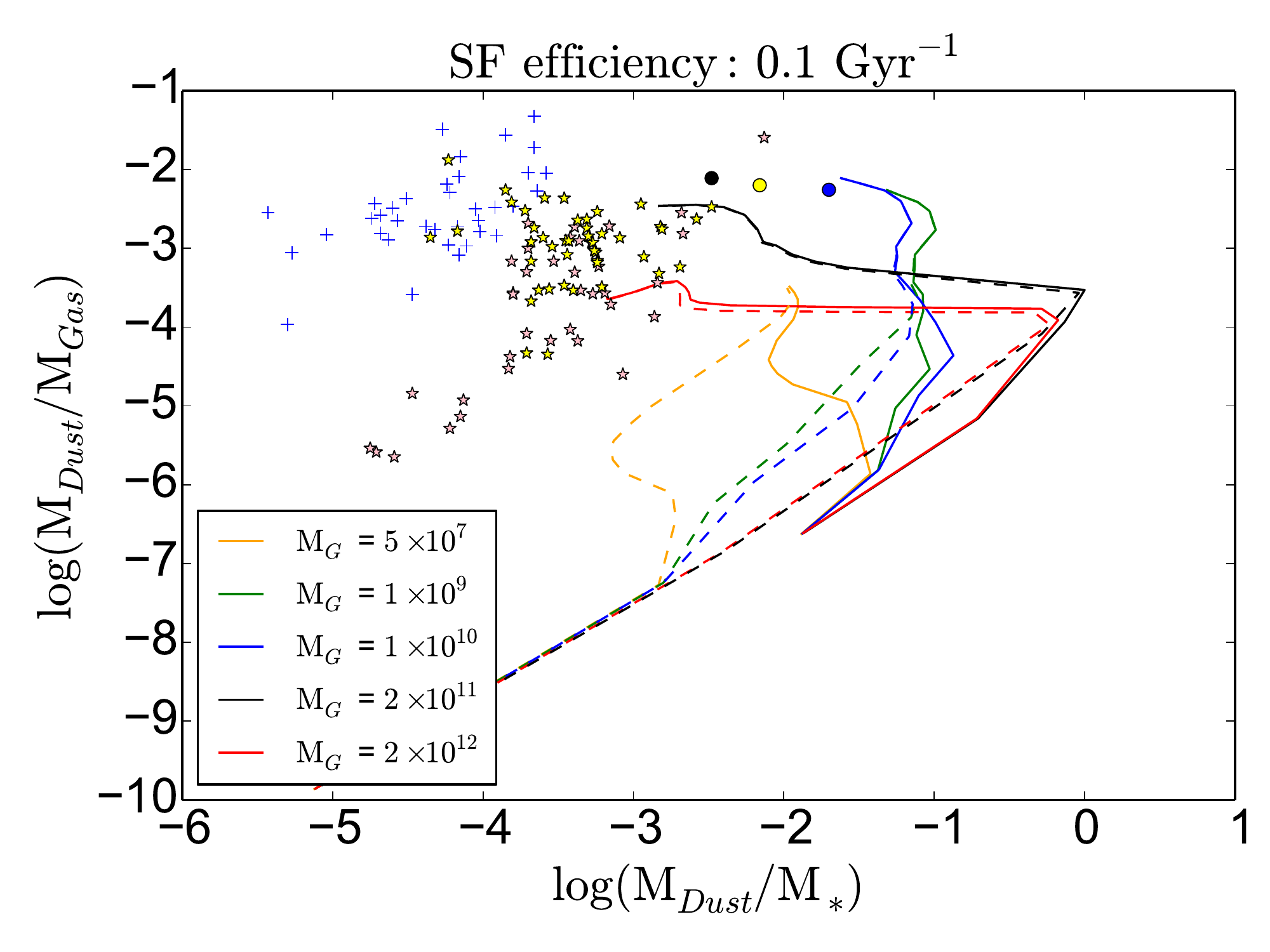}
\includegraphics[width=1. \columnwidth,angle=0]{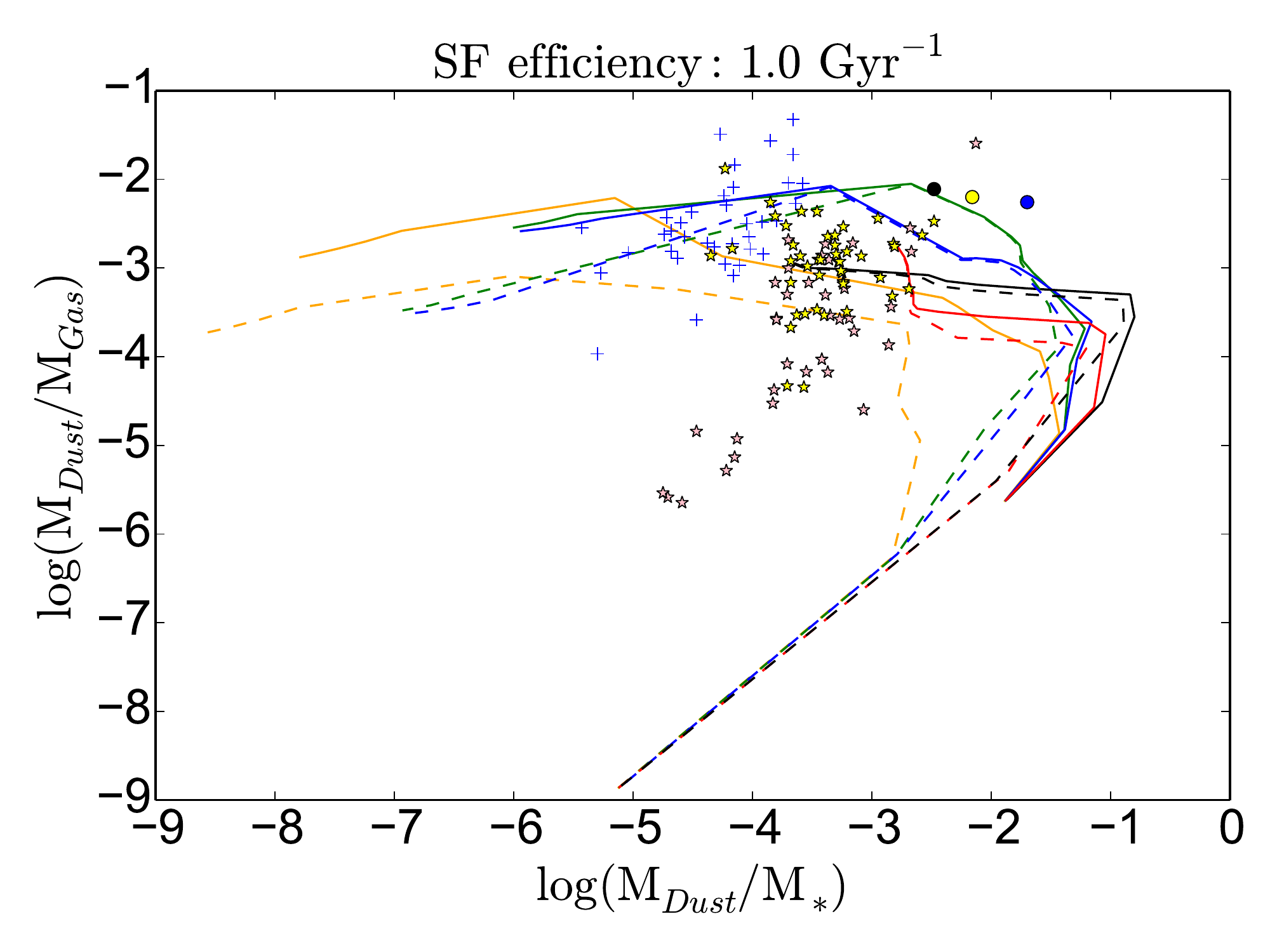}
\includegraphics[width=1. \columnwidth,angle=0]{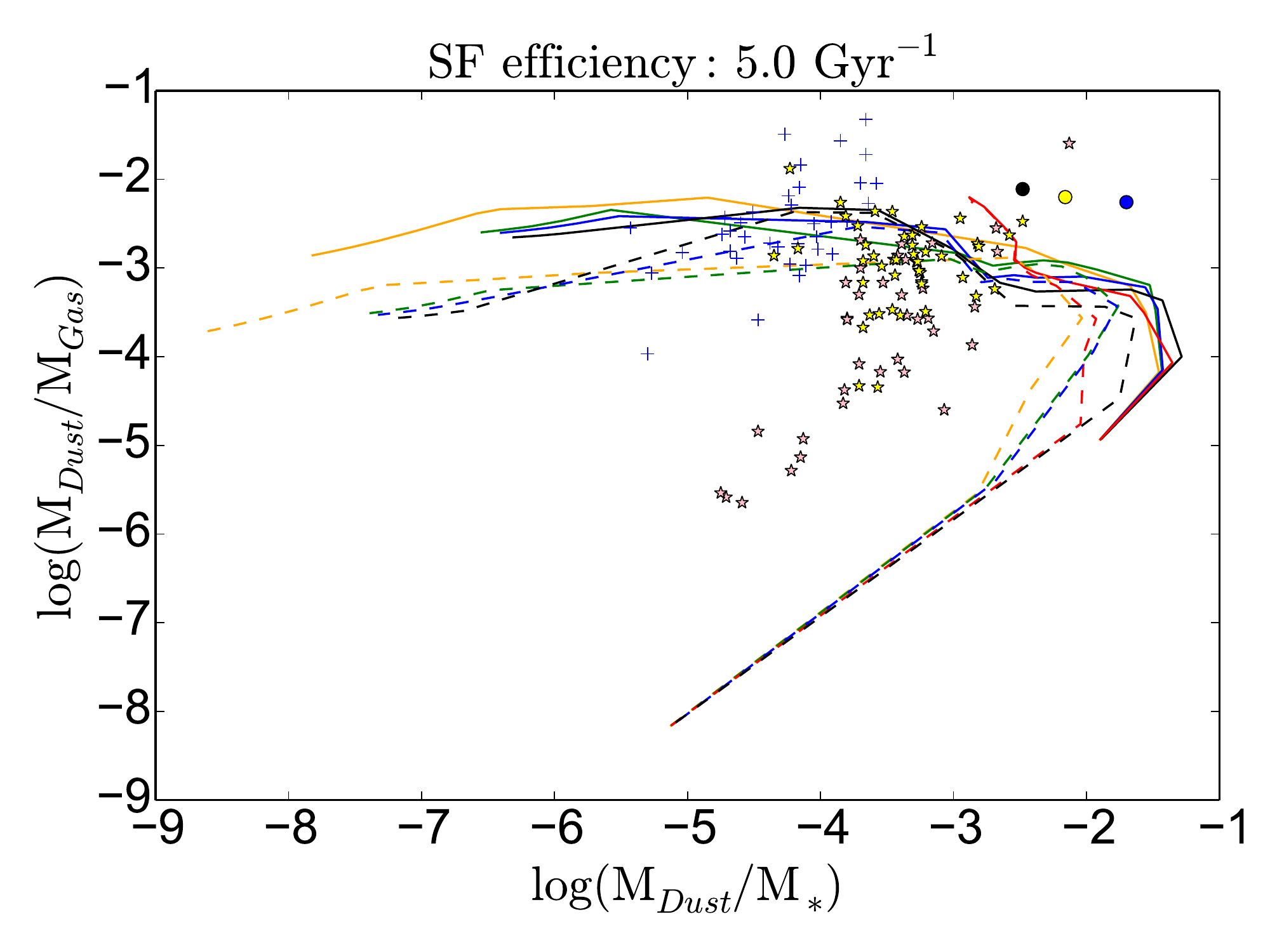}
\includegraphics[width=1. \columnwidth,angle=0]{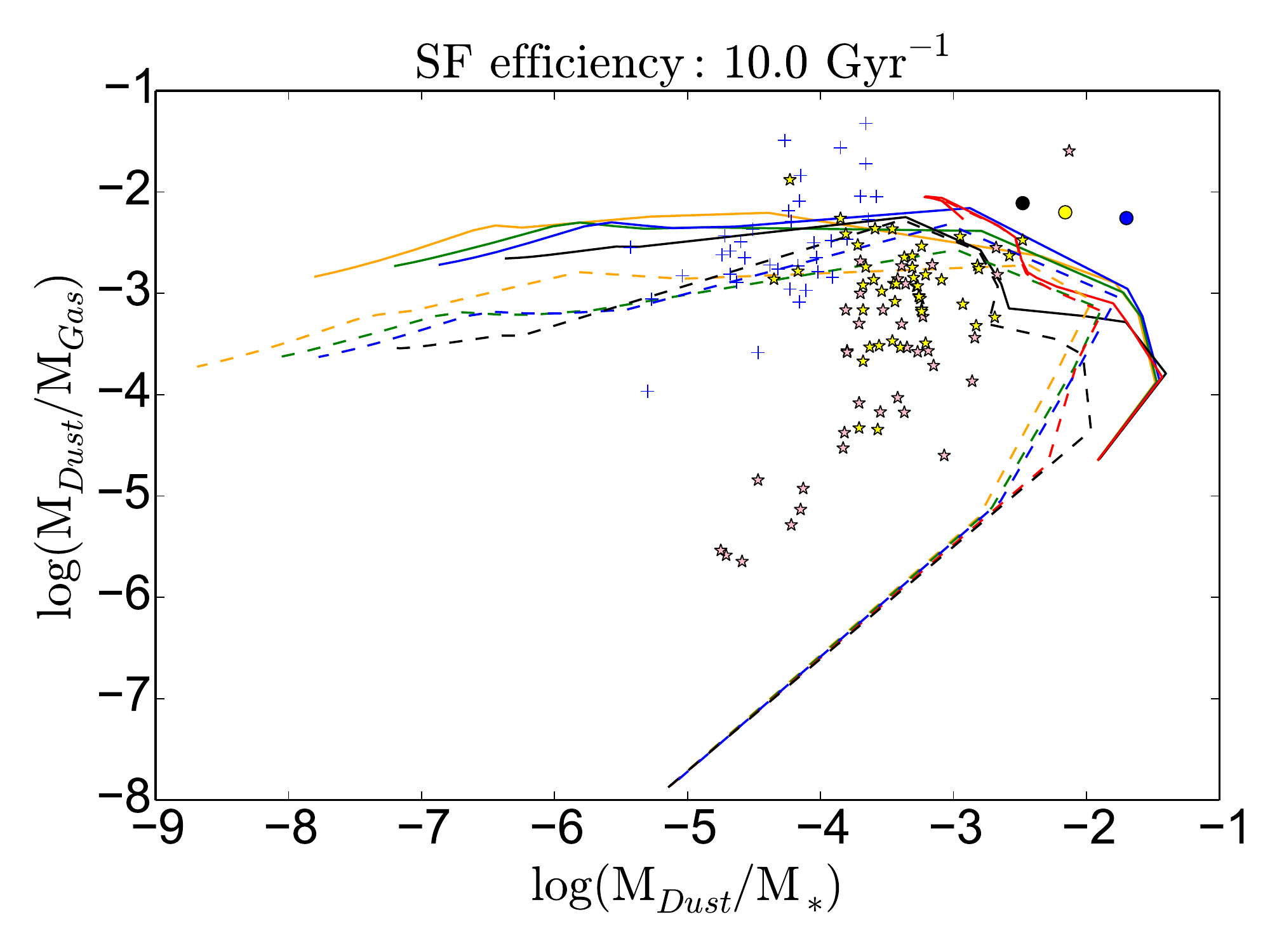}
\caption{
Dust-to-star ratio versus stellar formation rate predicted by the model. Line colours and styles of the model tracks have the same meanings as in Fig.~\ref{fig:dust-time}. The the data for galaxies of \citet{remy2014gas,remy2015linking}, by yellow and pink stars, of the ellipitical galaxies of \citet{lianou2016dustier}, by blue crosses, the data for the LBGs D49 and M28 of \citet{magdis2017dust}, by the yellow and black large dots, respectively, and for the reionization epoch DOG A1689-zD1 of \citet{knudsen2016merger}, by the large blue dots.} 
\label{Dusttogas-Dusttostar}
\end{center}
\end{figure*}

\begin{figure*}
\begin{center}
%\setcaptionmargin{1cm}
\includegraphics[width=1.0 \columnwidth,angle=0]{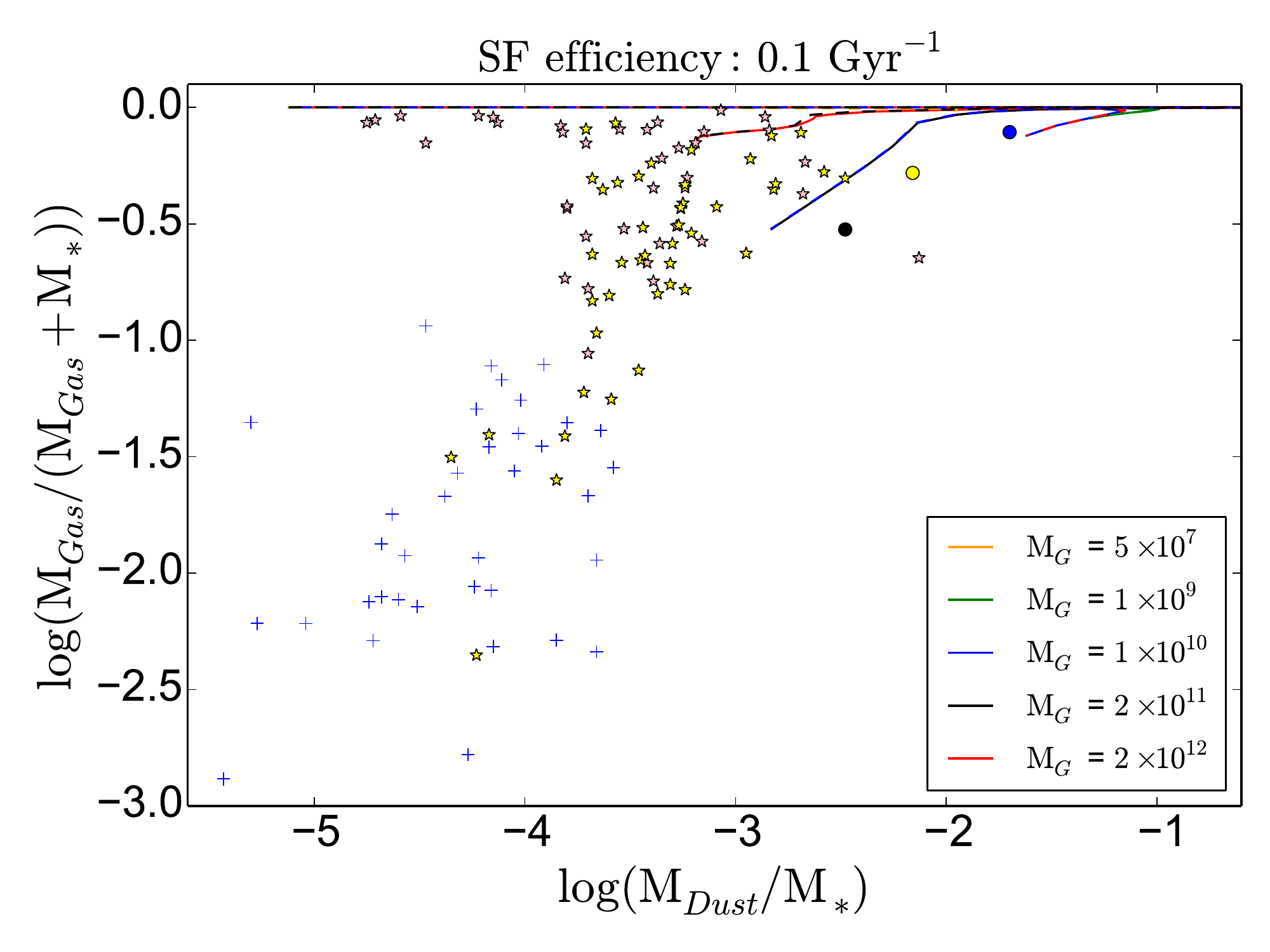}
\includegraphics[width=1.0 \columnwidth,angle=0]{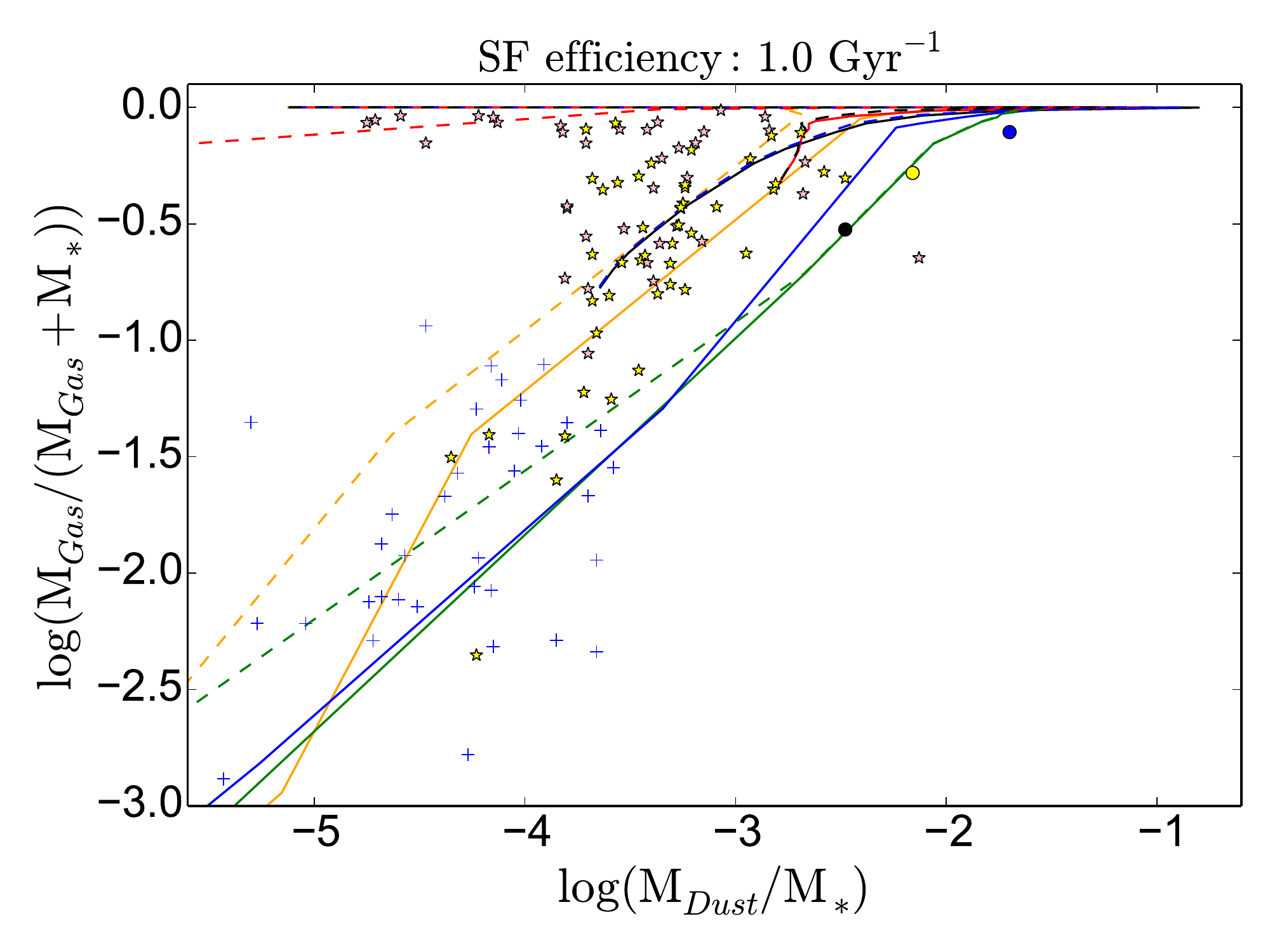}
\includegraphics[width=1.0 \columnwidth,angle=0]{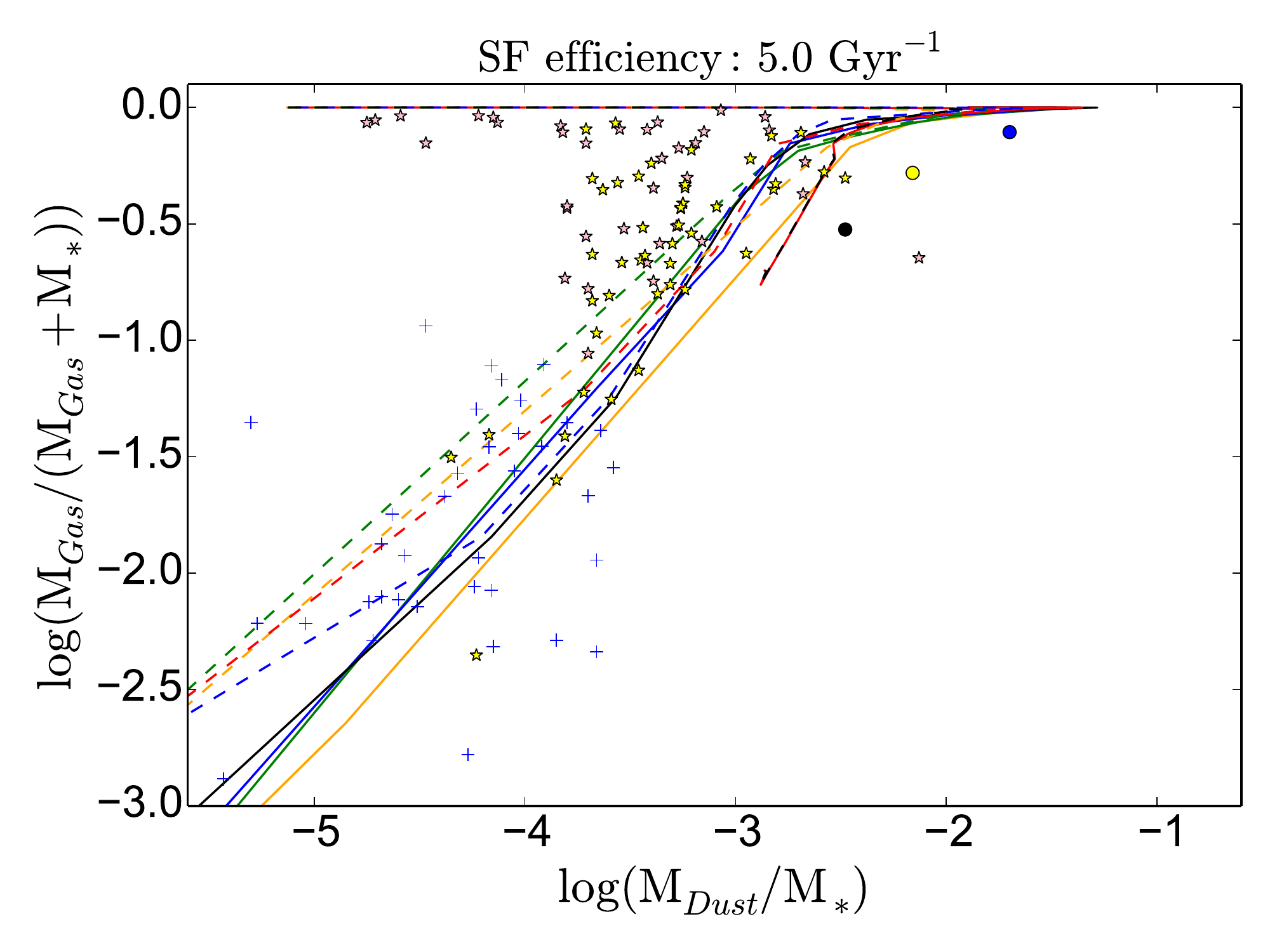}
\includegraphics[width=1.0 \columnwidth,angle=0]{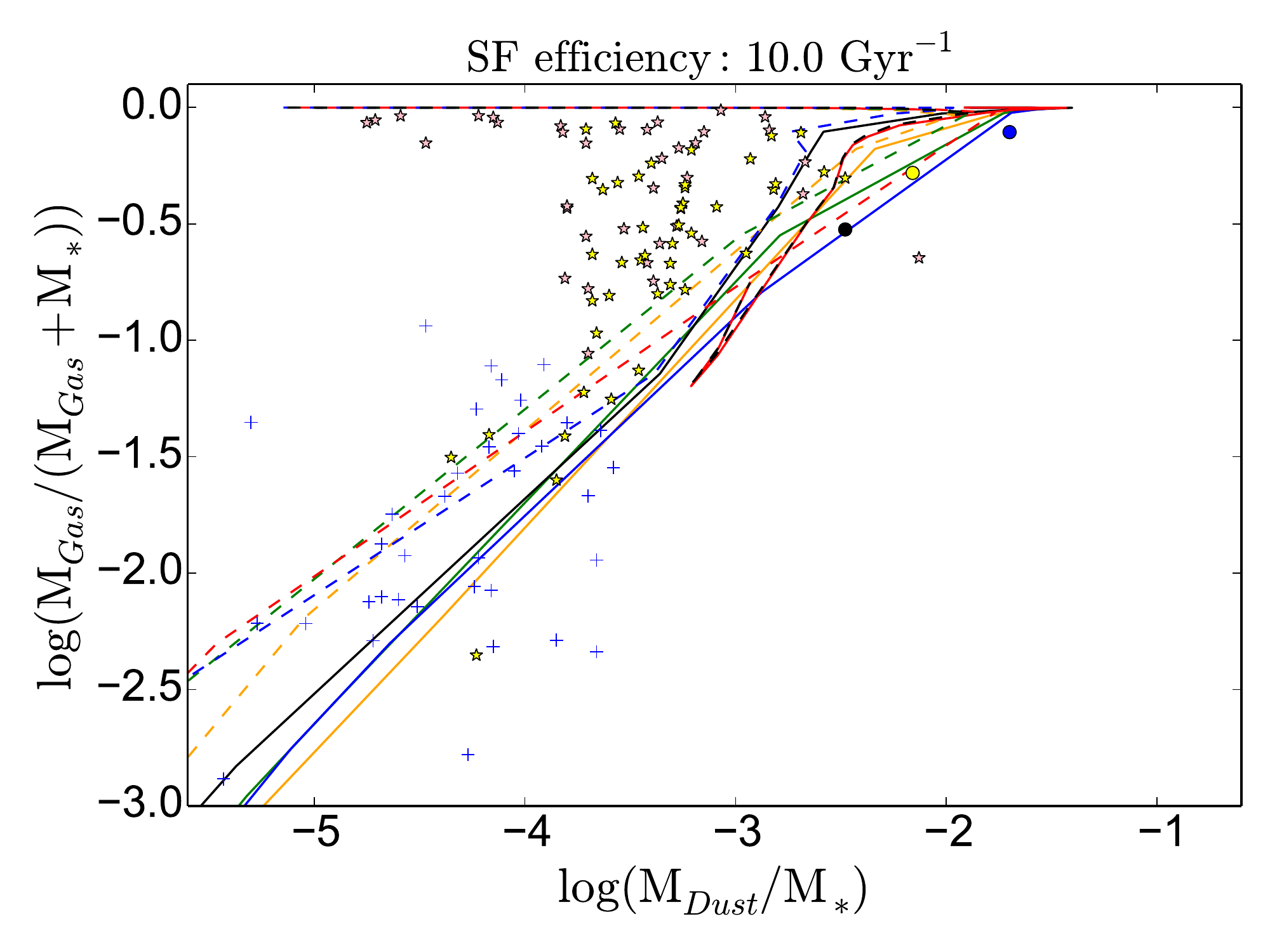}
\caption[]{ Galaxy gas fraction versus Dust-to-star ratio predicted by the model. Track line colours and styles have the same meaning as in Fig.~\ref{fig:dust-time}. The the data for galaxies from \citet{remy2014gas,remy2015linking} are yellow and pink stars, for elliptical galaxies from \citet{lianou2016dustier} are blue crosses, for the LBGs D49 and M28 from \citet{magdis2017dust} are the yellow and black large dots, respectively, and for the reionization epoch DOG A1689-zD1 from \citet{knudsen2016merger} is a large blue dot.} 
\label{fig:Barion-x-Dust-star-log}
\end{center}
\end{figure*}

In this work we carried out forty galaxy evolution simulations, varying the galaxy initial baryonic mass, $M_{G,0}$, star formation efficiency, $\nu_0$, and dust coagulation efficiency, $\Delta_A$, in order to investigate the main processes that drive dust mass evolution. We aim to explain the presence of high-$z$ DOGs and the relation between obscured star formation and galactic $M_*$. The simulations are compared with a collection of data available in the literature, covering a large range of mass and redshift.

In this section we discuss the main implication of our results, splitting the issues thematically. In sub-section \ref{ssec:parameters} we analyze the role of star formation and dust efficiency implication, in \ref{ssec:high-z-implication} the dust evolution constraint for high-$z$ galaxies and in \ref{ssec:obscuration} the implications for dusty obscured galaxies.

\subsection{$\nu_0$ and $\Delta_A$ effects}
\label{ssec:parameters}

The difference between dust produced in Cases A and B is more pronounced in small $\nu_0$ and small $M_{G,0}$ models (see figures \ref{fig:dust-time} and \ref{fig:dust/gas-time}). 
In models with strong outflows (high $\nu_0$), the dust mass is sensitive to $\Delta_A$ during the stellar mass assembly and passive phase, but they are almost insensitive to $\Delta_A$ between the maximum and the quenching of star formation. This is more clear in figures \ref{fig:dust/gas-sfr} and \ref{fig:dust-sfr}, where a difference of more than 3 dex is between Case A and B (in both figures) in the begin of the stellar mass assembly. For galaxies that pass by passive phase, dust mass (figure \ref{fig:dust/gas-sfr}) is more sensitive to $\Delta_A$ than dust-to-gas ratio (figure \ref{fig:dust-sfr}) due to the ejection of both dust and gas during outflow. 
As a consequence, high SFR systems and starbursts galaxies are not adequate to constrain $\Delta_A$ value. Passive and very evolved galaxies also have a great limitation due to the number of process involved in dust accumulation. In their place we suggest low star formation and low metallicity galaxies with young stellar population ($\leq$ 0.4 Gyr), like dwarf irregular galaxies and DLAs systems \citep{Gioannini2017}.

Coupling the equations \ref{eq:accretion} and \ref{eq:tau_g}, we notice that dust accretion in ISM is proportional to dust mass and metallicity. Stars and SNe are the primary source of dust and metals, so, in low $\nu_0$ models, the grain growth by accretion is not efficient enough to rule the dust mass evolution, as in \citet{aoyama2016galaxy} and \citet{Gioannini2017}, and the stellar production domains, making the dust amount in this systems more sensitive to $\Delta_A$, even for AGB sources, as IMS has time to evolve into AGB. In figure \ref{fig:dust-time}, in $M_{G,0} = 1 \times 10^{10}$ M$_\odot$ and $\nu_0 =$ 10.0 Gyr$^{-1}$ model, dust mass reaches the A2744\_YD4 value in $\sim 0.5$~Gyr, but it never reaches the A1689-zD1 dust mass. In figure \ref{fig:dust/gas-time} the same model needs $\sim 0.4$~Gyr to reach the A1689-zD1 dust-to-gas ratio.

Another advantage in investigate dust production in low mass galaxies is that low star formation galaxies, mainly the ones with younger stellar population, is not too sensible to star formation history and feedback recipes, although it can be more sensitive to the adopted IMF. A top-heavy IMF increases the CCSNe number, making the dust bulk production more efficient.

Search for dust in low star formation and metallicity galaxies can bias $\Delta_A$ estimation if their value is, at least, reasonably sensitive to metallicity \citep[as][values are]{Piovan2011}, and the detection and characterization of dust mass and composition in low star formation rate galaxies (like dwarf irregulars) is more difficult and expensive. But we argue that the dust mass is dominated by stellar sources and the number of variables is substantially reduced in these systems and balance some of the difficulties. Following \citet{calura2016dust}, the dust-to-stellar mass ratio can be a valuable tool to probe dust production in low star formation systems (see figure \ref{fig:DUST_Star-Star}) and to measure the stellar dust efficiency production. 

%From our results, is clear that dust, gas and star masses, and star formation have a complex interplay among themselves \rouge{acho que esta complexidade j\'a \'e bem conhecido h\' algum tempo, n\~ao?? R: Acho que sim, mas eu queria apenas introduzir o tema do par\'agrafo, do conte\'udo}. 
While the dust mass data have three regions in figure \ref{fig:DUST-Star}, passive early type, normal star forming and starburst like SMG and Lyman break galaxies. The connection between dust-to-gas ratio and $M_*$ data in figure \ref{fig:DUST_GAS-Star} is more spread than in figure \ref{fig:DUST-Star} and with less defined regions. This behavior probably is due to different evolutionary stage of each galaxy, with variations in gas reservoirs available to star formation. In fact, the models do not represent all these quantities at the same time, as can be noticed comparing figures \ref{fig:DUST-Star}, \ref{fig:dust-sfr} and \ref{fig:DUST_Star-Star}.

To balance evolutionary effects, we propose, as far as we know for the first time, a $M_{\rm Dust}/M_{\rm Gas}$ by $M_{\rm Dust}/M_{*}$ diagram (figure \ref{Dusttogas-Dusttostar}), allowing the link of all baryonic masses, $M_{\rm Dust}$, $M_{\rm Gas}$ and $M_{*}$, as well as the SFR. In figure \ref{Dusttogas-Dusttostar} the star forming galaxies follow a clear path, while the elliptical galaxies lie at higher dust-to-gas place, per dust-to-star, than the star forming ones. Unfortunately, the SMGs sample is not included, but the high-$z$ LBGs and A1689-zD1 are, lying with high dust-to-gas and high dust-to-stars. The distinction between star forming and passive galaxies is pronounced, but again the high-$z$ sample do not exhibit a distinguished pattern. The evolutionary tracks do not represent the data trends very well, mainly for low SFR models. For any $\nu_0$, the tracks has lower dust-to-gas ratio for dust-to-star ratio, than the data, suggesting to much gas in our model and the tracks turn on is near the high-$z$ galaxies. In the figure \ref{Dusttogas-Dusttostar}, Case B formulation suits better DGS than Case A, suggesting or a low $\Delta_A$ or its metallicity dependence. The dust Case B coupled with a top-heavy IMF could soften the tension but a more precise answer relies in more observational data (in high- and low-$z$, and for all galaxies types), and more theoretical works. 

The $M_{\rm Dust}/M_{\rm Gas}$ by $M_{\rm Dust}/M_{*}$ diagram (figure \ref{Dusttogas-Dusttostar}) is very similar to the fundamental relation among $M_*$, SFR and metallicity \citep[see][]{mannucci2010fundamental}. $M_{*}$ and $M_{\rm Gas}$ are linked by SFR, and $M_{\rm Dust}/M_{\rm Gas}$ relies in metallicity \citep[as discussed in][]{Galliano_2008, Galliano_2018, DeVis2019}. Even so, $M_{\rm Dust}/M_{\rm Gas}$ by $M_{\rm Dust}/M_{*}$ diagram facilitates the comparison between evolutionary tracks and observational data because it takes into account the efficiency of dust production ($M_{\rm Dust}/M_{*}$), the gas reservoir available to form stars ($M_{\rm Gas}$), and the past evolution $M_{\rm Dust}/M_{\rm Gas}$.

Another way to allow for evolutionary effects is by linking the gas fraction ($M_{\rm Gas}/(M_{*} + M_{\rm Gas})$) with the galactic dust production efficiency ($M_{*}$/$M_{\rm Dust}$). We show this plot as Fig.~\ref{fig:Barion-x-Dust-star-log}). 

In the $M_{\rm Gas}/(M_{*} + M_{\rm Gas})$ by $M_{*}$/$M_{\rm Dust}$ diagram, the observational data shows two mains patterns, one being a high gas fraction horizontal branch and the other a diagonal. The former branch is formed mainly by dwarf galaxies from DGS catalogue and is represented by low star formation efficiency and weak outflows models and, especially, by low $M_{\rm G,0}$ and Case~B models. In fact, the low $M_{*}$/$M_{\rm Dust}$ observed in DGS objects can only be reproduced by Case~B dust production, what can be interpreted as a $\Delta_A$ metallicity sensitivity. Our model seems to reproduce well the $M_{\rm Gas}/(M_{*} + M_{\rm Gas})$ by $M_{*}$/$M_{\rm Dust}$ properties of these systems. 

The latter one is composed by KINGFISH elliptical galaxies and high-$z$ galaxies, represented by high star formation efficiency and strong outflows models. The KINGFISH galaxies lie preferentially at $\log (M_{*}$/$M_{\rm Dust}) \gtrsim -4.0$ and $M_{\rm Gas}/(M_{*} + M_{\rm Gas})\gtrsim -1.0 $, while the elliptical galaxies are the opposite. The high-$z$ galaxies lie at the high dust production efficiency and high gas fraction locus, but they do not exhibit a clear different pattern.

The $M_{\rm Gas}/(M_{*} + M_{\rm Gas})$ by $M_{\rm Dust}/M_{*}$ diagram (Fig.~\ref{fig:Barion-x-Dust-star-log}) is also a reliable tool to investigate the galaxy and dust evolution connection, with the advantage of making clear how important it is the galaxy gas reservoir available to star formation. Nevertheless, the $M_{\rm Dust}/M_{\rm Gas}$ by $M_{\rm Dust}/M_{*}$ diagram (Fig.~\ref{Dusttogas-Dusttostar}) seems to be more sensitive to dust amount than the relation shown in Fig.~\ref{fig:Barion-x-Dust-star-log}.

\subsection{High$-z$ galaxies implication}
\label{ssec:high-z-implication}

Even though dust obscured high$-z$ galaxies can be reproduced by models with similar baryonic mass, $M_{G,0} = 10^{10}$~M$_\odot$, for A1689-zD1 \citep{knudsen2016merger}, with high SFR, $\nu_0 =$ 10.0 Gyr$^{-1}$, the comparison in section \ref{sec:Results} is quite simplistic. First, as discussed in \ref{ssec:parameters}, all galaxy evolutionary indicators are not satisfied simultaneously and both $M_{\rm Dust}/M_{*}$ vs.~$M_{*}$ (figure \ref{fig:DUST_Star-Star}) and  $M_{\rm Dust}/M_{\rm Gas}$ vs.~$M_{\rm Dust}/M_{*}$ (figure \ref{Dusttogas-Dusttostar}) are not in agreement with A1689-zD1. Second, the time needed to form and collapse of the primordial baryonic cloud was not took into account.

As discussed in section \ref{sec:Results}, for $M_{G,0} = 1 \times 10^{10}$ M$_\odot$ and $\nu_0 =$ 10.0 Gyr$^{-1}$ model, the time needed to reach dust mass in A2744\_YD4 is $\sim 0.5$~Gyr, while for A1689-zD1 dust mass it is about six times longer than this model, for both $\Delta_A$ adopted. The former has larger SFR than our model, while the latter has approximately the same. This may be due to a very strong starburst episode, differences in dust production or distribution into these objects, leading to different dust mass estimation. Another possibility is a non-universal IMF, as argued by \citet{calura2016dust}, being more top-heavy for low metallicity environment, that increases the number of CCSNe, metallicity and dust mass, speeding up the evolution the galaxy.
Even though, dust-to-gas ratio of the model agrees with $M_*$ and SFR of A1689-zD1, at a time of $\sim 0.4$~Gyr, compatible with the available cosmic time. Our model is better suited to explain dust-to-gas ratio than the total dust mass.

In order to estimate the dust mass of reionization objects and the DOGs number density, it is crucial to constrain both galaxy evolution and dust evolution models and also determine the cosmic star formation history.
The dominance of UV galaxies emitters can easily be due to  systematic selection effects, and the presence of DOGs at the cosmic dawn could be underestimated \citep{knudsen2016merger}. In this case, the cosmic star formation density census has to be updated and galaxy evolution theory upgraded. For a high DOGs number density during reionization, even reionization models must be revisited, due to dust high absorption in UV wavelength.

The simulated models show no tension with SMGs and LBGs at $z \sim 5$ or less. For high $\nu_0$ models, the high SFR (like SMG and LBGs), the galaxy has enough time to fully build its bulk mass, with a Salpeter's IMF, even considering the time needed to the formation and collapse of the cloud. The models also shows enough time to quench SMGs star formation and their change into passive galaxies \citet{article} until $z \sim 4$.

\subsection{Galaxy obscuration}
\label{ssec:obscuration}

As mentioned in section \ref{sec:Intro}, high mass galaxies have the major part of their star formation obscured by dust, reaching $\sim 90$\% in galaxies with $\log(M/{\rm M}_\odot) \simeq 10.5$ \citep[see][]{whitaker2017constant}. The low mass galaxies are the opposite case, the star formation is unobscured and the average between both is seen for galaxies with $\log(M/{\rm M}_\odot) =  9.4$, where half of the star formation is obscured \citep[see][]{whitaker2017constant}. This relation is observed, at least, until $z \sim 2.5$ \citep[see][]{whitaker2017constant}. 

A accurate treatment of obscured star formation in the framework of galaxy evolution models is a complex task. 
Stars are born in clusters inside giant molecular clouds (GMC), that are generally non symmetric and extended objects. 
The GMC intrinsic extinction relies in the local ISM composition, its geometry and its density.
The radiation emitted by massive newborn stars disrupt the cloud from inside out, creating HII bubbles into GMC and allowing the UV and visible photons to escape. 
A radiative transfer treatment is beyond the scope of this work. Here, we will deal only with dust mass production.

In figure \ref{fig:dust/gas-time}, for any $M_{G,0}$ and for both Cases A and B, the initial dust-to-gas ratio increases with $\nu_0$. The time spent to form the bulk of dust is also lower for high $\nu_0$ models. 
For low $\nu_0$ values, the molecular cloud will have a high fraction of UV photons escaping without interacting with dust (due to the low dust-to-gas ratio), making the star formation rate easily probed by UV and visible wavelengths. 
The smoother the star formation history (without bursts) is, the lower is the amount of dust in the environment where star formation takes place.
In high $\nu_0$ models, the fraction of UV photons escaping will be very small \citep[][model consider that all UV photons are processed by dust during the GMC first Myrs]{silva1998modeling}, due to the higher dust-to-gas ratio, so that the star formation rate must be probed in the IR and submillimetric bands. The grain accretion dependence with metallicity amplifying this difference.

As both SFR and metallicity are linked with galaxy evolution, depending on the galaxy mass \citep{mannucci2010fundamental}, we expect a slow evolution of the obscured star formation fraction for $z$ higher than the ``cosmic noon'', when galaxies are building the bulk of their stellar mass. The condensation efficiency can amplify the difference between massive and dwarf galaxies, due to the difference in the metallicity evolution time. It is clear that the time spent to form stars being lower, the metallicity evolves faster, which also turns the accretion more efficient. We conclude that $M_{G,0}$ is the main driver of the obscuration rate.

The $M_{\rm Dust}/M_{\rm Gas}$ by $M_{\rm Dust}/M_{*}$ diagram (figure \ref{Dusttogas-Dusttostar}) is very similar to fundamental relation among $M_*$, SFR and metallicity \citep[see][]{mannucci2010fundamental}. \citet{10.1093/mnras/stt190} also find a relationship between DM halo mass and $M_*$, as a SFR--$M_*$ relation. This is also expected and seen in differents star formation efficiency per $M_{G,0}$ and reflects in  the more massive halos have a shorter evolution time than to the less massive ones, being more efficient to form stars, can create the difference of unobscured to obscured not dependent (or with small dependence) of galaxy size relying, in ultimate instance, in the galaxy mass. 

The dust obscured star formation can be illustrated by figure \ref{Dusttogas-Dusttostar}. In any panel of this figure, the high $M_{\rm Dust}/M_{\rm Gas}$ and $M_{\rm Dust}/M_{*}$ corner corresponds to the largest dust fraction, where galaxies are most obscured. There is where we found the LBGs and A1689-zD1, all heavily obscured, evolved, high-$z$ galaxies. It is also the turnover place of the evolutionary tracks of $\nu_0 = 10$~Gyr$^{-1}$. This position also supports the idea that SMG can evolve into passive elliptical galaxies. 

DGS objects lie preferably in low $M_{\rm Dust}/M_{\rm Gas}$ and low $M_{\rm Dust}/M_*$ region. This subsample has low mean metallicity ($12 + \log {\rm O/H} = 7.93$), low stellar mass ($\log M_*/{\rm M_\odot} = 8.58$) and low dust mass ($\log M_{\rm Dust}/{\rm M_\odot} = 5.69$) \citep{remy2015linking}. Only dust production Case B can reach the $M_{\rm Dust}/M_{\rm Gas}$ and the $M_{\rm Dust}/M_{\rm Gas}$ values from DGS objects.

The $M_{G,0} = 5 \times 10^{7} \mathrm{M}_\odot$  was chosen to be the inferior threshold of dwarf galaxies, although it is bellow the lower observational data limit.
For this $M_{G,0}$ value, only $\nu_0 = 0.1 \, \mathrm{Gyr}^{-1}$ model makes a normal star forming galaxy (see~\ref{fig:Galcomp-1}), taking almost 10~Gyr to reach $10^{-4}$ dust-to-gas ratio \citep[the same order as][sample of dwarf galaxies, see also figure \ref{fig:dust/gas-time}]{lisenfeld1998dust}, in both dust production cases (this is also almost the time needed to Case B to reach Case A). 
For this same $M_{G,0}$, any other $\nu_0$ makes a starburst, followed by a passive evolution. Despite the low $M_*$, the normal star forming model reaches $M_{\rm Dust}$ and dust-to-gas ratio of small $M_*$ objects from \citet{remy2014gas,remy2015linking}, as can be seen in figures \ref{fig:DUST-Star} and \ref{fig:DUST_GAS-Star}.

The extremely metal poor galaxy IZw 18 has dust-to-gas ratio between 3.2--$13 \times 10^{-6}$ \citep{fisher2014rarity}, or about $3.0 \times 10^{-6}$ derived by \citet{remy2014gas,remy2015linking}, the $M_{G,0} = 5 \times 10^{7} \mathrm{M}_\odot$ normal star forming galaxy model takes about 0.1 Gyr, in Case A, and 1.0 Gyr, in Case B, to reach IZw 18 dust-to-gas ratio (see figures \ref{fig:dust-time} and \ref{fig:dust/gas-time}).
IZw 18 stellar population has an age of approximately 0.5 Gyr \citep[for further discussion, see][]{papaderos2002Izw18},  $M_* = 9 \times 10^7 \mathrm{M}_\odot$ and $M_{\rm Gas} = 2.8 \times 10^8 \mathrm{M}_\odot$ ($M_{\rm Gas} = M_{\rm HI}+M_{\rm H_2}$), this galaxy is in a fast evolution (starburst like) and its SFR is about 0.05 $\mathrm{M_\odot}/\mathrm{year}$  \citep{fisher2014rarity}. If we compare IZw 18 with any starburst model we see that only Case B can reproduce it, so we conclude that $\Delta_A$ is sensitive to metallicity.

\section{Conclusions}
\label{sec:Conclusion}

In this work we have carried out forty semi-analytical simulations of galaxy evolution, varying the galaxy initial mass and both star formation and dust coagulation efficiency, to probe the dust enhancement process that leads to the obscuration of galaxies, in particular the shrouding of star formation in high mass galaxies. The simulations are compared with data available in the literature, covering a large range of mass and redshift. Here, we will briefly summarize out main results: 

\begin{itemize}
    \item Our results corroborate the scenario where SMGs evolve to elliptical galaxies \citep{article}. Despite the small number of LBGs in our sample, we also find that LGB galaxies have similar dust, gas and stellar masses, and SFR compared to SMGs,  probably implying in similar evolutionary phase.
    
    \item During intense star formation episodes, the dust mass and the dust-to-gas ratio of galaxies is almost unaffected by dust coagulation efficiency due to grain accretion dominance, making low SFR system, with low metallicity, a more suitable place to study $\Delta_A$, even if it can lead to bias toward low metallicity $\Delta_A$. Our results also suggest a $\Delta_A$ metallicity dependent scenario.
    
    \item With Salpeter IMF, high SFR systems build the bulk of their dust mass in $\sim 0.6$~Gyr, the time-scale required to built dust obscured galaxies during the reionization epoch. However, there is some tensions with A1689-zD1 dust mass and to explain the presence of passive evolved galaxies at $z \sim 4$. The tension is relieved with a top-heavy IMF, as suggested by \citet{calura2016dust}, but we find that a high star formation with grain accretion in ISM may be enough.
    
    \item Our models do not represent a complete scenario of dust and galaxy evolution. The simulated quantities $M_{\rm Dust}$, $M_{\rm Gas}$ and $M_{*}$ do not represent all the data simultaneously. We suggest that this is a problem from both galactic and dust models. An adequate dust model needs to constrain all the quantities. Due to better agreement with dust-to-gas ratio, we suggest that the adopted dust model is more suitable to represent the dust-to-gas ratio than dust mass alone.
    
    \item The $M_{\rm Dust}/M_{\rm Gas} \times M_{\rm Dust}/M_{*}$ diagram is a powerful tool to study obscuration, dust production model and galaxy evolution, because this diagram constrains all baryons phases. Despite the reduced data, star forming and passive galaxies lie in a clear different locus on this kind of diagram.
    
    \item The dust-to-gas ratio relies strongly on star formation, while the SFR relies on the gravitational potential well. Our results point to an almost Universal obscured star formation fraction even in $z$ larger than 3, independently of $\Delta_A$.
\end{itemize}

Dust obscuration is an important issue to understand galaxy evolution and to constrain the star formation density across the cosmic time, in particular during the cosmic peak of star formation. 
The new generation facilities should shed new light to this question. James Webb Space telescope (JWST), Extremely Large Telescope (ELT), Giant Magellan Telescope (GMT), and SPICA together with ALMA will bring new constrain in the investigation of high-$z$ galaxies. SPICA will be especially useful to investigate dust production in metal poor galaxies and constrain stellar dust production.

\section*{Acknowledgements}

JHBS acknowledges CNPq institutional scholarship and CAPES/PROEX program for the financial support. GBLN is grateful for the financial support from FAPESP (grant 2018/17543-0) and CNPq. We thank ALMA programme 2011.0.00294.S for the ALESS SMG catalogue availability.

%%%%%%%%%%%%%%%%%%%%%%%%%%%%%%%%%%%%%%%%%%%%%%%%%%



\bibliographystyle{mnras}
\bibliography{bibliografia} % if your bibtex file is called example.bib

%\begin{thebibliography}{99}
%\end{thebibliography}

%%%%%%%%%%%%%%%%%%%%%%%%%%%%%%%%%%%%%%%%%%%%%%%%%%

%%%%%%%%%%%%%%%%% APPENDICES %%%%%%%%%%%%%%%%%%%%%

\appendix

\section{Galaxy evolution in simulations}
\label{App:galaxy-ev}

Here we provide the time evolution of gas (blue line), star (red line), dust (brown continuous line for Case A and brown dashed line for Case B), and total mass (black continuous line) of all 40 simulated models. Each figure is related to one $M_{G,0}$ model (table~\ref{tab:galaxymod}).
%
%% Isto vai nas legendas das figuras.
% where 
% figure \ref{fig:Galcomp-1} is related to $5 \times 10^{7}$ M$_\odot$, 
% figure \ref{fig:Galcomp-2} to $10^{9}$ M$_\odot$, 
% figure \ref{fig:Galcomp-3} to $10^{10}$ M$_\odot$, 
% figure \ref{fig:Galcomp-4} to $2 \times 10^{11}$ M$_\odot$, and 
% figure \ref{fig:Galcomp-5} to $2 \times 10^{12}$ M$_\odot$. 

Each panel in the figures represents a star formation efficiency in crescent order, from left to right and from top do bottom. The labels and tags are the same in all figures.

\begin{figure*}
\begin{center}
%\setcaptionmargin{1cm}
\includegraphics[width=0.8 \columnwidth,angle=0]{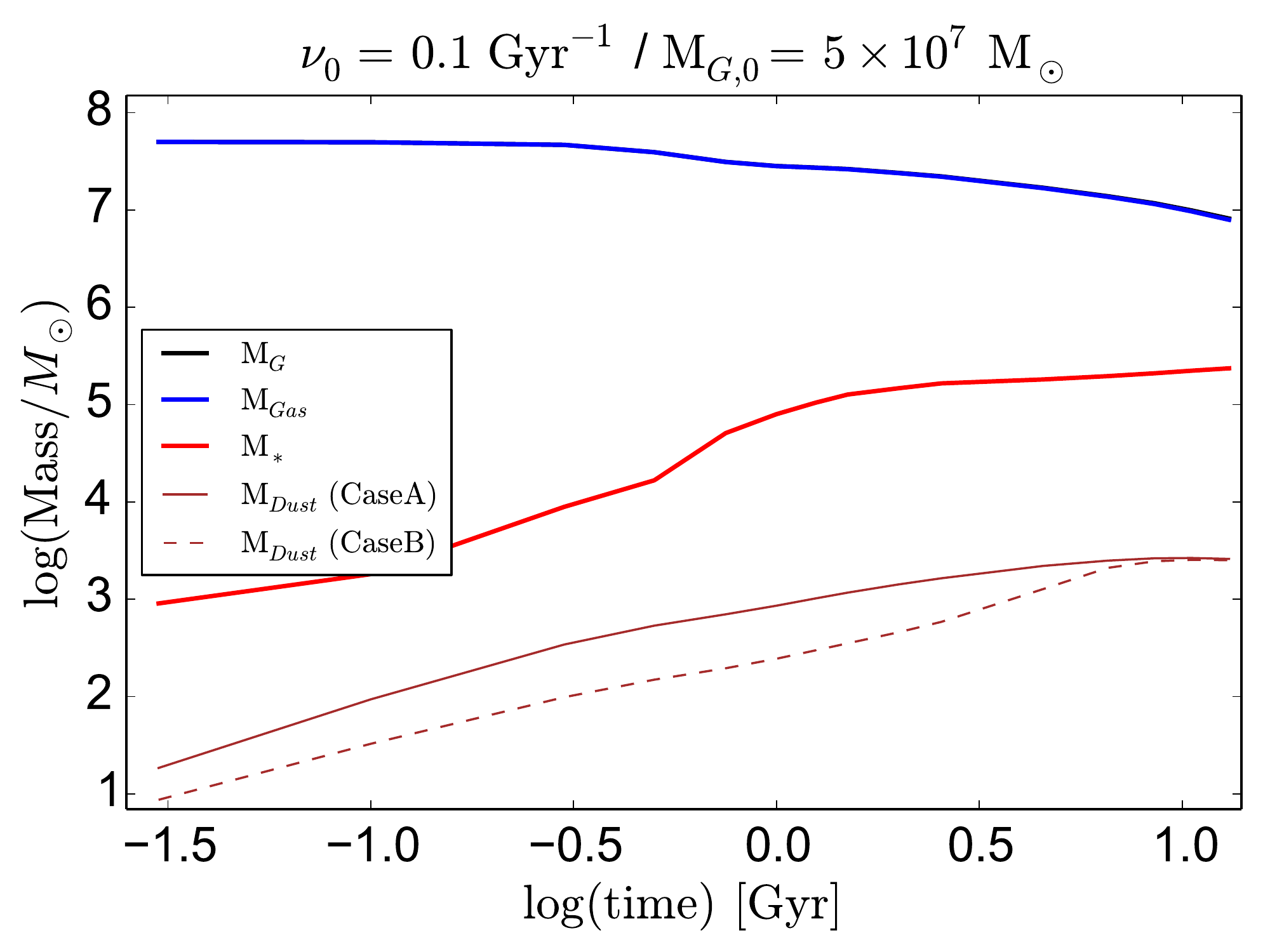}
\includegraphics[width=0.8 \columnwidth,angle=0]{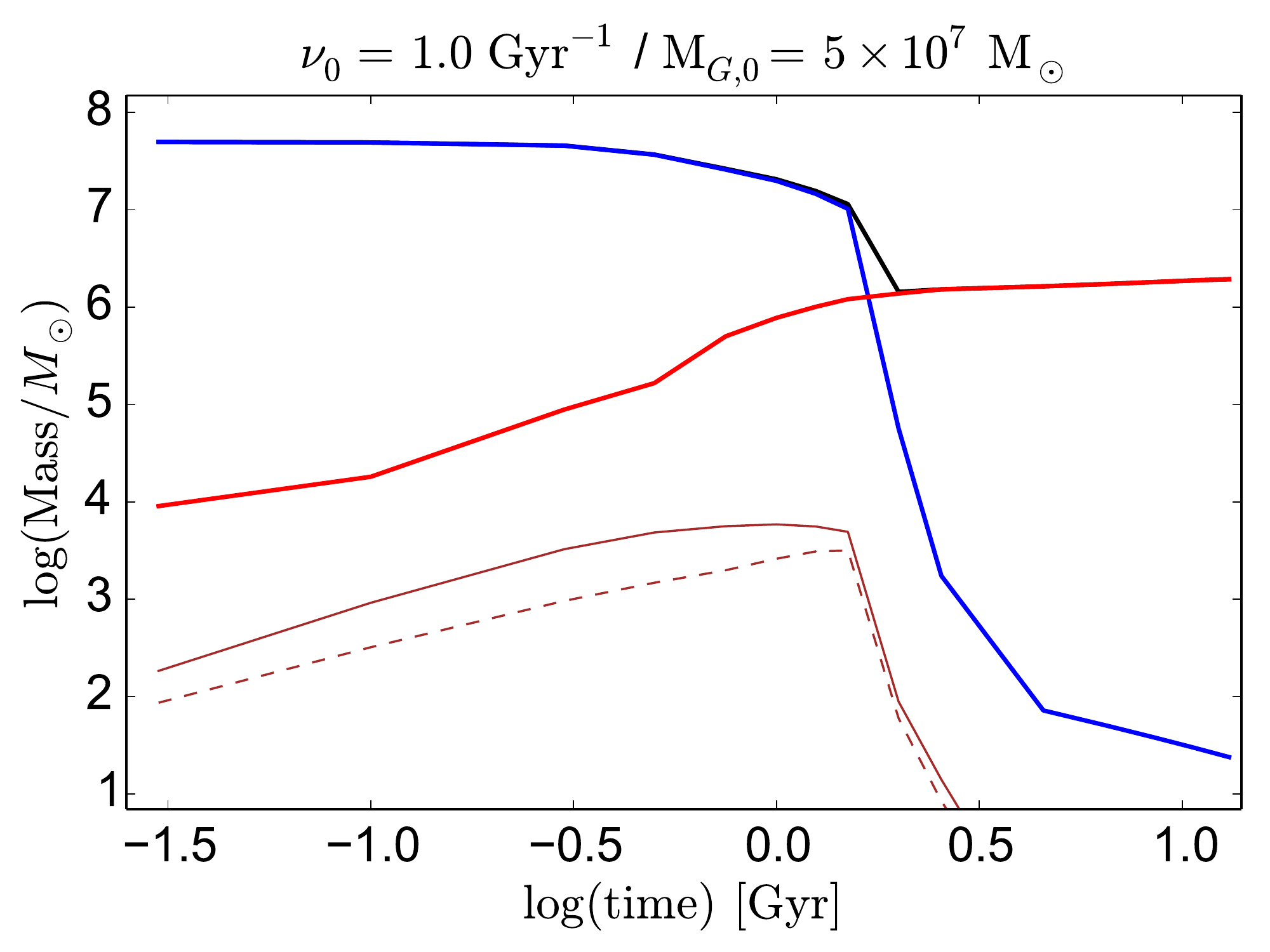}
\includegraphics[width=0.8 \columnwidth,angle=0]{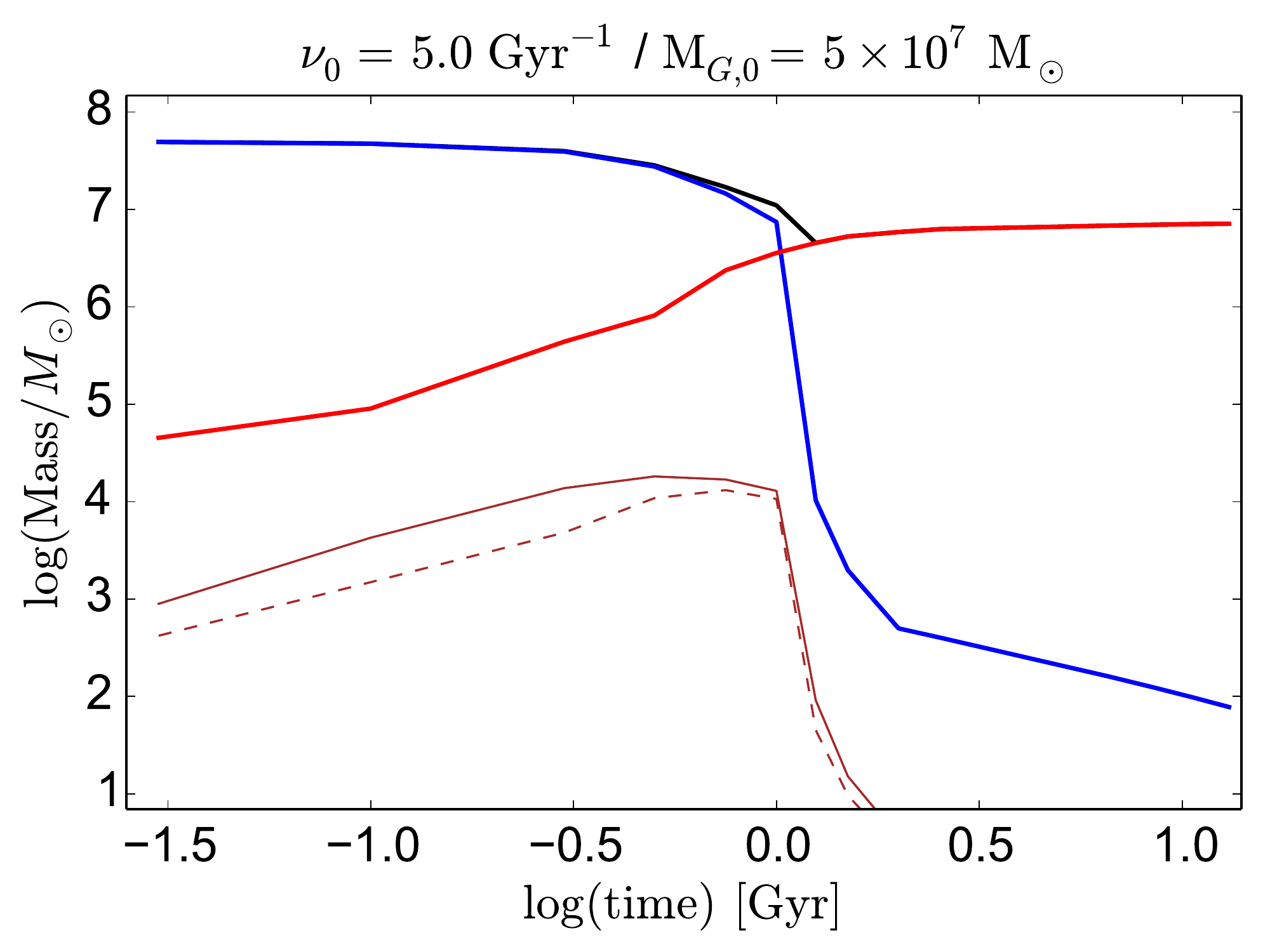}
\includegraphics[width=0.8 \columnwidth,angle=0]{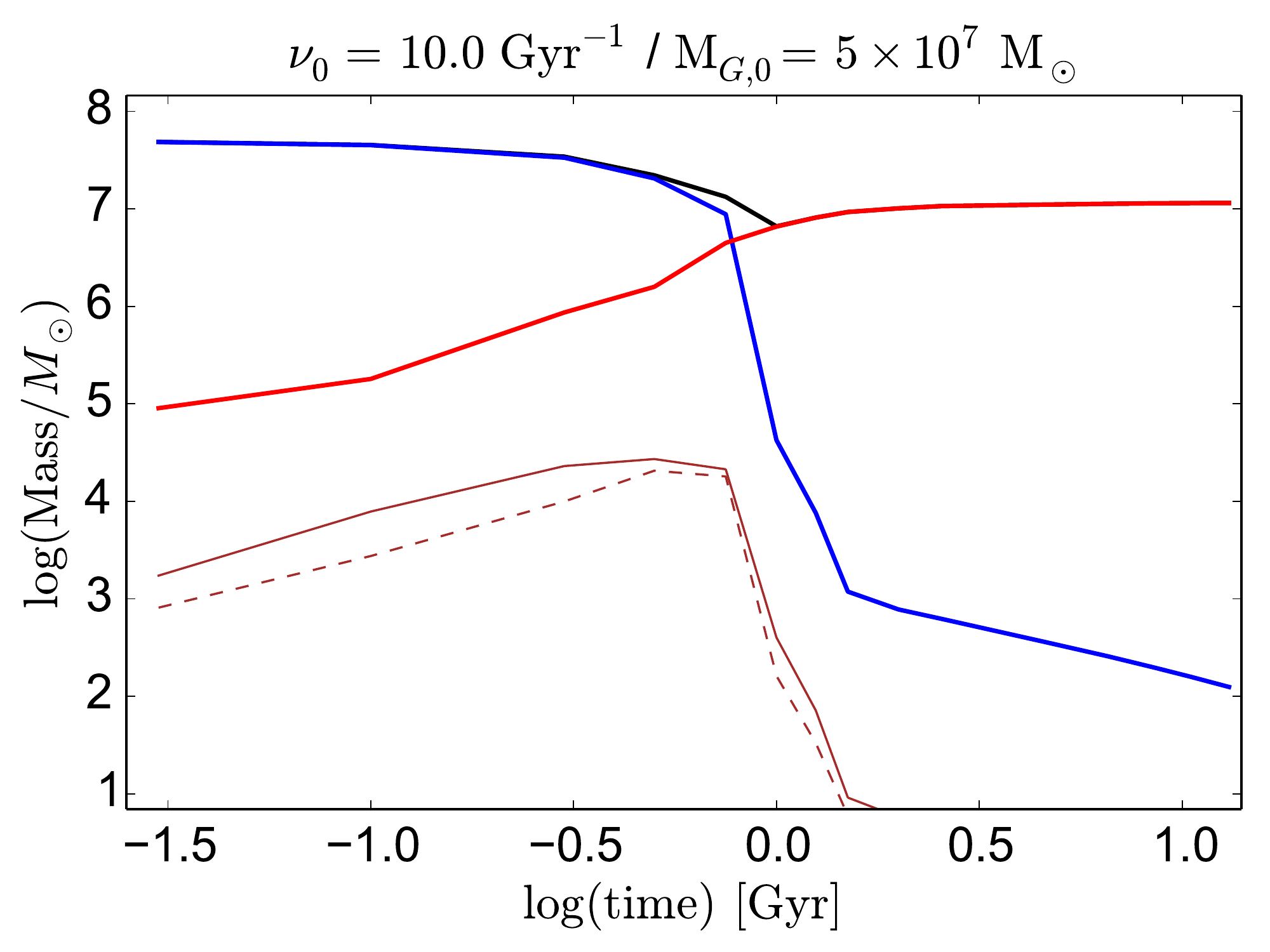}
\caption[]{Time evolution of gas (blue line), stellar (red line), dust (brown continuous line Case A and brown dashed line Case B) and total mass (black continuous line) of $M_{G,0} = 5 \times 10^{7} \, \mathrm{M}_\odot$ model. The panels are ordered from left to right and top to bottom, been related to $\nu_0 = 0.1 \, \mathrm{Gyr}^{-1}$, $1.0 \, \mathrm{Gyr}^{-1}$, $5.0 \, \mathrm{Gyr}^{-1}$, and $10.0 \, \mathrm{Gyr}^{-1}$, respectively.} 
\label{fig:Galcomp-1}
\end{center}
\end{figure*}

\begin{figure*}
\begin{center}
%\setcaptionmargin{1cm}
\includegraphics[width=0.8 \columnwidth,angle=0]{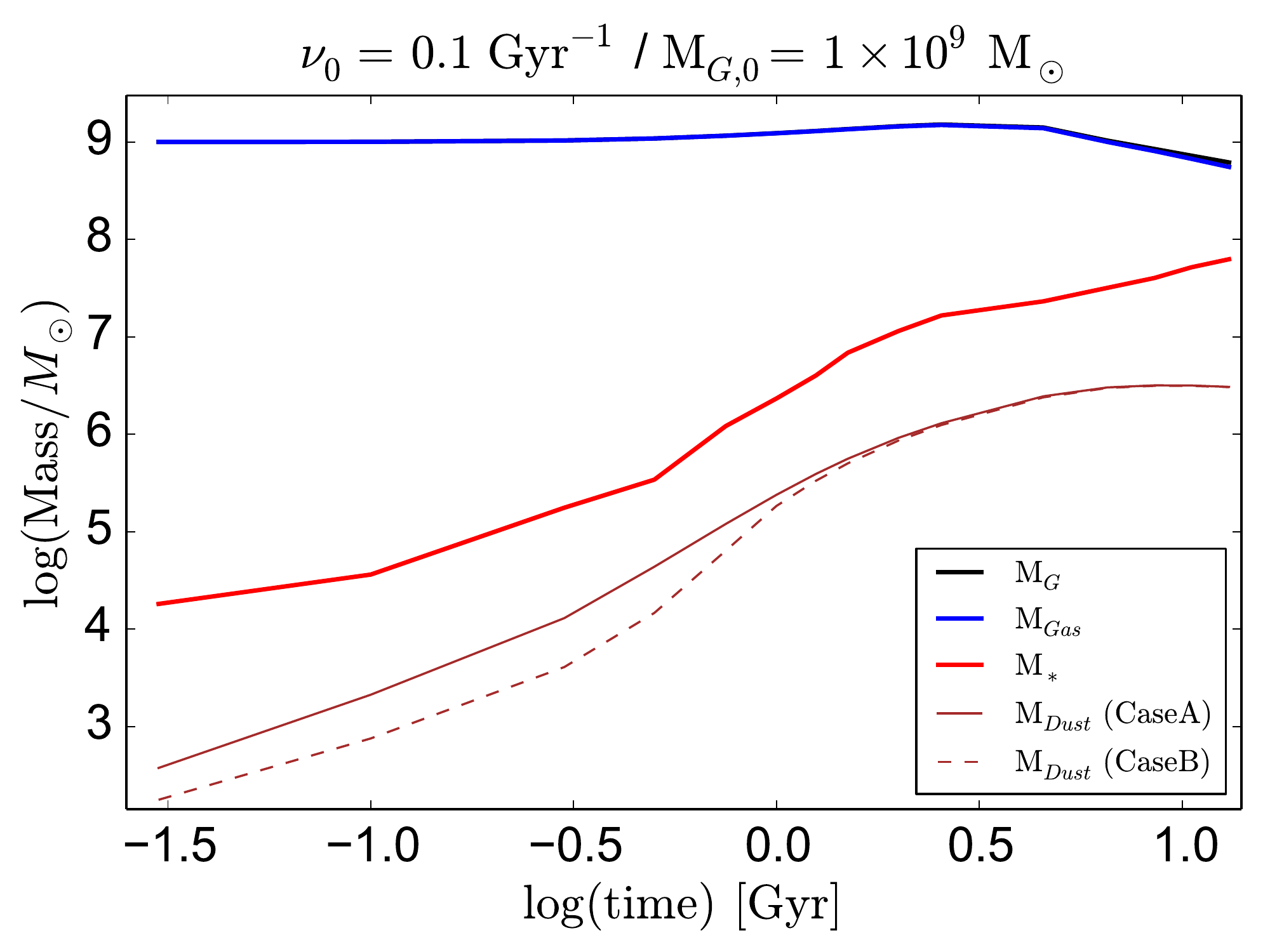}
\includegraphics[width=0.8 \columnwidth,angle=0]{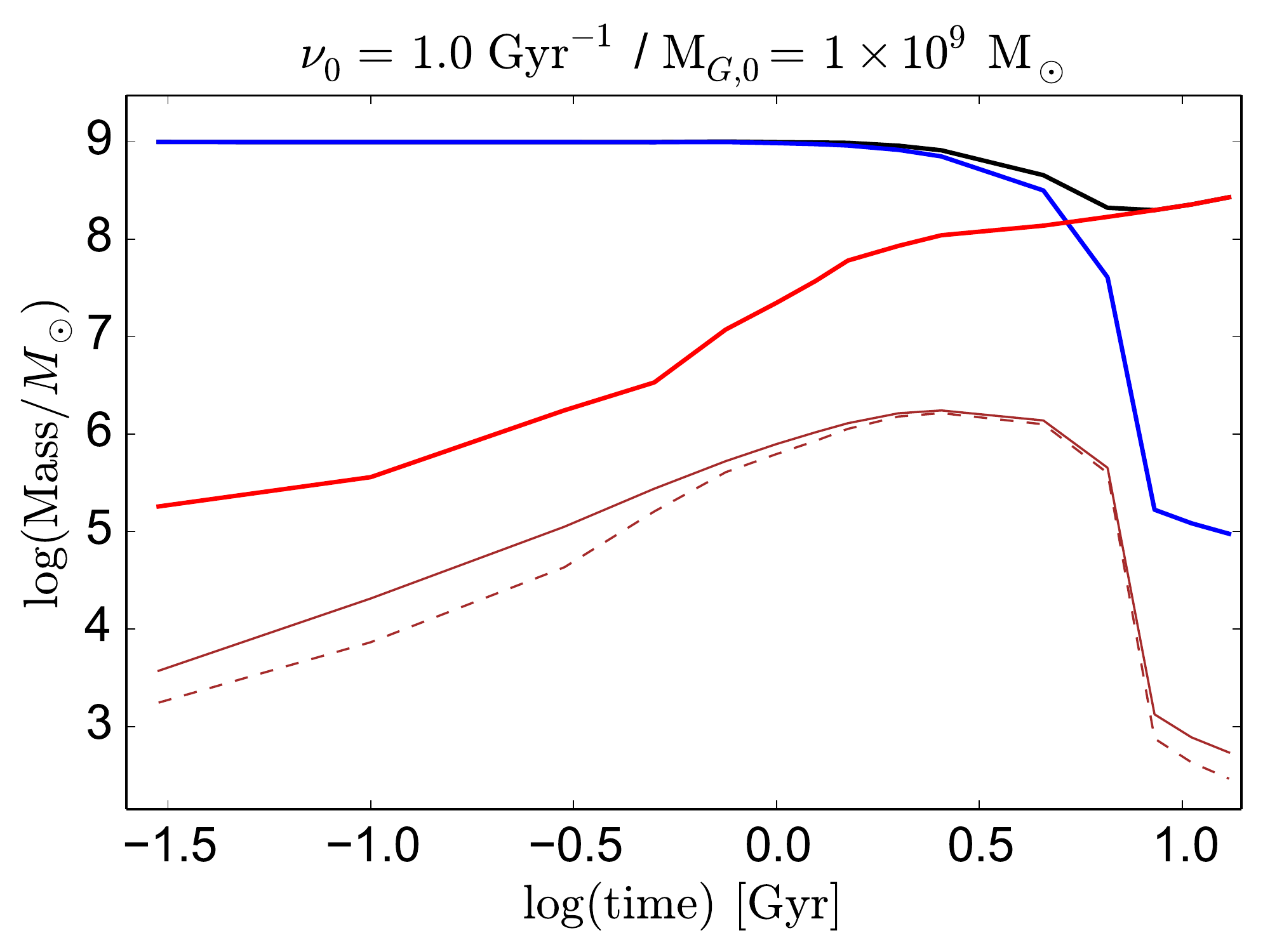}
\includegraphics[width=0.8 \columnwidth,angle=0]{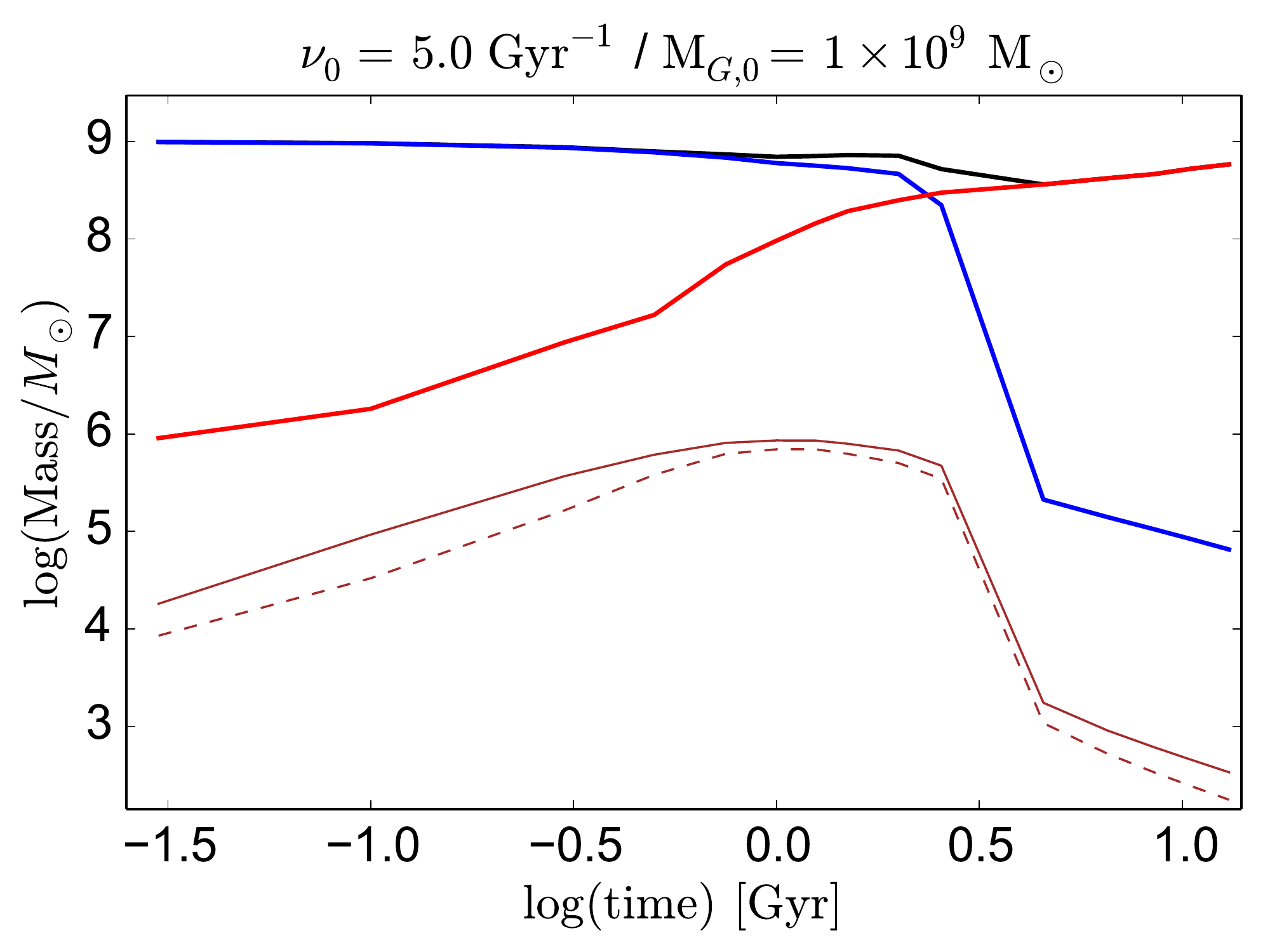}
\includegraphics[width=0.8 \columnwidth,angle=0]{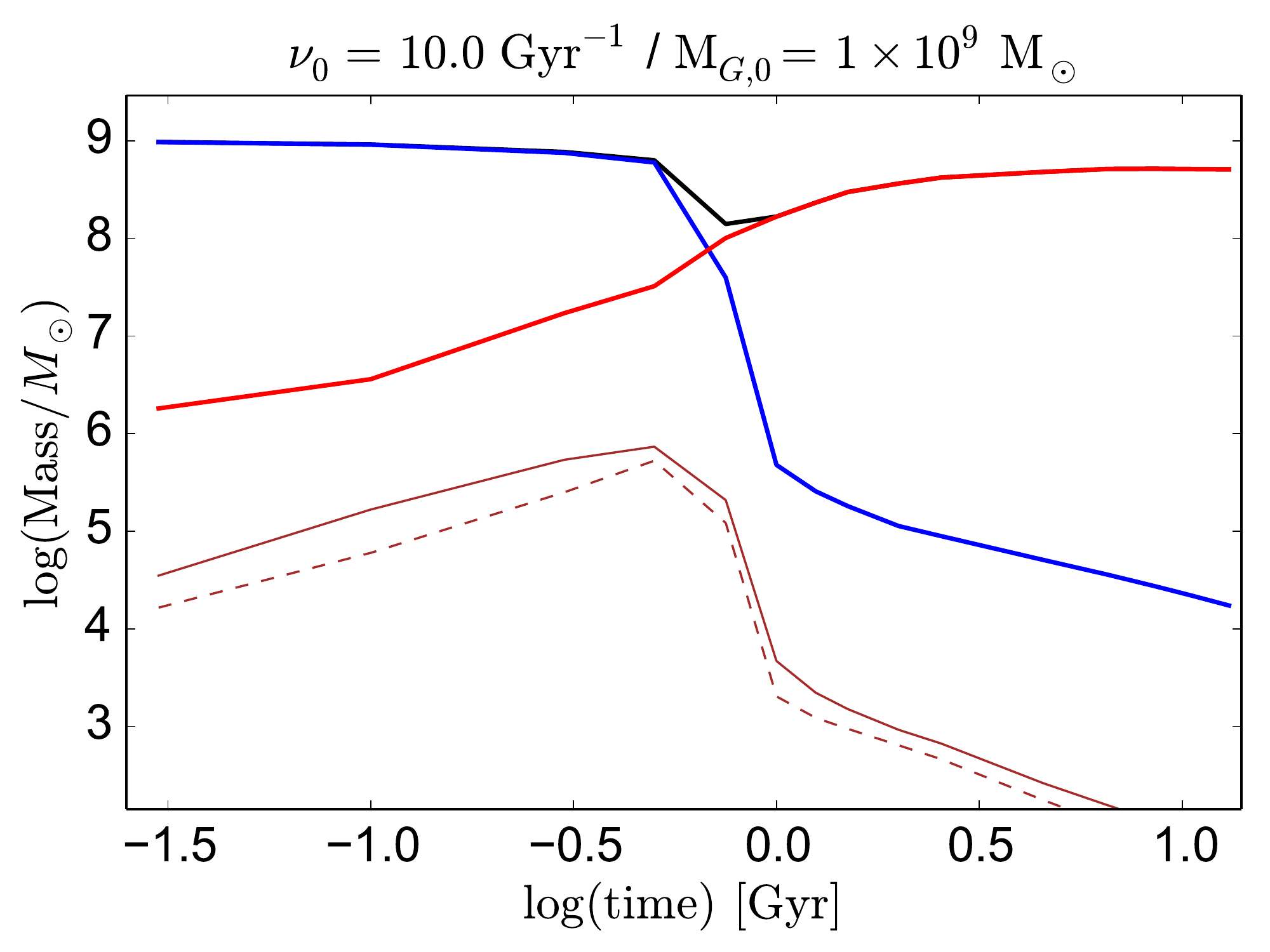}
\caption{The same as \ref{fig:Galcomp-1}, but for $M_{G,0} = 10^{9} \, \mathrm{M}_\odot$ model.} 
\label{fig:Galcomp-2}
\end{center}
\end{figure*}

\begin{figure*}
\begin{center}
%\setcaptionmargin{1cm}
\includegraphics[width=0.8 \columnwidth,angle=0]{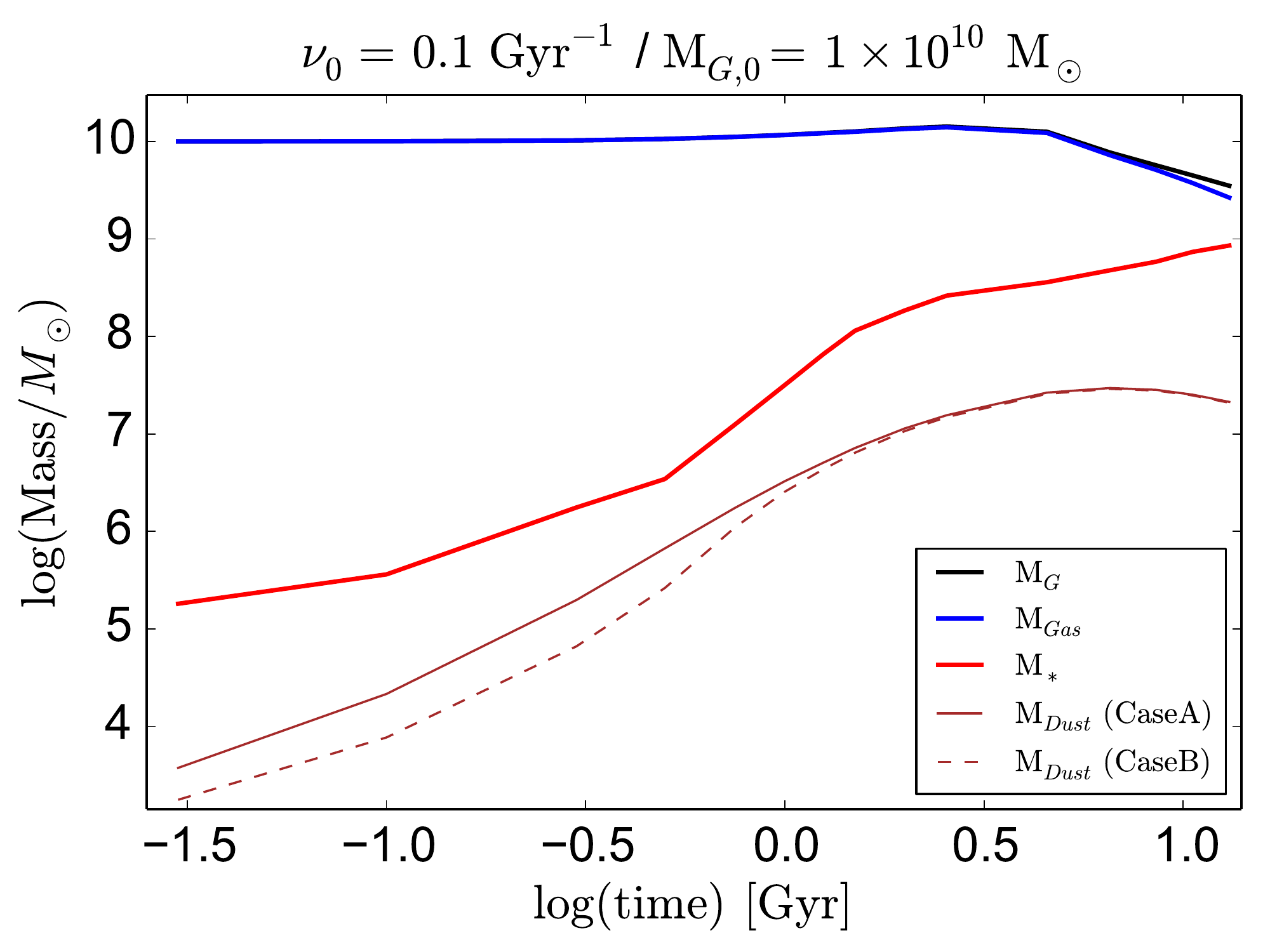}
\includegraphics[width=0.8 \columnwidth,angle=0]{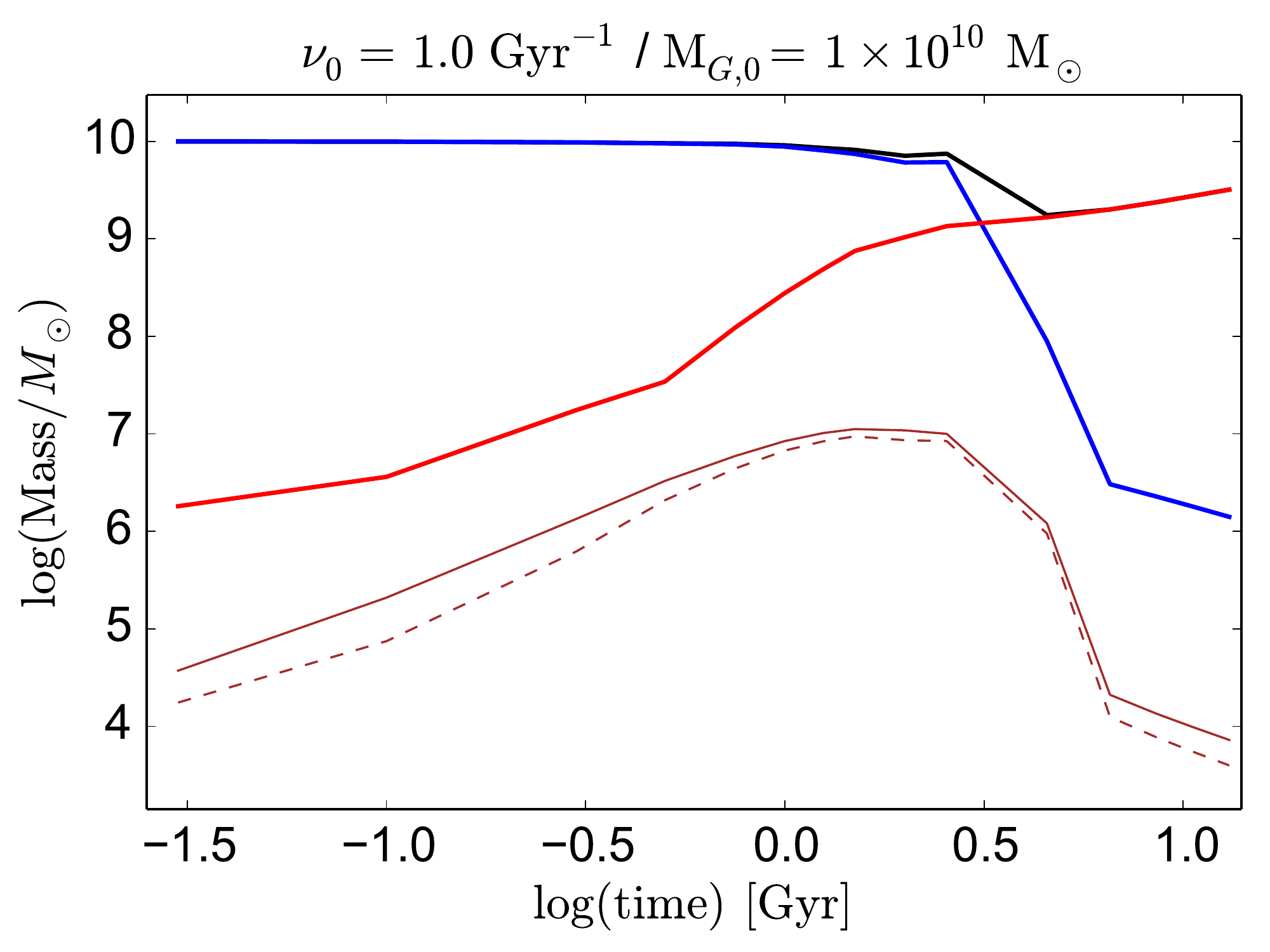}
\includegraphics[width=0.8 \columnwidth,angle=0]{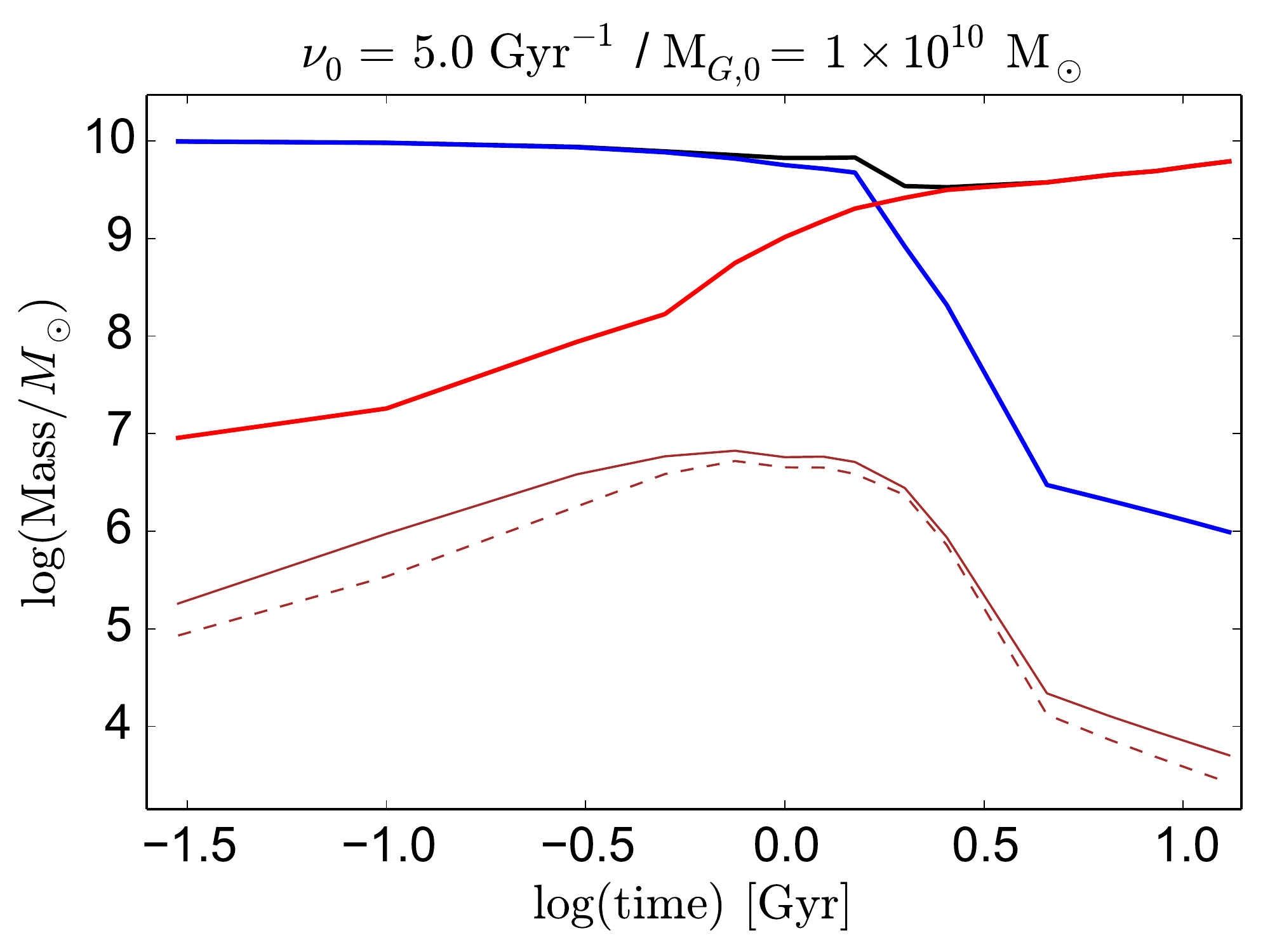}
\includegraphics[width=0.8 \columnwidth,angle=0]{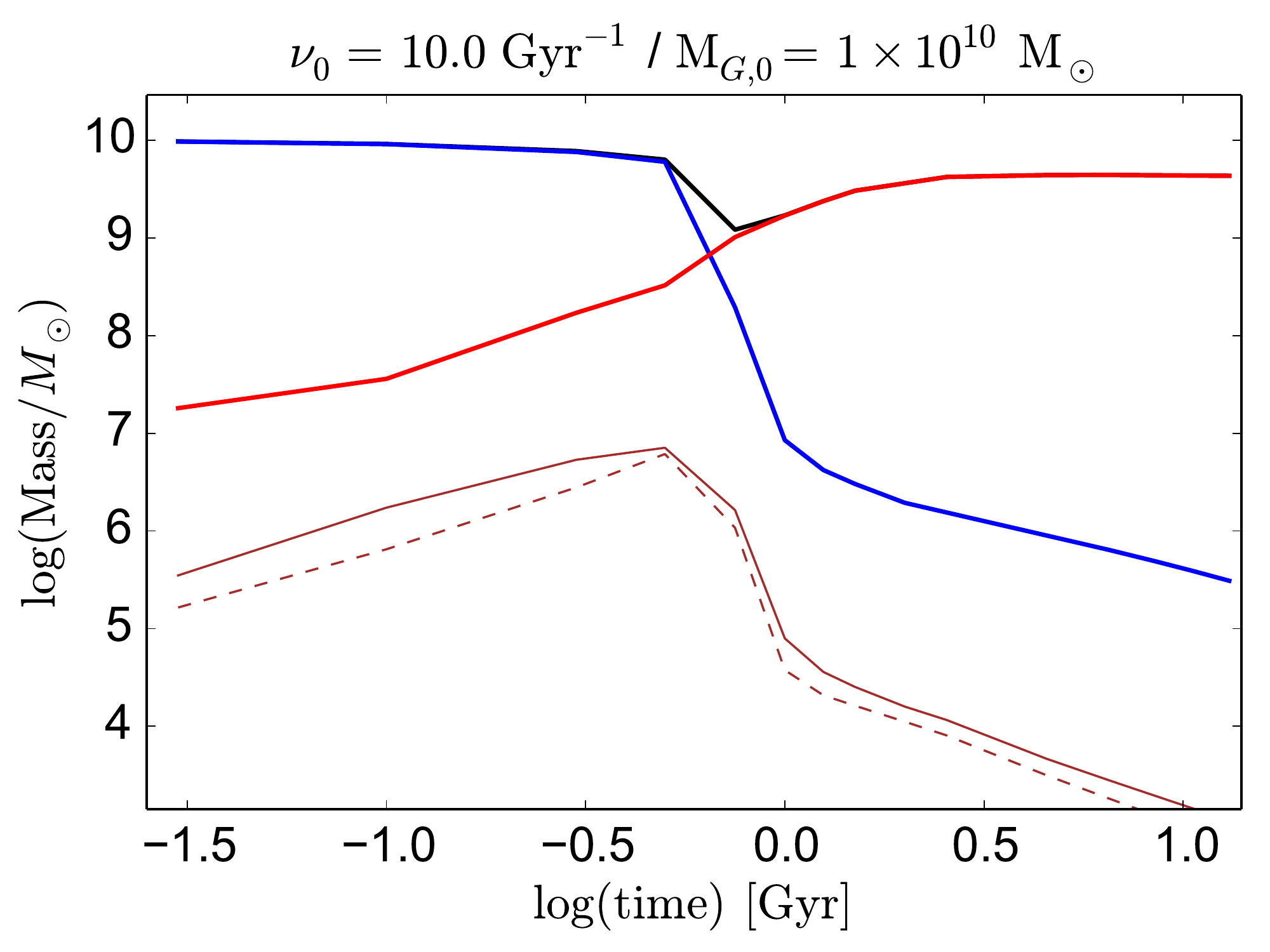}
\caption{The same as \ref{fig:Galcomp-1}, but for $M_{G,0} = 10^{10} \, \mathrm{M}_\odot$ model. } 
\label{fig:Galcomp-3}
\end{center}
\end{figure*}

\begin{figure*}
\begin{center}
%\setcaptionmargin{1cm}
\includegraphics[width=0.8 \columnwidth,angle=0]{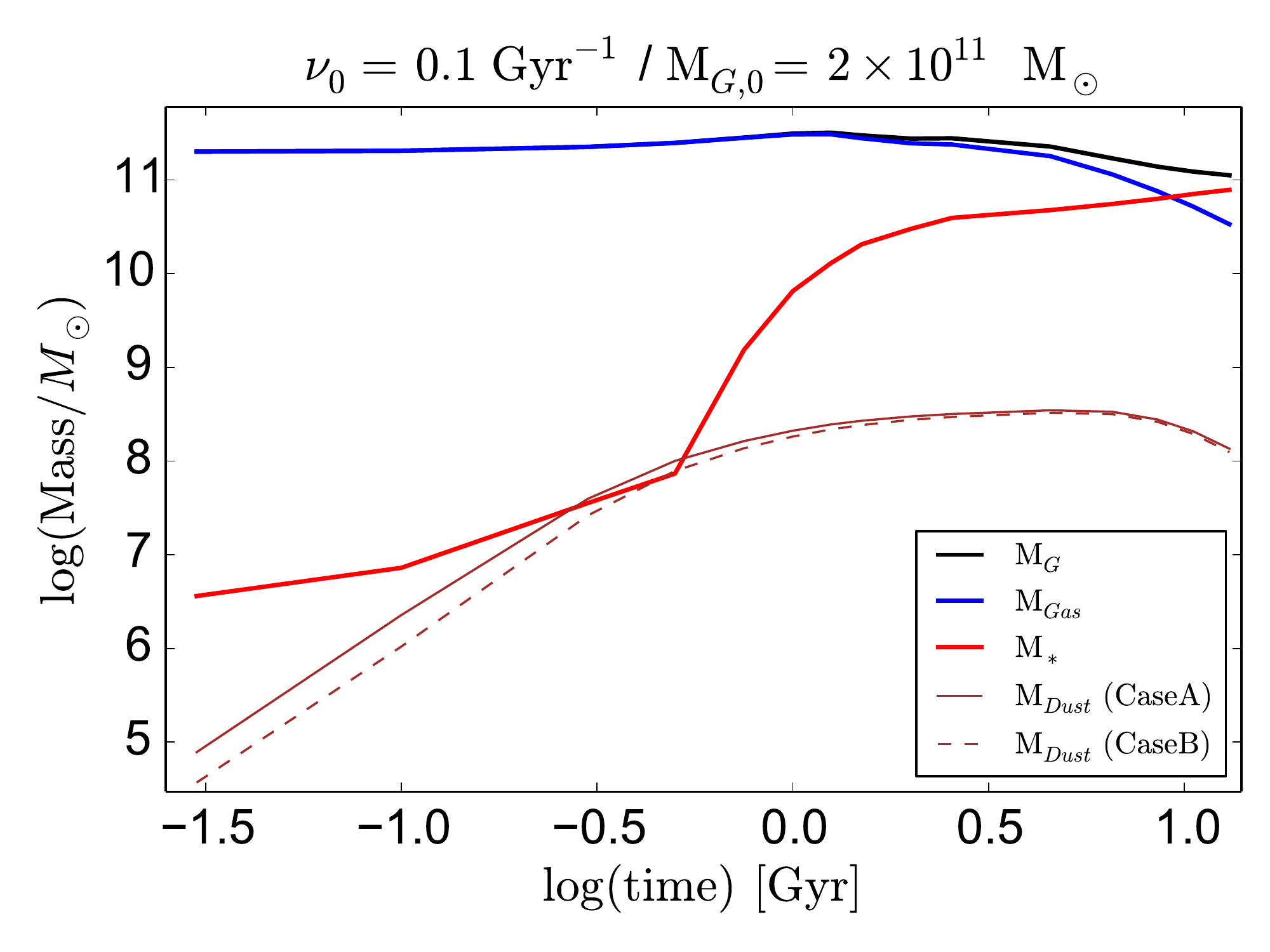}
\includegraphics[width=0.8 \columnwidth,angle=0]{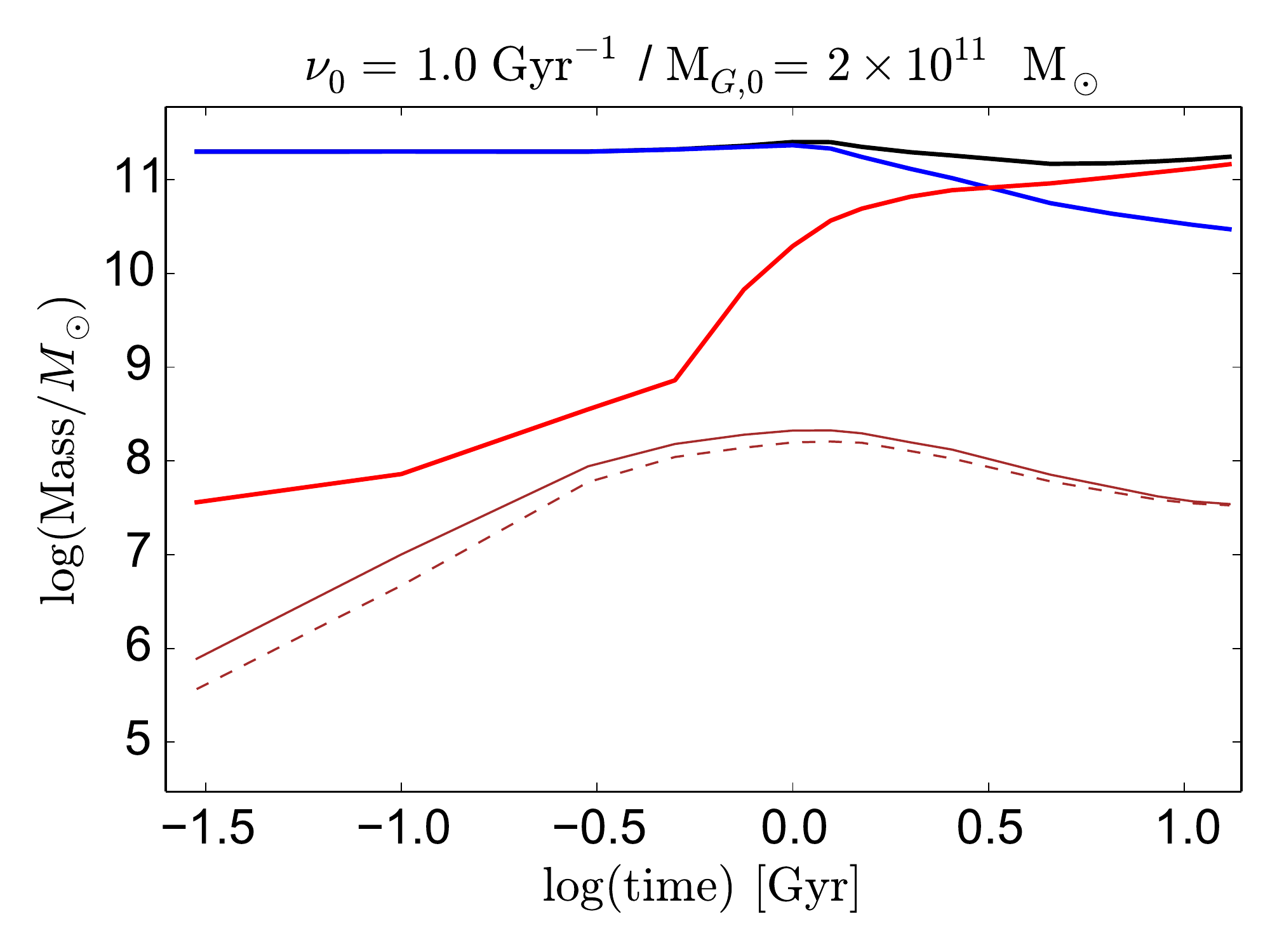}
\includegraphics[width=0.8 \columnwidth,angle=0]{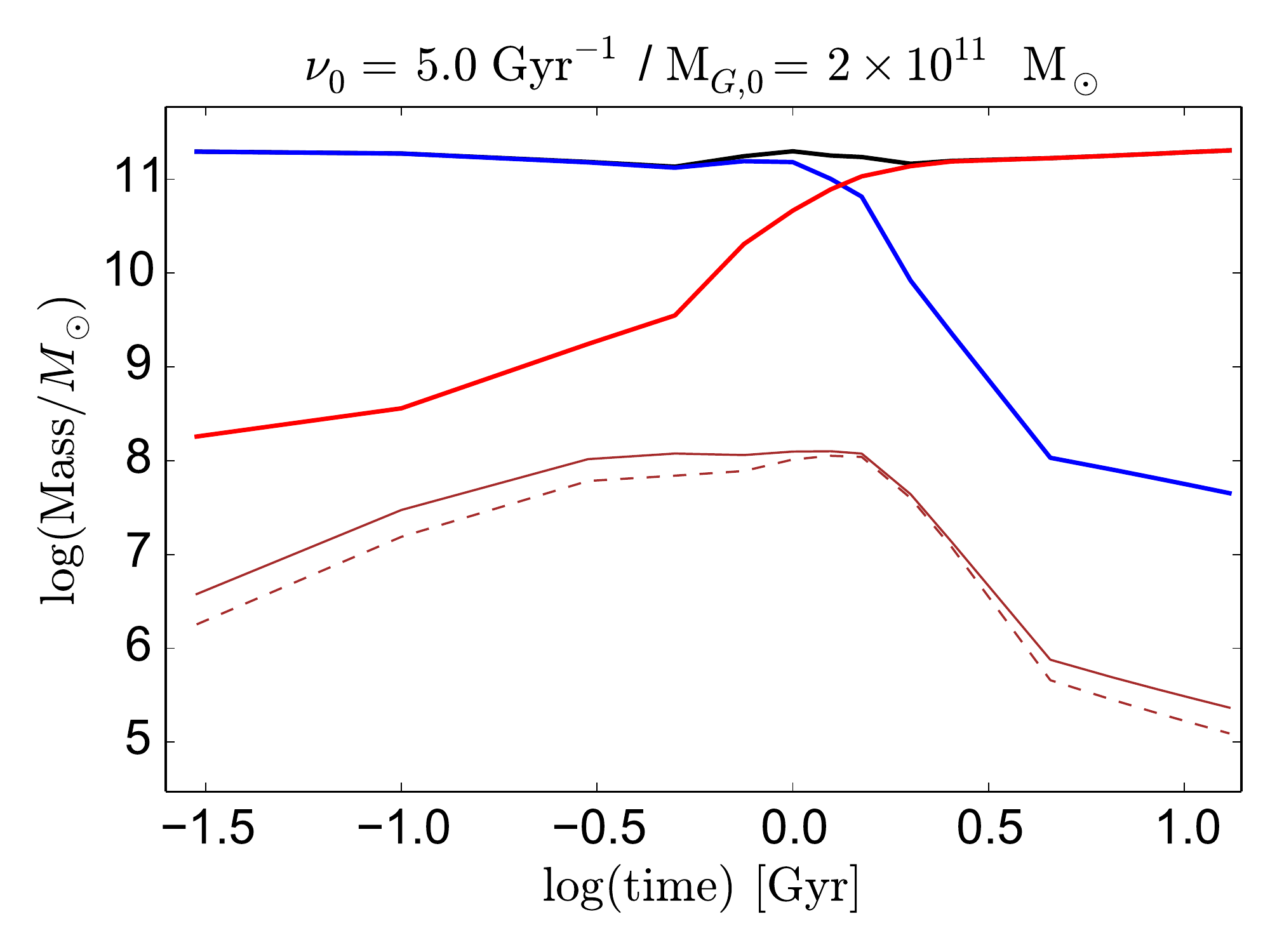}
\includegraphics[width=0.8 \columnwidth,angle=0]{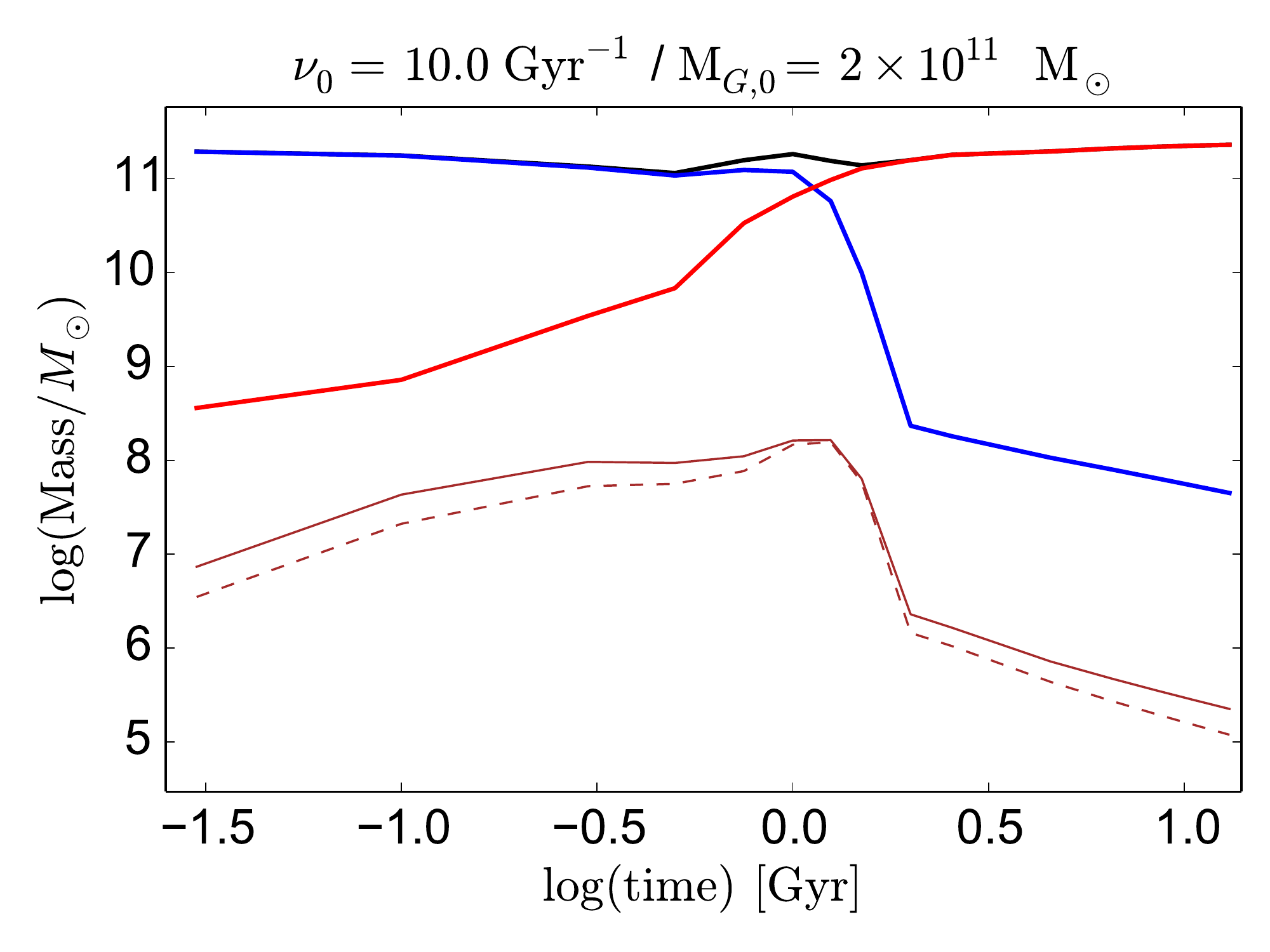}
\caption{The same as \ref{fig:Galcomp-1}, but for $M_{G,0} = 2 \times 10^{11} \, \mathrm{M}_\odot$ model.} 
\label{fig:Galcomp-4}
\end{center}
\end{figure*}
		
\begin{figure*}
\begin{center}
%\setcaptionmargin{1cm}
\includegraphics[width=0.8 \columnwidth,angle=0]{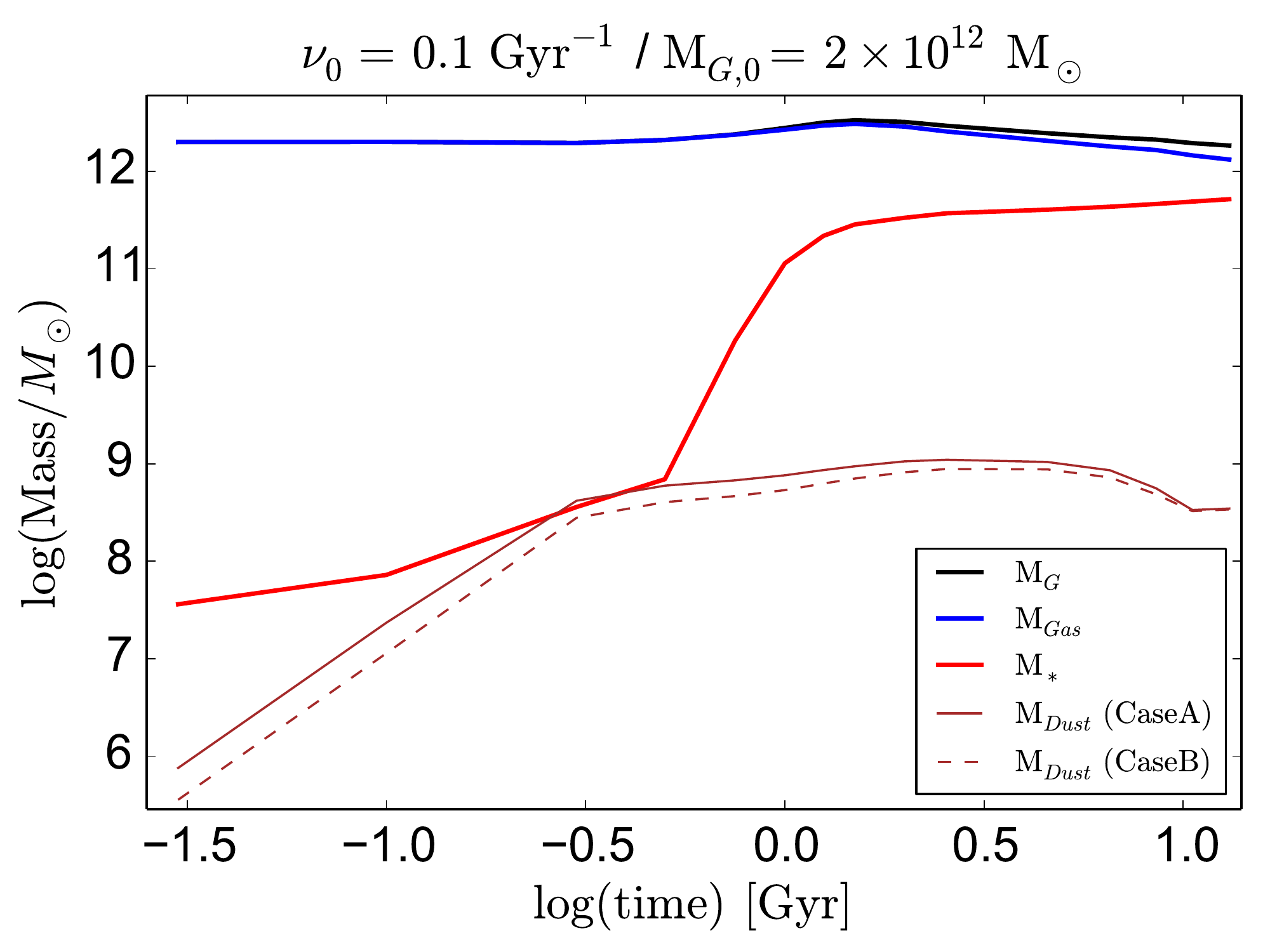}
\includegraphics[width=0.8 \columnwidth,angle=0]{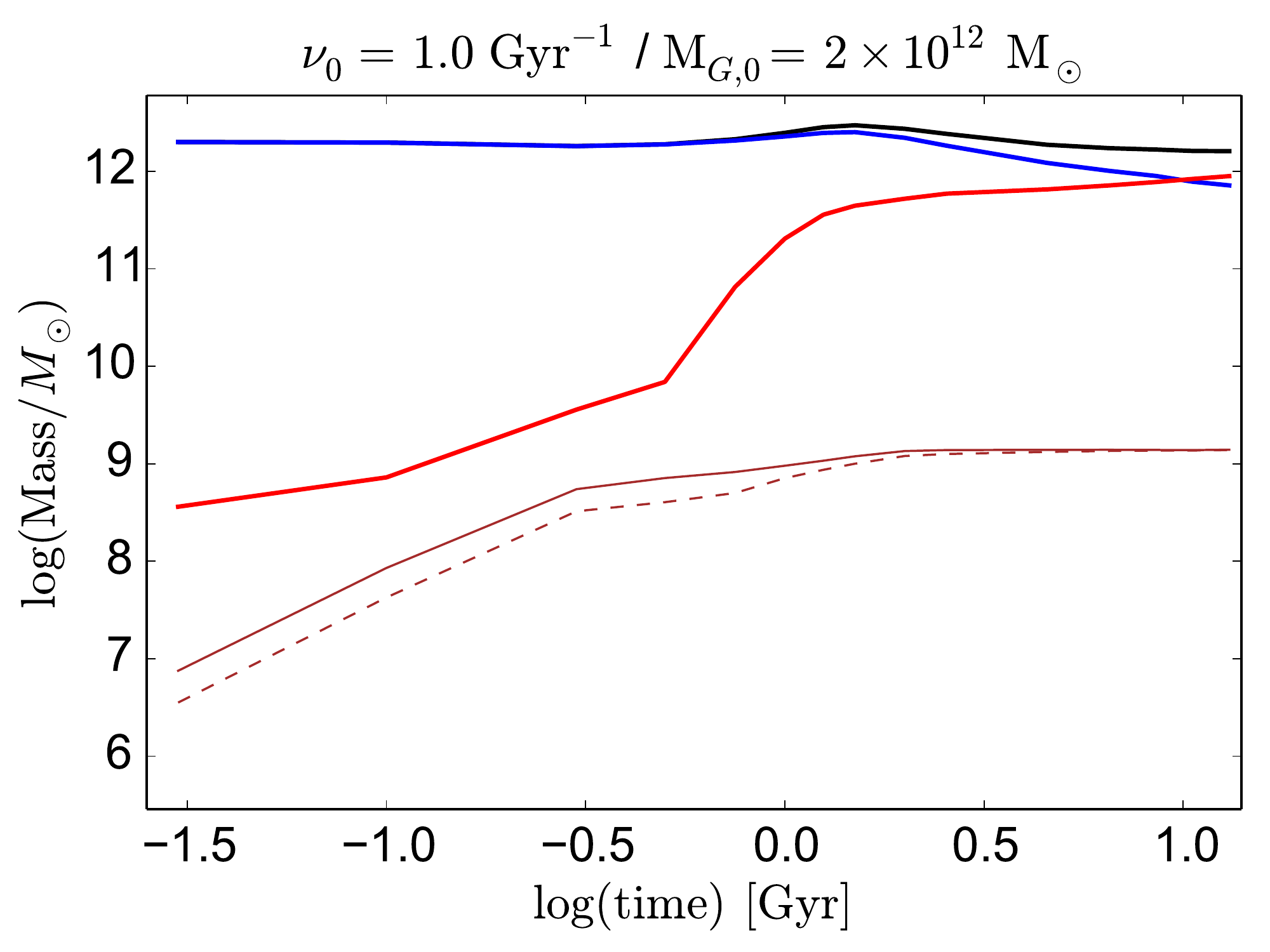}
\includegraphics[width=0.8 \columnwidth,angle=0]{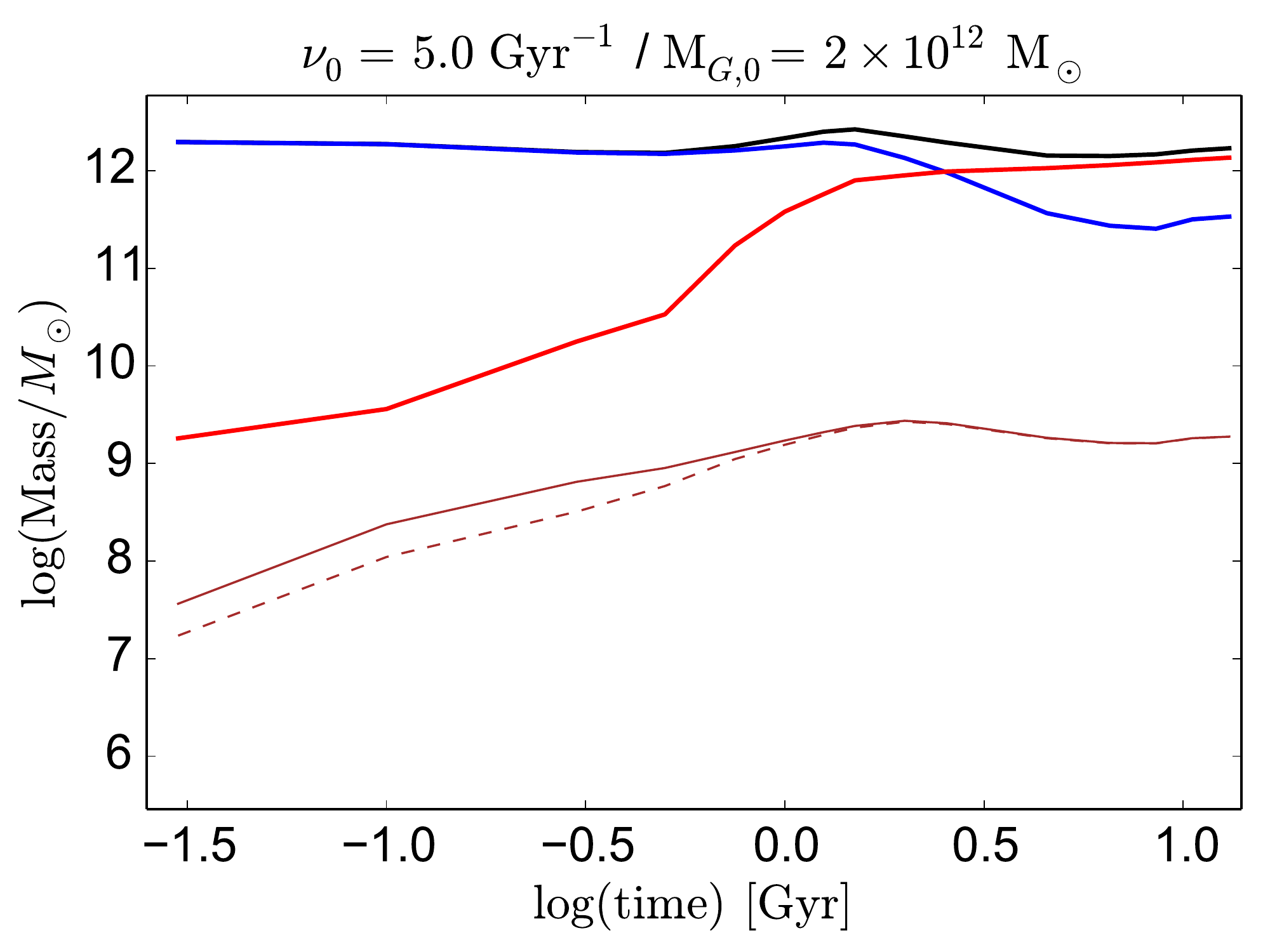}
\includegraphics[width=0.8 \columnwidth,angle=0]{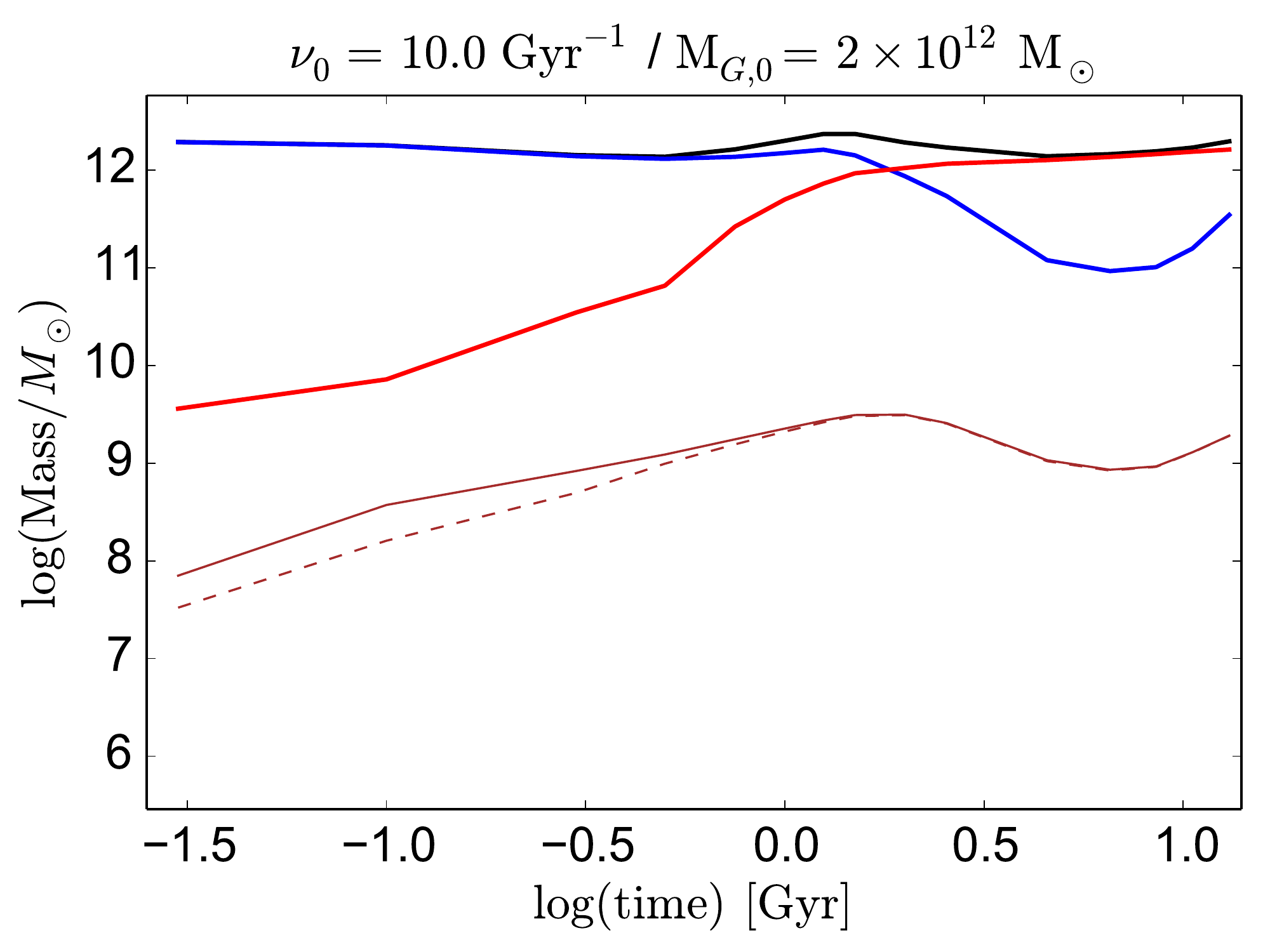}
\caption{The same as \ref{fig:Galcomp-1}, but for $M_{G,0} = 2 \times 10^{12} \, \mathrm{M}_\odot$ model.} 
\label{fig:Galcomp-5}
\end{center}
\end{figure*}

%%%%%%%%%%%%%%%%%%%%%%%%%%%%%%%%%%%%%%%%%%%%%%%%%%

% Don't change these lines
\bsp	% typesetting comment
\label{lastpage}
\end{document}